\documentclass[
    a4paper,
    reprint,
    superscriptaddress,
    amsmath,amssymb,
    aip,
]{revtex4-2}

\usepackage{cmap}
\usepackage[T1]{fontenc}
\usepackage{newtxtext}
\usepackage[varvw]{newtxmath}
\usepackage{microtype}
\usepackage{graphicx}
\graphicspath{{figures/}}
\usepackage{dcolumn}
\usepackage{bm}
\usepackage{hyperref}
\hypersetup{colorlinks=true,allcolors=blue}
\usepackage{mathtools}
\usepackage{upgreek}
\usepackage{cancel}
\usepackage{braket}
\usepackage{makecell}
\usepackage{orcidlink}
\allowdisplaybreaks

\renewcommand{\geq}{\geqslant}

\usepackage[margin=13.5mm,top=20mm,bottom=20mm]{geometry}
\usepackage[font=small,labelfont=bf]{caption}

\DeclareCaptionLabelFormat{customfig}{Fig.~#2.}
\DeclareCaptionLabelFormat{customtable}{Table~#2.}
\usepackage{xpatch}
\makeatletter
\xpatchcmd{\@ssect@ltx}{\@xsect}{\protected@edef\@currentlabelname{#8}\@xsect}{}{}
\xpatchcmd{\@sect@ltx}{\@xsect}{\protected@edef\@currentlabelname{#8}\@xsect}{}{}
\makeatother

\DeclareCaptionLabelFormat{extendeddata}{Extended Data Fig.~#2.}

\begin{document}

\title{
    Distinct weak antisymmetric interactions shape human brain functions
    as probability fluxes
}

\author{Yoshiaki Horiike\,\orcidlink{0009-0000-2010-2598}}
\email{yoshi.h@nagoya-u.jp}
\affiliation{
    Department of Applied Physics, Nagoya University, Nagoya, Japan
}
\affiliation{
    Department of Neuroscience, University of Copenhagen, Copenhagen, Denmark
}

\author{Shin Fujishiro\,\orcidlink{0000-0002-0127-0761}}
\affiliation{
    Fukui Institute for Fundamental Chemistry, Kyoto University, Kyoto, Japan
}

\date{21 October 2026}

\begin{abstract}
    The functional computation of the human brain is hypothesized to arise
    from the collective behaviour of the underlying neural network.
    A method based on an idea from equilibrium statistical mechanics has
    been applied to infer the underlying interaction network among the
    brain regions from whole-brain imaging,
    but recent studies have challenged the equilibrium assumption.
    Without equilibrium,
    how does the human brain encode its function in biased state transitions?
    Here we show that the probability fluxes,
    which quantify the irreversibility of state transitions,
    exhibit unique, task-dependent patterns in spontaneous and
    task-induced whole-cerebral-cortex activity.
    We then fit an Ising model with asymmetric interactions and reveal
    that the symmetric interactions among the brain regions are strong
    and task-independent,
    whereas the antisymmetric interactions are weak and task-dependent,
    and that the fitted model reproduces the observed fluxes for most
    tasks.
    Our results indicate that the human brain performs its functional
    computation by subtly modifying the antisymmetric part of the
    interactions among the brain regions,
    offering a new explanation for how a similar interaction network
    supports diverse brain functions.
    The method can be applied to the time-series data of other
    high-dimensional many-body systems to reveal the probability fluxes
    and infer the underlying interactions among components.
\end{abstract}

\maketitle

How the human brain works is mysterious
even with the accumulation of detailed knowledge~\cite{Crick1979}.
Interacting neurons exhibit emergent collective
behaviour~\cite{Anderson1972},
and for biological systems,
such collective behaviour is the \emph{functional}
property~\cite{Hopfield1994,Hopfield1999}:
the functional computation is performed as state transition
pathways in the state space of the dynamical
system~\cite{Hopfield1994,Hopfield1999}.
It might therefore be possible to understand the brain function as a
manifestation of the state transition dynamics emerging from the network
of interacting neurons.
Recent technological advancements have enabled one to record
the population activity of
neurons~\cite{VanEssen2013,Insel2013},
offering an opportunity to examine the functional computation of the brain
from observed neural activity.
Nevertheless, such large, complex, and high-dimensional data
are challenging to analyse,
and an explanatory rather than descriptive model is
demanded~\cite{Sejnowski2014}.

Ideas from statistical mechanics~\cite{Sethna2021} have been applied to analyse
neural data with the aim of explaining them from the underlying
interaction networks~\cite{Meshulam2025,Masuda2025}.
A series of pioneering studies~\cite{Schneidman2006,Watanabe2013}
infers the interaction parameters of the Ising spin system---a
paradigmatic model of collective behaviour---from neural data,
assuming the principle of maximum entropy,
which is valid for systems in equilibrium~\cite{Jaynes1957a}.
Maximum entropy modelling has later evolved into energy landscape
analysis~\cite{Masuda2025},
a systematic method to explain the neural data using the
intuitive metaphor of the landscape~\cite{Hopfield1982}.

In spite of the successes,
the fundamental assumption of these methods,
the equilibrium,
has been shown to be invalid~\cite{Nartallo-Kaluarachchi2026}
across a wide range of scales from neurons~\cite{Lynn2022a}
to the whole brain~\cite{Lynn2021,SanzPerl2021,Kringelbach2023,Deco2023}.
It has been hypothesized that because of the asymmetry of the underlying
interactions~\cite{Lynn2021,Nartallo-Kaluarachchi2024},
neural systems exhibit irreversible state transitions,
which break the time-reversal symmetry and cannot be extracted by
methods based on equilibrium.
In addition to the energy landscape,
a rotational field must also be considered
to describe the nonequilibrium dynamics of neural
systems~\cite{Battle2016,Nartallo-Kaluarachchi2026}.
Furthermore, one need to re-examine the correspondence between the brain
function and the brain activity pattern.
In the energy landscape picture~\cite{Masuda2025},
brain states fluctuate around the minima that correspond to the
patterns in which specific regions of the cortex are active.
Nonequilibrium dynamics drive the system away from such minima;
thus, no single pattern of brain activity is stable in the state space.
Then, is there a way to represent brain function beyond
static activity patterns?
Moreover, how does the brain perform its wide repertoire of functions with
nonequilibrium dynamics?

Here, we reveal that the state transition dynamics of the human brain
exhibit unique sequential patterns depending on tasks,
and such patterns emerge from a subtle modification in the antisymmetric
part of the interaction network among the brain regions.
Analysing the whole-cerebral-cortex activity data recorded through
blood-oxygen-level-dependent (BOLD)
functional magnetic resonance imaging (fMRI),
we show that the spatiotemporal patterns of human brain activity
exhibit irreversible state transitions and are unique depending on tasks.
The task-dependent pattern of state transition dynamics enables us to
characterize brain function through high-dimensional vectors.
To understand the structural origin of these task-dependent
state transition dynamics,
we investigate the underlying interaction network among the brain regions
by developing a method to infer the Ising spin system with asymmetric
interactions---the prototypical model to study nonequilibrium
collective behaviour on
networks~\cite{Hopfield1982,Lynn2021,Nartallo-Kaluarachchi2024}---from the data.
As a results,
we find that the symmetric part of the interaction matrix is similar
across tasks,
but the antisymmetric part of that is task-dependent:
the human brain function is shaped by the distinct weak antisymmetric
interactions among the brain regions.
We confirm that our model captures the task-dependent dynamics
observed in the data by reconstructing the state transition dynamics.

\section*{Task-dependent irreversible dynamics of the human brain}
We begin by developing the procedure to analyse the irreversible dynamics of
the human brain.
The data we employ (Methods)
contain the whole-cerebral-cortex activity
(divided into 100 cortical parcels~\cite{Schaefer2018})
of 590 healthy adults at rest and during seven cognitive and motor tasks,
recorded through BOLD fMRI as a part of the Human Connectome
Project~\cite{VanEssen2013}.
We analyse approximately the first three minutes of each recording,
over which the probability distribution is approximately stationary
(Extended Data Fig.~\ref{fig:fig-s01}).

To examine the task-dependent irreversible dynamics of human brains,
we perform the hypercubic probability flux analysis.
The probability flux is a measure of microscopic irreversibility---the
degree to which the system breaks the detailed balance---and
characterizes the nonequilibrium state transition~\cite{Zia2007};
probability flux analysis reveals the
nontrivial probability flux, that is, cycles, from data~\cite{Battle2016}.
Formally, the probability flux from state $\nu$ to $\mu$,
$\mathcal{J}(\mu, \nu)$, is defined as the difference between the
forward (from state $\nu$ to $\mu$) and backward
(from state $\mu$ to $\nu$) joint transition rates,
\begin{equation}
    \mathcal{J}(\mu, \nu)
    \coloneqq
    w(\mu \mid \nu) p(\nu) - w(\nu \mid \mu) p(\mu)
    \label{eq:probability-flux}
    ,
\end{equation}
where $w(\mu \mid \nu)$ is the (conditional) transition rate from
state $\nu$ to $\mu$,
and $p(\mu)$ is the probability of finding the system in state $\mu$.
To estimate the probability flux from the empirical data,
we reduce the spatial dimensionality of the data,
temporally binarize the data,
and estimate the probability flux on the resulting hypercubic state
space (\nameref{sec:hypercubic-probability-flux-analysis} in
\nameref{sec:methods})
as briefly described below.

\begin{figure*}[tb]
    \centering
    \includegraphics{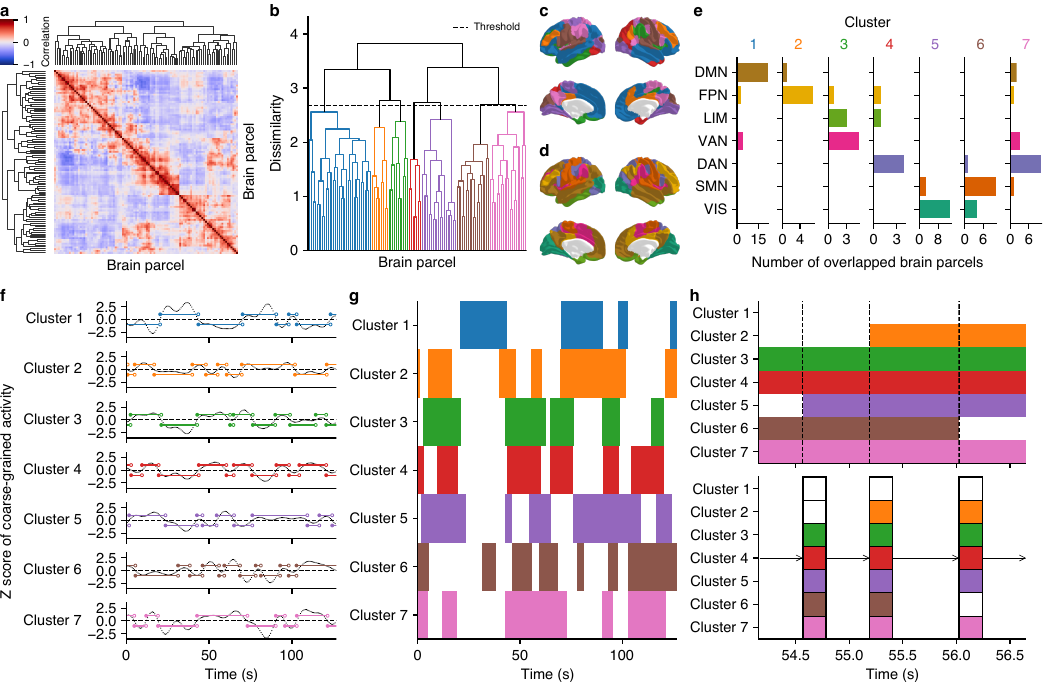}
    \caption{\textbf{Spatiotemporal coarse-graining of brain activity
        through hierarchical clustering and binarization.}
        \textbf{a},
        Hierarchically clustered correlation matrix (connectivity matrix)
        of the neural activity of brain parcels over all participants and
        tasks.
        \textbf{b},
        Dendrogram of panel a.
        The dashed line indicates the cluster threshold,
        and the leaf colour indicates the cluster assignment.
        Neural activity is averaged within each cluster to define the
        coarse-grained activity.
        \textbf{c},
        The resulting clusters of brain regions, coloured as in panel b.
        \textbf{d},
        Known functional clusters of brain regions, coloured as in panel e.
        \textbf{e},
        Alignment between our clusters (panels b and c) and the known
        functional clusters (panel d).
        DMN\@, default mode network;
        FPN\@, frontoparietal network;
        VIS\@, visual;
        SMN\@, somatomotor network;
        DAN\@, dorsal attention network;
        VAN\@, ventral attention network;
        LIM\@, limbic.
        \textbf{f},
        Example binarization of coarse-grained activity by thresholding at
        the mean.
        \textbf{g},
        Binarized activity of the coarse-grained brain regions of panel f.
        \textbf{h},
        Top: magnified view of panel g between 54 and 57~s,
        where dashed vertical lines indicate state transitions.
        Bottom: vector representation of the corresponding state
    transitions.}
    \label{fig:fig-01}
\end{figure*}

As the first step, we define the discrete state space of data by
spatially coarse-graining the brain regions following the correlation
among them~\cite{Meshulam2019}.
We perform hierarchical clustering~\cite{Mullner2011} of the
correlation matrix (connectivity matrix) over all participants and tasks
(Fig.~\ref{fig:fig-01}a)
and define seven clusters of brain regions by thresholding the dendrogram
(Fig.~\ref{fig:fig-01}b,c).
Our seven clusters roughly correspond to
the seven known functional clusters~\cite{ThomasYeo2011},
with minor misalignment
(Fig.~\ref{fig:fig-01}d,e).
We then average the neural activity within each cluster to obtain a time
series of coarse-grained activity for each cluster.

As the second step, we binarize the seven-dimensional time series
by applying a threshold to the time series itself or its
derivative with respect to time (Fig.~\ref{fig:fig-01}f),
so that each brain region is either active ($+1$) or inactive ($-1$)
at any time point
(Fig.~\ref{fig:fig-01}g).
The probability flux is then estimated
from the number of observed transitions between states~\cite{Battle2016}
(Fig.~\ref{fig:fig-01}h).

\begin{figure*}[tb]
    \centering
    \includegraphics{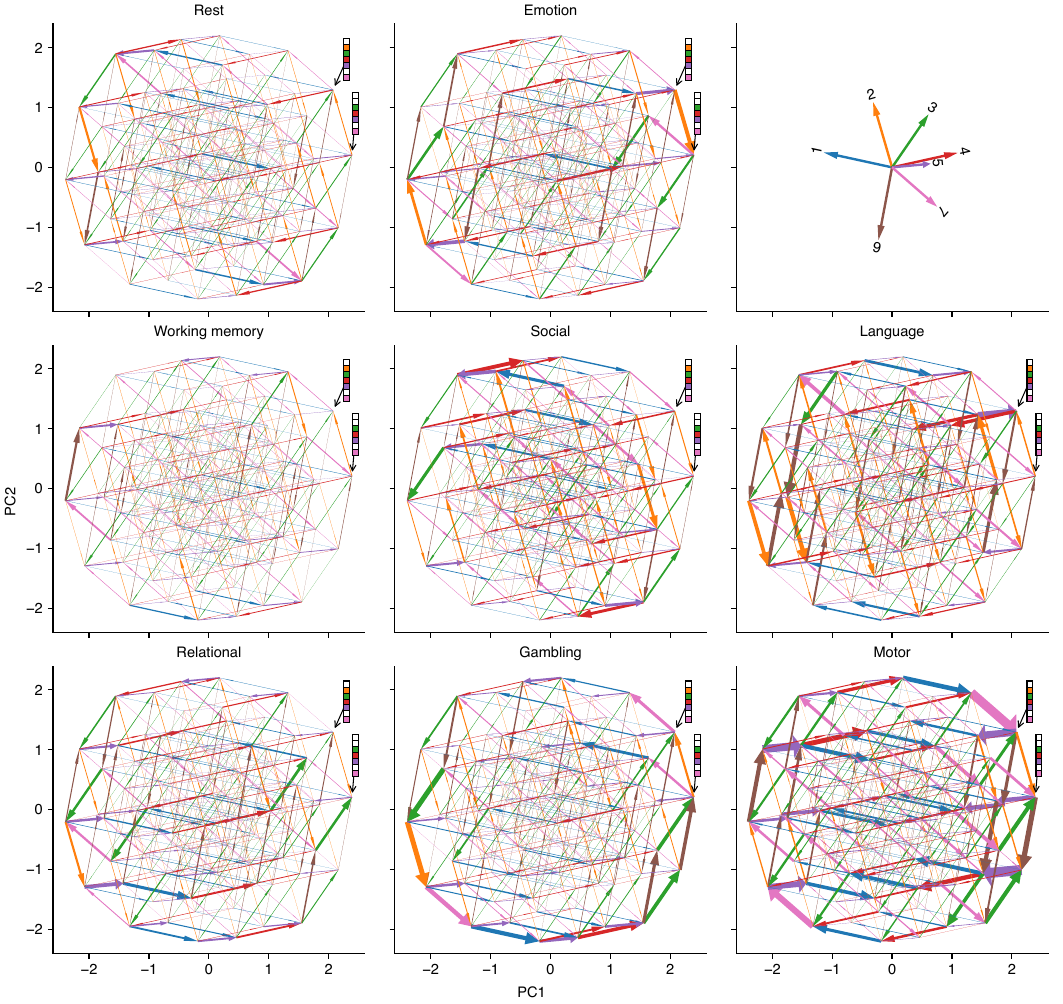}
    \caption{\textbf{Estimated probability fluxes of the human brain.}
        Each panel, except for the top right,
        shows the hypercubic probability flux diagram of one task.
        The top right panel shows the biplot vectors,
        the projected unit vectors of the original dimensions,
        coloured as the coarse-grained brain regions in
        Fig.~\ref{fig:fig-01}b,c.
        Each hypercubic arrow represents a probability flux:
        its width is proportional to the magnitude
        $|\mathcal{J}(\mu, \nu)|$ and its direction indicates the sign
        $\operatorname{sgn}\bm{(}\mathcal{J}(\mu, \nu)\bm{)}$;
        the arrow in the top right panel indicates $0.001$~s$^{-1}$.
        An arrow (anti)aligned with the biplot vector of the same colour
        represents a transition of that brain region from the inactive
        (active) to the active (inactive) state.
        Annotations show the Ising state vectors,
        where filled (empty) squares denote active (inactive) regions,
        coloured by cluster.
        The hypercubic vertices (binary states) are projected by PCA of the
    task-averaged probability distribution.}
    \label{fig:fig-02}
\end{figure*}

We begin with visually examining the resulting probability flux diagrams
(Fig.~\ref{fig:fig-02}).
Because the state space of binarized brain activity consists of the
vertices of a high-dimensional cube, or hypercube,
the state transitions correspond to the edges of the hypercube.
We visualize the probability flux on the hypercubic edges by projecting
the hypercube onto a two-dimensional plane with principal component
analysis (PCA)~\cite{Horiike2025}
(Fig.~\ref{fig:fig-02} and \nameref{sec:probability-flux-diagram}
in \nameref{sec:methods}).
Resulting diagrams show unique sequential patterns of brain activity
depending on the task being performed.
They exhibit cycles of probability flux,
which are characteristic of nonequilibrium steady
states~\cite{Hill1966,Schnakenberg1976,Zia2007},
consistent with previous
studies~\cite{Vidaurre2017,Cornblath2020,Lynn2021,VanEs2025}.
The tasks differ in the size, strength, and number of cycles of
probability flux.
As suggested previously~\cite{Lynn2021},
our results indicate that the task-dependent unique pattern of probability flux
is closely related to the cognitive and motor processes involved in each task.

These hypercubic probability flux diagrams indicate that the
biased state transition pattern arises from the
task-dependent sequential pattern of brain activity.
For example,
the probability flux involving cluster 2 in
Fig.~\ref{fig:fig-01}b,c (related to frontoparietal)
exhibits a strong magnitude in the social and language tasks,
while the probability flux involving cluster 1
(related to default mode)
exhibits a strong magnitude in the motor task.
The sequence of brain state transitions, or order of
activation (or inactivation) of brain regions, is thus distinct across the
tasks being performed,
which suggests that such a pattern represents human brain function.
We confirm that
the task-specific patterns are seen in the PCA-projected state
space (Extended Data Fig.~\ref{fig:fig-s02}),
and the task-dependent unique pattern emergence is independent of the method:
usign the known functional
clusters~\cite{ThomasYeo2011} (Extended Data Fig.~\ref{fig:fig-s03}),
changing the number of clusters (Extended Data Fig.~\ref{fig:fig-s04}),
or changing the discretization method (Extended Data Fig.~\ref{fig:fig-s05}).
Diagrams using other principal components (PCs) are shown in
Extended Data Fig.~\ref{fig:fig-s06}.

\begin{figure*}[tb]
    \centering
    \includegraphics{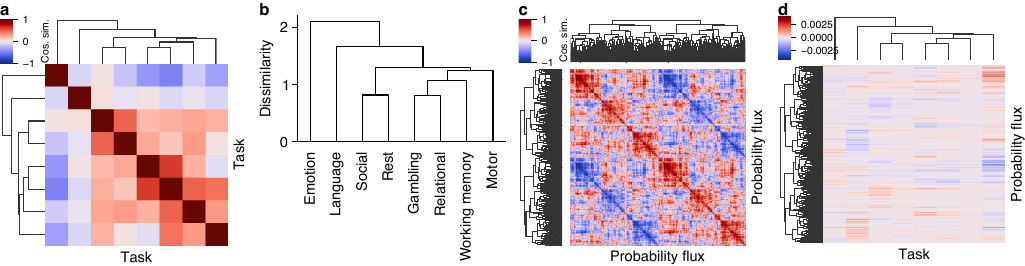}
    \caption{\textbf{Clustering brain function using probability flux patterns.}
        \textbf{a},
        Hierarchically clustered cosine similarity among tasks based on
        probability flux.
        \textbf{b},
        Dendrogram of panel a.
        \textbf{c},
        Hierarchically clustered cosine similarity among probability fluxes
        across tasks.
        \textbf{d},
        Probability flux matrix sorted by the hierarchical clustering of
        panels a and c.
        Panels c and d show the fluxes of both directions;
        because the flux is antisymmetric,
        $\mathcal{J}(\mu, \nu) = - \mathcal{J}(\nu, \mu)$,
    panel c contains two identical blocks and panel d is antisymmetric.}
    \label{fig:fig-03}
\end{figure*}

\section*{Human brain functions as patterns of probability flux}
We then compare the task-specific patterns of probability flux
to investigate the relation among tasks.
Hierarchical clustering of the tasks based on their probability flux
(Fig.~\ref{fig:fig-03}a,b)
reveals a large correlated group consisting of rest and five
tasks (social, gambling, relational, working memory, and motor),
whereas two tasks (emotion and language) are weakly or negatively
correlated with the others.
This result indicates that the emotion and language tasks may involve
distinct information processing compared to the other tasks.
Results for different numbers of clusters are given in
Extended Data Fig.~\ref{fig:fig-s07} and
Supplementary Note~1.

To characterize each state transition, or directed hypercubic edge,
we define an eight-dimensional vector
whose elements are the probability fluxes of the tasks,
and we perform hierarchical clustering of these vectors
(Fig.~\ref{fig:fig-03}c).
There are several correlated clusters,
indicating that there are directed hypercubic edges sharing the
probability flux pattern, which might be the fundamental
state transitions of the human brain.

In Fig.~\ref{fig:fig-03}d, we visualize the
probability fluxes of each task as a list sorted by the results of the
hierarchical clustering of Fig.~\ref{fig:fig-03}a,c.
Its columns show that the probability fluxes exhibit patterns
characteristic of the tasks as visualized in Fig.~\ref{fig:fig-02},
with minor similarities between the tasks,
and its rows show that some groups of probability fluxes are similar
across tasks.
Together, these results suggest that there are shared dynamics in the
probability fluxes across tasks, although the major features of the
probability fluxes are still task-dependent---which might reflect the
separation of fundamental task-independent dynamics and specific
task-dependent dynamics.

\section*{Structural origin of task-dependent irreversible dynamics}
To reveal the mechanism of such task-dependent dynamics,
we investigate the structural origin of the observed probability flux.
Because methods assuming equilibrium are incapable of capturing the
probability flux, we present a method
(\nameref{sec:model-fitting} in \nameref{sec:methods})
to infer the Ising spin system with asymmetric (or nonreciprocal)
pairwise interactions from the transition rates
$\{w(\mu \mid \nu)\}$,
assuming the stationarity of the probability distribution and a
stochastic model of state transition based on a
(pseudo-)Hamiltonian.

\begin{figure*}[tb]
    \centering
    \includegraphics{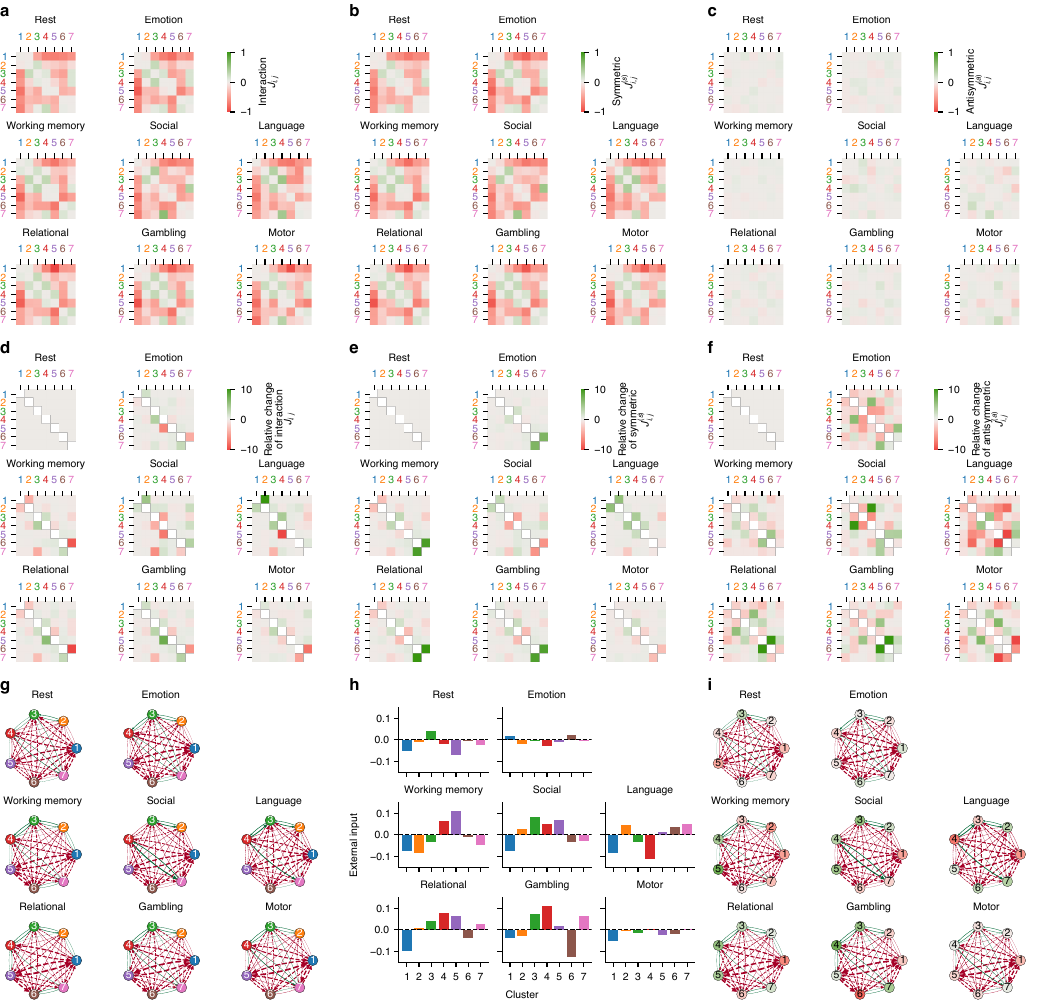}
    \caption{\textbf{Inferring the Ising spin system from the
        estimated transition rates.}
        \textbf{a},
        Inferred effective interaction matrix $\beta\bm{J}$,
        where the inverse temperature $\beta$ is a nuisance parameter.
        \textbf{b},
        Its symmetric part
        $
        \beta
        \bm{J}^{\mathrm{(s)}}
        \coloneqq
        \beta
        (1/2)
        (\bm{J} + \bm{J}^\top)
        $.
        \textbf{c},
        Its antisymmetric part
        $
        \beta
        \bm{J}^{\mathrm{(a)}}
        \coloneqq
        \beta
        (1/2)
        (\bm{J} - \bm{J}^\top)
        $.
        \textbf{d--f},
        Relative change from rest to task of the interaction matrix (d),
        its symmetric part (e), and its antisymmetric part (f),
        for example,
        $
        (
            \beta J_{i, j}^{\mathrm{(task)}}
            -
            \beta J_{i, j}^{\mathrm{(rest)}}
        )
        /
        \beta J_{i, j}^{\mathrm{(rest)}}
        $
        for panel d.
        \textbf{g},
        Effective interaction network of panel a.
        Node colours match the clusters in Fig.~\ref{fig:fig-01}b,c;
        edge width is proportional to $|J_{i, j}|$;
        green solid and red dashed lines indicate
        $\operatorname{sgn}(J_{i, j})=+1$ and $-1$, respectively.
        \textbf{h},
        Inferred effective external input $\beta \bm{h}$,
        coloured as the clusters in Fig.~\ref{fig:fig-01}b,c.
        \textbf{i},
        Effective external input shown as node colour on the network of
        panel g;
        darker green (red) indicates a stronger positive (negative) input.
        The inferred rate constants $A$ are
        0.096 (rest), 0.116 (emotion), 0.099 (working memory), 0.101
        (social), 0.108 (language), 0.112 (relational), 0.105
    (gambling), and 0.115 (motor) $\mathrm{s}^{-1}$.}
    \label{fig:fig-04}
\end{figure*}

In Fig.~\ref{fig:fig-04},
we infer the (effective) interaction networks
$\{\bm{J} \in \mathbb{R}^{N\times N}\}$ and
effective external inputs $\{\bm{h} \in \mathbb{R}^N\}$ from the
estimated transition rates of each task,
assuming the Arrhenius-type transition rate
(\nameref{sec:model-fitting} in \nameref{sec:methods}).
The element $J_{i, j}$ of the inferred effective interaction matrix
(Fig.~\ref{fig:fig-04}a) represents the effective interaction strength
from cluster $j$ to cluster $i$, and its sign is the type of interaction;
positive means excitatory and negative means inhibitory.
There is no apparent difference among the inferred
effective interactions across tasks.
The majority of the interactions are negative (Fig.~\ref{fig:fig-04}a),
indicating that the brain clusters tend to inhibit each other's activity,
which is consistent with the hierarchical clustering of the
original unbinarized data (Fig.~\ref{fig:fig-01}a).

The task-dependent differences in the inferred effective interactions are
revealed by the decomposition of the interaction matrix into its
symmetric and antisymmetric parts:
\begin{equation}
    \bm{J}
    =
    \bm{J}^{\mathrm{(s)}}
    +
    \bm{J}^{\mathrm{(a)}}
    ,
\end{equation}
where
$
\bm{J}^{\mathrm{(s)}}
\coloneqq
(1/2)
(\bm{J} + \bm{J}^\top)
$
and
$
\bm{J}^{\mathrm{(a)}}
\coloneqq
(1/2)
(\bm{J} - \bm{J}^\top)
$
(Fig.~\ref{fig:fig-04}b,c).
The symmetric part does not exhibit significant differences across tasks,
while the antisymmetric part shows more variability across tasks.
The antisymmetric part of the interaction matrix is the source of the
nonequilibrium steady state~\cite{Lynn2021,Nartallo-Kaluarachchi2024},
indicating that the antisymmetric interactions are modified by the
tasks being performed.
Moreover, the magnitude of the elements of the antisymmetric part seems
considerably smaller than that of the symmetric part,
indicating that the antisymmetric interactions may be easily modified to meet
task demands:
the brain might achieve efficient information processing by slightly
modifying the antisymmetric part of the interaction matrix based on
the task being performed.

The interaction network (Fig.~\ref{fig:fig-04}g)
shows there are many closed loops of interactions,
which leads to geometrical frustration~\cite{Mezard1986a}.
Although a ground state of a pseudo-Hamiltonian is ill-defined for an
asymmetric interaction matrix,
the probability fluxes appear to circulate around low-pseudo-energy states
(Fig.~\ref{fig:fig-02}).
There are particularly strong negative interactions
involving cluster 1 (related to the default mode),
which is expected to inhibit the activity of other brain regions,
consistent with the known features of the default mode~\cite{Raichle2006a}.

The effective external input (Fig.~\ref{fig:fig-04}h,i)
differs largely across tasks,
which may reflect the task-dependent external stimulus from outside the
cortex.
Nevertheless, the majority of tasks share a strong negative field on
brain cluster 1
(related to the default mode; cf. Fig.~\ref{fig:fig-01}c--e).
Together with the negative interaction involving cluster 1
(Fig.~\ref{fig:fig-04}a,g),
this indicates that the default mode is suppressed by the external
stimulus,
which leads to the activation of other brain regions.
Note that the Glauber-type transition rate~\cite{Glauber1963}
can also reproduce the probability flux,
but the rate constant differs largely among tasks
(Extended Data Figs.~\ref{fig:fig-s08} and~\ref{fig:fig-s10},
    Supplementary Fig.~1
and Supplementary Note~2).
Allowing the rate constant to depend on the brain cluster
leaves the inferred interaction network virtually unchanged
for both transition rates
(Supplementary Note~8
and Supplementary Figs.~2 and~3).

\begin{figure*}[tb]
    \centering
    \includegraphics{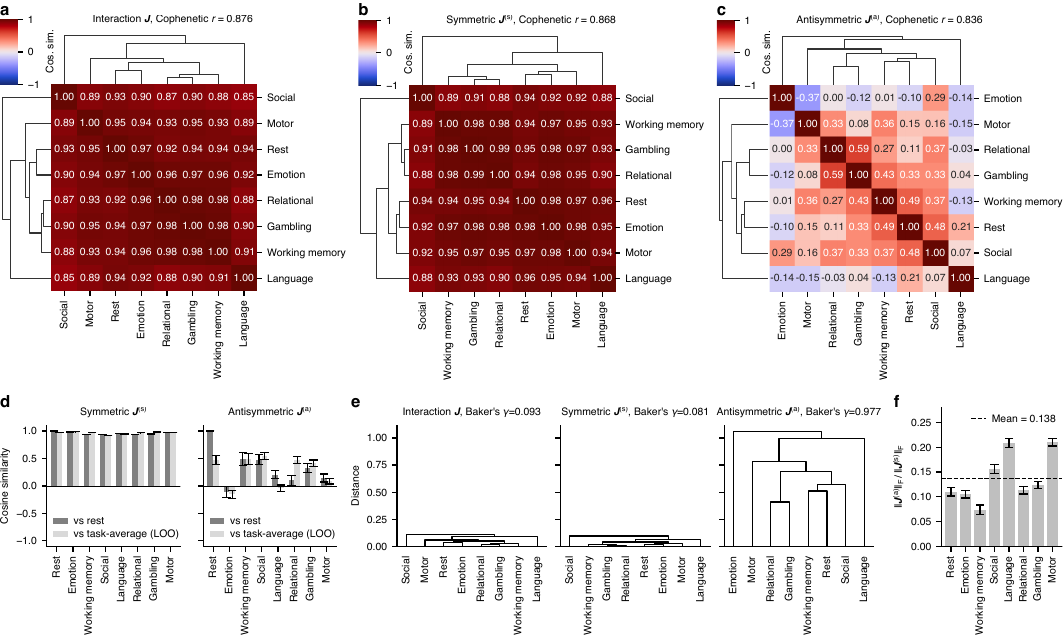}
    \caption{\textbf{Comparing the inferred effective interaction
        network structure among tasks.}
        \textbf{a--c},
        Pairwise cosine similarity between tasks of the vectorized
        interaction matrices (a), their symmetric parts (b),
        and their antisymmetric parts (c), sorted by hierarchical
        clustering.
        \textbf{d},
        Cosine similarity between the interaction matrices of each task and
        rest, and of each task and the task average (leave-one-out).
        \textbf{e},
        Dendrograms of panels a--c.
        Baker's gamma coefficient between each dendrogram and that of the
        probability flux (Fig.~\ref{fig:fig-03}b) is written next to each
        title.
        \textbf{f},
        Ratio of the Frobenius norm of the antisymmetric part to that of the
        symmetric part,
        $
        \|\bm{J}^{\mathrm{(a)}}\|_{\mathrm{F}}
        /
        \|\bm{J}^{\mathrm{(s)}}\|_{\mathrm{F}}
        $,
        for each task.
        In panels d and f,
    error bars indicate the bootstrap standard deviation.}
    \label{fig:fig-05}
\end{figure*}

To quantify the observation on the interaction network,
we compute the pairwise cosine similarity between the vectorized
interaction matrices of different tasks~\cite{Cole2014}
(\nameref{sec:matrix-comparison} in \nameref{sec:methods}).
The interaction network is highly similar across tasks,
with cosine similarity greater than 0.8 for all pairs (Fig.~\ref{fig:fig-05}a),
and this similarity arises from the symmetric part (Fig.~\ref{fig:fig-05}b).
On the other hand, the antisymmetric part shows low similarity for most
pairs, with cosine similarity less than 0.5 (Fig.~\ref{fig:fig-05}c).
The symmetric part is also highly similar
between rest and each task, and between the task average and rest,
with cosine similarity greater than 0.95 (Fig.~\ref{fig:fig-05}d).
The cophenetic correlation coefficient,
which measures how well a dendrogram preserves the pairwise distances,
is relatively high for the interaction matrix (0.876) and its
symmetric part (0.868)
but lower for the antisymmetric part (0.836),
indicating that the antisymmetric part is more task-dependent.
Moreover, Baker's gamma coefficient between the dendrograms of
Fig.~\ref{fig:fig-05}e and that of the probability flux
(Fig.~\ref{fig:fig-03}b) is 0.093 for the interaction matrix and 0.081
for the symmetric part,
but 0.977 for the antisymmetric part,
indicating that the structure of the probability flux diagrams is
governed by the antisymmetric part.
The ratio of the Frobenius norms
$
\|\bm{J}^{\mathrm{(a)}}\|_{\mathrm{F}}
/
\|\bm{J}^{\mathrm{(s)}}\|_{\mathrm{F}}
$
(Fig.~\ref{fig:fig-05}f) has a mean of approximately 0.14
and is largest, approximately 0.2, for the language and motor tasks:
the antisymmetric part of the interaction matrix is indeed weaker
than the symmetric part.
Although coarse-graining reduces the degree of
nonequilibrium~\cite{Egolf2000,Esposito2012,Nartallo-Kaluarachchi2026a},
and the asymmetry ratio decreases as we decrease the number of brain clusters
(Supplementary Fig.~4),
the system exhibits nonequilibrium dynamics at the scale of brain regions,
which is the biologically meaningful scale to discuss brain structure
and function.

\begin{figure*}[tb]
    \centering
    \includegraphics{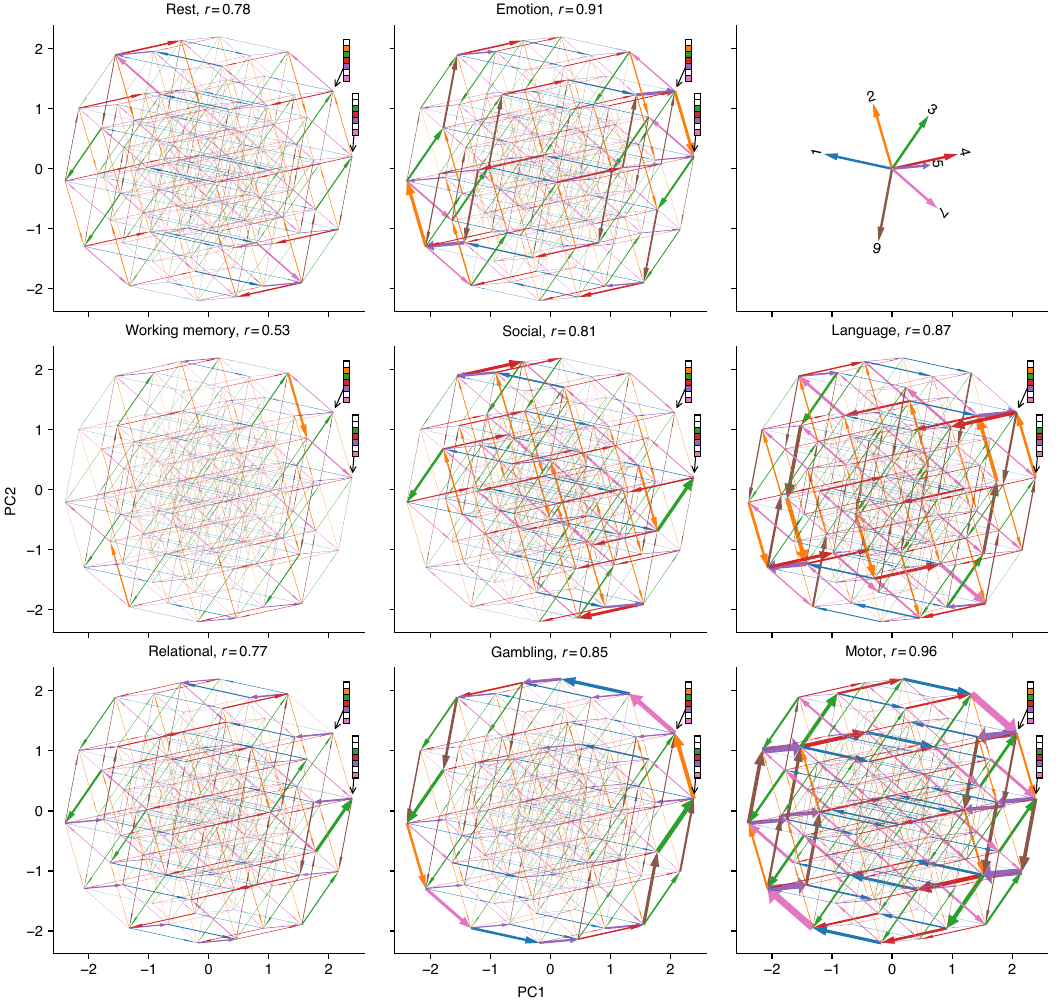}
    \caption{\textbf{Probability fluxes reconstructed from the
        inferred Ising spin system.}
        Same as Fig.~\ref{fig:fig-02},
        but the probability fluxes are calculated from the inferred
        interaction matrix $\beta \bm{J}$, effective external input
        $\beta \bm{h}$, and rate constant $A$,
        using the same PC loadings as in Fig.~\ref{fig:fig-02}.
        The correlation coefficient $r$ between the original and
    reconstructed probability fluxes is written next to each task name.}
    \label{fig:fig-06}
\end{figure*}

To validate our inferred model,
we reconstruct the probability flux diagrams (Fig.~\ref{fig:fig-06})
and compare them with the empirical observations.
The empirical and reconstructed probability fluxes agree quantitatively
for most tasks,
with Pearson coefficient larger than 0.7,
except for the working memory task.
This may be improved by introducing more complexity to the model
(Supplementary Note~3),
but the overall features of the probability flux diagrams
are captured by the method we present in this study.
By performing the parameter inference while fixing the interaction network
among tasks
(Supplementary Figs.~5 and~6),
we confirm that the probability flux diagrams are largely reconstructed from
the inferred effective interaction network rather than the effective
external input.

\section*{Brain function as irreversible state-transition patterns}
Large-scale recordings of the human
brain~\cite{VanEssen2013,Insel2013} demand
explanatory methods to understand the neural data~\cite{Sejnowski2014},
and energy landscape analysis~\cite{Watanabe2013,Masuda2025} has
been a good candidate to explain brain function and structure from data.
Recent studies, however, show that the human brain operates out of
equilibrium~\cite{Lynn2021,Nartallo-Kaluarachchi2026},
challenging the fundamental assumptions of such analysis.
In this study, through hypercubic probability flux analysis of fMRI data,
we show that the human brain function is characterized by the
sequential pattern of state transitions,
and that such patterns emerge from the antisymmetric part of the interaction
network between the brain regions.
The results indicate that the human brain function is defined by
a dynamic sequential pattern rather than a single static pattern,
consistent with the idea of wave-like motifs~\cite{Foster2024} or
cyclical patterns~\cite{Vidaurre2017,Cornblath2020,Lynn2021,VanEs2025}.

Our findings may provide a different view of cognition as
sequential and dynamic rather than a single static computation and
representation~\cite{Barack2021}.
The qualitative feature of the inferred interaction networks is consistent with
previous studies~\cite{Cole2014,Cole2021},
which show that functional connectivity during tasks is a
combination of strong intrinsic connectivity and weak task-evoked
connectivity.
Moreover, as irreversible dynamics enable the error-correction
mechanism in kinetic proofreading~\cite{Hopfield1974},
they might also enable the functional computation of the human brain.
The human brain constantly consumes a large portion of the
total body energy~\cite{Rolfe1997}
and operates out of equilibrium~\cite{Schrodinger1992},
yet it does not significantly increase its energy demand when it
engages in tasks~\cite{Raichle2006a}.
Subtle modification of the interaction network to represent the
brain function as irreversible dynamics might be a way to operate within
such constraints of energy consumption.

As a problem of physics,
we show that the probability flux (how the system breaks detailed balance)
rather than the entropy production rate
(which quantifies the breaking of detailed balance over the entire
state space; \nameref{sec:entropy-production-rate} in \nameref{sec:methods})
carries richer organization and information about the nonequilibrium dynamics
(Fig.~\ref{fig:fig-02}, Supplementary Note~4
and Supplementary Fig.~7).
The probability traffic,
the total rate of transitions between two states regardless of their
direction,
also carries little task-specific information
(Extended Data Fig.~\ref{fig:fig-s17}, Supplementary Note~5
and Supplementary Figs.~8--14).
The diagrammatic description of nonequilibrium steady states may provide
not only an alternative technique to calculate the entropy
production rate~\cite{Hill1966,Schnakenberg1976}
but also unexplored aspects of stochastic
thermodynamics~\cite{VanDenBroeck2015}.
Indeed, the results of quantifying the degree of nonequilibrium
depend largely on the method, and there is no consensus
on the ranking of irreversibility of the seven
tasks~\cite{Nartallo-Kaluarachchi2026} of Human Connectome
Project~\cite{VanEssen2013},
so observing the state transition dynamics~\cite{Battle2016} may
capture something ignored by such quantities~\cite{Anscombe1973,Matejka2017}.
Unlike maximum entropy
modelling~\cite{Schneidman2006,Meshulam2025} or
energy landscape analysis~\cite{Watanabe2013,Masuda2025},
our approach focuses on the
probability flux---the hallmark of nonequilibrium steady
states~\cite{Zia2007}---rather than on static correlations,
and it can be applied to reveal nonequilibrium aspects of other
high-dimensional or many-body systems.

\bibliography{references}

\section*{Methods}
\label{sec:methods}
\subsection*{BOLD fMRI data}
\label{sec:data}
We analysed the previously collected~\cite{VanEssen2013,Smith2013,Barch2013} and
preprocessed~\cite{Lynn2021} BOLD fMRI data.
The data were collected from 590 healthy adults, and for each
participant,
the recordings were performed at rest and during seven tasks
(emotion, working memory, social, language, relational, gambling, and motor).
The cortex is parcelled into 100 regions following
previously published work~\cite{Schaefer2018}.
The BOLD fMRI data contain two phase encoding directions,
specifically left-to-right (LR) and right-to-left (RL).
The data duration for each task is different,
so we analyse the first 176 time points of each task, matching
the shortest task duration (emotion task), to
avoid systematic differences due to varying task lengths~\cite{Lynn2021}.
The data were collected by the Human Connectome Project under protocols
approved by the institutional review board of Washington University
in St.~Louis,
and all participants provided written informed consent~\cite{VanEssen2013}.
We used the data in accordance with the WU-Minn HCP Open Access Data Use
Terms.

\subsection*{Hypercubic probability flux analysis (HyPFA)}
\label{sec:hypercubic-probability-flux-analysis}
HyPFA combines the `Probability flux analysis',
`Spatial coarse-graining through hierarchical clustering', `Temporal
coarse-graining through time series binarization',
and `Hypercubic probability flux diagram' below.

\subsection*{Probability flux analysis}
\label{sec:probability-flux-analysis}
Under the assumption of stationary dynamics,
the probability flux can be estimated from the time series of
state (phase) of the target system~\cite{Battle2016}.
The procedure of probability flux analysis~\cite{Battle2016} is as follows.
First, the discrete state of the system is defined by coarse-graining,
resulting in a coarse-grained state space.
Thus, the original trajectory data are converted to
a discrete state time series.
Next, the joint transition rate is estimated by assuming a steady state,
that is,
$(\mathrm{d}/\mathrm{d} t) p(\mu; t) = 0$ for all $\mu$.
The probability of finding the system in state $\mu$ at time $t$ is
denoted as $p(\mu; t)$, which satisfies the normalization
$\sum_\mu p(\mu; t) = 1$.
The joint transition rate
$w(\mu \mid \nu) p(\nu; t)$
from state $\nu$ to $\mu$ at time $t$
is the product of the probability of existence
$p(\nu; t) \in [0, 1]$
and the transition rate from state $\nu$ to $\mu$,
$w(\mu \mid \nu) \in \mathbb{R}_{\geq 0}$.
Thus, in the steady state, the joint transition rate corresponds to
the number of transitions from one state to the other per unit time,
that is,
\begin{equation}
    w(\mu \mid \nu) p(\nu; t)
    \approx
    \frac{1}{\tau}
    n(\mu, \nu)
    ,
\end{equation}
where $n(\mu, \nu)$ is the number of transitions from state $\nu$ to $\mu$,
and $\tau$ is the time duration of the observation.
In coarse-grained state space,
the original data are discretized into coarse-grained states, and
the number of transitions from one state to the other can be counted.
Finally, the probability flux in the steady state is estimated as the
difference between the forward joint transition rate and backward
joint transition rate (equation~\eqref{eq:probability-flux}),
that is,
\begin{equation}
    \mathcal{J}(\mu, \nu)
    \approx
    \frac{1}{\tau}
    [n(\mu, \nu) - n(\nu, \mu)]
    \label{eq:probability-flux-estimated}
    .
\end{equation}
In a similar manner,
the probability traffic can also be estimated as the sum of the
forward and backward joint transition rates
(Supplementary Note~5).

For a finite observation time $\tau$, the probability flux can be affected by
finite sampling effects; thus,
it is vital to consider the statistical significance of
the estimated flux values.
Bootstrapping~\cite{Efron1979} is performed to assess the variability of
the estimated flux values.
Assuming a Markov process,
we perform trajectory bootstrapping~\cite{Battle2016},
which resamples the observed state transitions.
Consider the coarse-grained state time series with $L$ time points,
$\{\mu_1, \mu_2, \ldots, \mu_L\}$.
The state transition is recorded as a state transition matrix of size
$3\times (L-1)$,
\begin{equation}
    \bm{K}
    \coloneqq
    \begin{bmatrix}
        \mu_1 & \mu_2 & \cdots & \mu_{L-1} \\
        \mu_2 & \mu_3 & \cdots & \mu_{L} \\
        \Delta t_{2, 1} & \Delta t_{3, 2} & \cdots & \Delta t_{L, L-1}
    \end{bmatrix}
    \label{eq:state-transition-matrix}
    ,
\end{equation}
where $\mu_i$ is the coarse-grained state at $i$th time point,
and $\Delta t_{i, j} \coloneqq t_i - t_j$ is the time difference
between the $i$th and $j$th time points,
that is, the time spent in state $\mu_j$.
Each column of the state transition matrix corresponds to
the state transition from a specific time point.
From this matrix, the joint transition rate is calculated as
\begin{equation}
    w(\mu \mid \nu) p(\nu)
    \approx
    \frac{1}{\sum_{j=1}^{L-1} K_{3, j}}
    \sum_{j=1}^{L-1}
    \delta_{\mu, K_{2, j}}
    \delta_{\nu, K_{1, j}}
    ,
\end{equation}
where $\delta_{x, y}$ is the Kronecker delta.
The probability flux is calculated from
equation~\eqref{eq:probability-flux-estimated}.
Similarly,
the stationary distribution is estimated from the state transition matrix as
\begin{equation}
    p(\nu)
    \approx
    \frac{1}{\sum_{j=1}^{L-1} K_{3, j}}
    \sum_{j=1}^{L-1}
    K_{3, j} \delta_{\nu, K_{1, j}}
    \label{eq:stationary-distribution-estimated}
    .
\end{equation}
Because of the Markov assumption, each column of the state transition
matrix is independent of the others.
Therefore, we can resample each column of the state transition
matrix $\bm{K}$
randomly to create a bootstrapped state transition matrix of
the same size.
The value of interest is calculated from the bootstrapped state
transition matrix,
and the error is estimated from the ensemble of realizations.
The error of the probability flux is estimated as the standard
deviation over
the bootstrapped trajectories.

\subsection*{Examining the stationarity of probability distribution}
\label{sec:examining-stationarity}
The assumption behind the probability flux analysis
(`Probability flux analysis') is
that the probability distribution is stationary.
To validate this assumption,
we calculate the probability change of state $\mu$,
$\Delta p(\mu)$,
from the estimated probability fluxes~\cite{Lynn2021}:
\begin{equation}
    \Delta p(\mu)
    \coloneqq
    \sum_{\nu=1}^{2^N}
    [
        w(\mu \mid \nu) p(\nu)
        -
        w(\nu \mid \mu) p(\mu)
    ]
    \label{eq:probability-change}
    .
\end{equation}
This is zero in the steady state; thus, we investigate the distribution of
$\Delta p(\mu)$ to validate the stationarity of the probability
distribution $p(\mu)$.
In Extended Data Fig.~\ref{fig:fig-s01}, we show the distribution of
$\Delta p(\mu)$ for each task.
The distribution is sharply peaked around zero for all tasks,
indicating that the stationarity assumption is reasonable.

\subsection*{Spatial coarse-graining through hierarchical clustering}
\label{sec:coarse-graining}
The neural data have high dimensionality, and defining the coarse-grained
state space is not straightforward.
Previous work used PCA and $k$-means clustering~\cite{Lynn2021} to
reduce the dimensionality of the data and that of the state space.
We instead reduce the dimensionality of the data and the state space
at once, starting from empirical correlations rather than the actual
interaction network, which is unknown
beforehand~\cite{Meshulam2019}
(Supplementary Note~3).

To coarse-grain the human brain neural network system,
we perform hierarchical clustering~\cite{Mullner2011} of the
brain regions.
Note that, as mentioned in ref.~\cite{Meshulam2019},
there are many possible methods to perform
coarse-graining~\cite{Gabrielli2025},
and we emphasize that hierarchical clustering is just one of them.
We begin with the correlation matrix between the brain regions
calculated over all types of scans, tasks, participants, and time points.
The data dimension is $N = 100$ and its length is
$M = 2 \times 8 \times 590 \times 176$.
Then, using the correlation matrix,
we perform hierarchical clustering of the brain regions
using the unweighted pair group method with arithmetic mean (UPGMA) with
the Euclidean distance metric.
To obtain a clear leaf structure of the dendrogram, we perform optimal leaf
ordering~\cite{Bar-Joseph2001} of the linkage matrix.
We then threshold the dendrogram by its dissimilarity to define the
clusters of brain regions.
The neural signal is averaged over the brain regions within each cluster.

\subsection*{Temporal coarse-graining through time series binarization}
\label{sec:binarization}
To discretize and coarse-grain the state space,
we apply a binarization technique to the time series data.
We employ the symbolic string transformation~\cite{Kurths1996} to
convert the continuous time series into the discrete symbol sequence.
We first approximate the original $N$-dimensional data of
$M$ time points, with each row standardized,
$\bm{X} \in \mathbb{R}^{N\times M}$ by
$N$ temporally continuous spline functions written as a vector
$
\bm{f}(t)
\coloneqq
[
    f_1(t) \; \cdots \; f_N(t)
]^\top
\in \mathbb{R}^{N}
$.
Cubic spline interpolation, whose spline functions can
be differentiated twice, is employed for the time series of each dimension.
Using the spline functions, we define three types of
binarization distinguished by the order of differentiation.
We obtain the event time points
(intersection points, stationary points, or inflection points)
of the $i$th spline function through differentiation,
\begin{equation}
    \{t_l^\ast\}_{l=1}^{L}
    =
    \operatorname*{arg\,where}_{t}
    \bigg[
        \frac{\mathrm{d}^{\alpha}}{\mathrm{d}t^{\alpha}}
        f_i(t)
        =
        0
        ,
        \forall i
    \bigg]
    ,
    \quad
    \alpha \in \{0, 1, 2\}
    ,
\end{equation}
where the index of event time points is sorted in ascending order,
$ t_1^\ast < t_2^\ast < t_3^\ast < \cdots < t_l^\ast < \cdots < t_{L}^\ast $,
and $L$ is the number of event time points.
Note that when $\alpha = 0$, the points are the intersection points,
when $\alpha = 1$, the points are the stationary points,
and when $\alpha = 2$, the points are the inflection points.
We then discretize the spline function
into a binary time series $s_i (t)$ by one of three types of
transformation,
\begin{equation}
    s_i(t)
    =
    \begin{cases}
        +1
        & \mathrm{if }
        \quad
        \sigma_\alpha
        \frac{\mathrm{d}^{\alpha}}{\mathrm{d}t^{\alpha}}
        f_i(t)
        >0
        \\
        -1
        & \mathrm{if }
        \quad
        \sigma_\alpha
        \frac{\mathrm{d}^{\alpha}}{\mathrm{d}t^{\alpha}}
        f_i(t)
        <0
        \\
        -s_i(t - 0)
        & \mathrm{if }
        \quad
        \frac{\mathrm{d}^{\alpha}}{\mathrm{d}t^{\alpha}}
        f_i(t)
        =0
    \end{cases}
    ,
\end{equation}
where $\alpha = 0$, $1$, and $2$ give the static, dynamic, and curve
transformations, respectively,
with $\sigma_0 = \sigma_1 = +1$ and $\sigma_2 = -1$.
Here,
$
s_i (t-0)
\coloneqq
\lim_{\epsilon \to +0}
s_i (t - \epsilon)
$
is the limit approaching $t$ from below,
and
the state transition occurs as
$
s_i (t^\ast)
=
\lim_{\epsilon \to +0} s_i (t^\ast + \epsilon)
=
- s_i (t^\ast - 0)
$
at time $t^\ast$.
The static and dynamic transformations were proposed
previously~\cite{Kurths1996},
and we add the curve transformation.
See also Extended Data Fig.~\ref{fig:fig-s05} for examples of
each binarization method.
Finally, we temporally discretize the $N$-dimensional binarized time series
$\bm{s} (t)$
into a state transition matrix:
\begin{equation}
    \bm{K}
    =
    \begin{bmatrix}
        \mu (0) &
        \mu (t_1^\ast) &
        \cdots &
        \mu (t_{L-1}^\ast) &
        \mu (t_{L}^\ast) \\
        \mu (t_1^\ast) &
        \mu (t_2^\ast) &
        \cdots &
        \mu (t_{L}^\ast) &
        \mu (\tau) \\
        t_{1}^\ast         &
        \Delta t_{2, 1}^\ast &
        \cdots &
        \Delta t_{L, L-1}^\ast &
        \tau - t_{L}^\ast
    \end{bmatrix}
    ,
\end{equation}
where
$
\mu (t)
=
1
+
\sum_{i=1}^{N}
2^{N-i}
(1/2)
[1 + s_i (t)]
$
is the integer index of $N$-dimensional Ising state vector
$
\bm{s} (t)
\coloneqq
[
    s_1 (t) \; \cdots \; s_N (t)
]^\top
\in
\{+1, -1\}^N
$
at time $t$,
$\mu(0)$ is the initial state index,
$\mu(\tau)$ is the final state index,
and
$\Delta t_{l, l-1}^\ast \coloneqq t_l^\ast - t_{l-1}^\ast$
is the time difference between the $l$th and $(l-1)$th event time points.
With the state transition matrix, we perform the probability flux
analysis
(`Probability flux analysis').
For the neural data,
we create a state transition matrix for each task by combining the
state transition matrices of all scans and subjects.

\subsection*{Hypercubic probability flux diagram}
\label{sec:probability-flux-diagram}
The $N$-dimensional Ising state vector $\bm{s}$ corresponds to a vertex
of the hypercube, and a state transition involving a single spin flip
corresponds to an edge of the hypercube.
To visualize the probability flux in the hypercubic state space---or
hypercubic probability flux---we need to project the hypercube onto
a two-dimensional plane.
We employ PCA to obtain reproducible, interpretable, and automatic
projections of hypercubes~\cite{Horiike2025}.
With the empirical stationary distribution
(equation~\eqref{eq:stationary-distribution-estimated})
of each hypercubic state
$
\bm{s}
\coloneqq
[
    s_1 \; \cdots \; s_N
]^\top
\in
\{+1, -1\}^N
$,
we calculate the covariance matrix
\begin{equation}
    \bm{\varSigma}
    \coloneqq
    \langle
    (
        \bm{s} - \langle \bm{s} \rangle
    )
    (
        \bm{s} - \langle \bm{s} \rangle
    )^\top
    \rangle
    ,
\end{equation}
where the angle brackets
$\langle x \rangle\coloneqq \sum_{\mu=1}^{2^N} p(\mu) \, x$
denote the average over the stationary distribution.
After diagonalizing the covariance matrix,
we obtain the PC loadings $\{\bm{v}_i\}_{i=1}^{N}$ and
PC scores $\{\bm{r}_i\}_{i=1}^{N}$.
We then introduce the biplot vectors
$\{\tilde{\bm{e}}_i\}_{i=1}^{N}$ of PC$j$ and PC$k$ as
\begin{equation}
    \tilde{\bm{e}}_i
    \coloneqq
    2
    \begin{bmatrix}
        v_{i; j} \\
        v_{i; k}
    \end{bmatrix}
    ,
\end{equation}
where $v_{i; j}$ and $v_{i; k}$ are the $j$th and $k$th elements of
the $i$th PC loading $\bm{v}_i$, respectively.
The biplot vectors are the projections of the unit vectors of the
original high-dimensional space onto the chosen PCs.
Typically, we employ the first two PCs.
Using the biplot vectors, the probability fluxes are visualized as
hypercubic arrows~\cite{Horiike2025},
where the width is proportional to the magnitude
$| \mathcal{J}(\mu, \nu) |$ and the direction corresponds
to the sign of the probability flux
$
\operatorname{sgn}
\bm{(}
    \mathcal{J}(\mu, \nu)
\bm{)}
=
- \operatorname{sgn}
\bm{(}
    \mathcal{J}(\nu, \mu)
\bm{)}
$.
The direction of the arrow aligns with the direction of the
probability flux.

\subsection*{Inferring the Ising spin system from transition rates}
\label{sec:model-fitting}
Here, we describe the procedure for inferring the Ising spin system from the
transition rates of the master equation.
To begin with, we derive the transition rates of the Ising spin system
with symmetric interactions.
We then apply the derived transition rate to infer the Ising spin
system with asymmetric interactions.

The (pseudo-)Hamiltonian (or energy function) of the Ising spin
system is given by the interaction and the external input:
\begin{align}
    \mathcal{H} (\mu)
    &\coloneqq
    -
    \frac{1}{2}
    \sum_{i=1}^{N} \sum_{j=1}^{N}
    s_{\mu; i} J_{i,j} s_{\mu; j}
    -
    \sum_{i=1}^{N}
    s_{\mu; i}
    h_i
    \\&=
    -
    \frac{1}{2}
    \bm{s}_\mu^\top \bm{J} \bm{s}_\mu
    -
    \bm{s}_\mu^\top \bm{h}
    \label{eq:ising-hamiltonian}
    ,
\end{align}
where
$
\bm{s}_\mu
\coloneqq
[
    s_{1; \mu} \; \cdots \; s_{N; \mu}
]^\top
\in
\{+1, -1\}^N
$
is the Ising state of the system consisting of $N$ components,
$\bm{J} \in \mathbb{R}^{N \times N}$ is the
(not necessarily symmetric) interaction matrix where element
$J_{i, j}$ is the interaction from component $j$ to $i$, and
$
\bm{h}
\coloneqq
[
    h_1 \; \cdots \; h_N
]^\top
\in
\mathbb{R}^{N}
$
is the external input where element
$h_i$ is the external input to component $i$.
The self-interaction is set to zero, that is, $J_{i, i} \coloneqq 0$
for all $i$.
A positive interaction, $J_{i, j}>0$, is
excitatory (or ferromagnetic),
and a negative interaction, $J_{i, j}<0$, is inhibitory
(or antiferromagnetic).
The same goes for the external input $\bm{h}$:
a positive external input, $h_i>0$, is an excitatory bias,
and a negative external input, $h_i<0$, is an inhibitory bias.
If the interaction matrix is symmetric, $\bm{J} = \bm{J}^\top$,
equation~\eqref{eq:ising-hamiltonian} is the Hamiltonian, but if the
interaction matrix is asymmetric $\bm{J} \neq \bm{J}^\top$,
equation~\eqref{eq:ising-hamiltonian} is the pseudo-Hamiltonian
because the energy is ill-defined.
Below, we assume a symmetric interaction matrix
to derive the transition rate of the Ising spin system.
Our goal is to infer the interaction matrix $\bm{J}$ and
external input $\bm{h}$ from the estimated joint transition rates
$\{w(\mu \mid \nu)\}$
rather than from the empirical probability distribution $\{p(\mu)\}$.

From the probability flux analysis,
we obtain the transition rate from state $\nu$ to $\mu$ by
dividing the joint transition rate by the stationary distribution $p(\nu)$:
\begin{equation}
    w(\mu \mid \nu)
    \approx
    \frac{1}{\tau}
    \frac{n(\mu, \nu)}{p(\nu)}
    \label{eq:transition-rate-estimated}
    .
\end{equation}
The stationary distribution is estimated from
equation~\eqref{eq:stationary-distribution-estimated}.
Thus, we can estimate the transition rate from the empirical data.

We then build a stochastic process model and fit it to the observations.
Assuming a continuous-time Markov process, we introduce
the time evolution of the
probability distribution by employing the master
equation for single-spin flips introduced by Glauber~\cite{Glauber1963},
\begin{align}
    \frac{\mathrm{d}}{\mathrm{d} t}
    p(\mu; t)
    &=
    \sum_{\nu \in \{\mu^{(k)}\}_{k=1}^{N}}
    [
        w(\mu \mid \nu)
        p(\nu; t)
        -
        w(\nu \mid \mu)
        p(\mu; t)
    ]
    \label{eq:master-equation}
    \\&=
    \sum_{\nu \in \{\mu^{(k)}\}_{k=1}^{N}}
    \mathcal{J}(\mu, \nu; t)
    ,
\end{align}
where $p(\mu; t)$ is the probability of finding the
system in state $\mu$ at time $t$,
$w(\mu \mid \nu)$ is the (conditional) transition rate from state
$\nu$ to $\mu$,
and
$\mu^{(k)}$ is the state index obtained by flipping
the $k$th spin of state $\mu$, that is,
$
\bm{s}_{\mu^{(k)}}
\coloneqq
\bm{F}_{k}
\bm{s}_\mu
=
[
    s_{1; \mu} \; \cdots \; -s_{k; \mu} \; \cdots \; s_{N; \mu}
]^\top
$,
where
$\bm{F}_{k}
\coloneqq
\bm{I}
-
2
\bm{e}_k
\bm{e}_k^\top
$
is the spin flip matrix of spin $k$,
where
$\bm{I} \coloneqq \operatorname{diag} (\bm{1})$ is the identity matrix
with $\bm{1} \in \{1\}^{N}$ being the vector of ones,
and $\bm{e}_k$ is the $k$th standard unit vector of $\mathbb{R}^N$.

The transition rates
$\{w(\mu^{(k)} \mid \mu)\}$
are given by one of two transition rates:
the Arrhenius transition rate
\begin{align}
    w(\mu^{(k)} \mid \mu)
    =
    A
    \exp
    \bm{(}
        -
        \beta
        s_{k; \mu}
        g_{k}(\mu)
    \bm{)}
    \label{eq:transition-rate-arrhenius}
    ,
\end{align}
and the Glauber transition rate~\cite{Glauber1963}
\begin{align}
    w(\mu^{(k)} \mid \mu)
    =
    A
    \frac{1}{
        1
        +
        \exp
        \bm{(}
            2
            \beta
            s_{k; \mu}
            g_{k}(\mu)
        \bm{)}
    }
    \label{eq:transition-rate-glauber}
    ,
\end{align}
where $A \in \mathbb{R}_{>0}$ is the rate constant,
$\beta \coloneqq 1/k_\mathrm{B} T$ is the inverse temperature with
$k_\mathrm{B}$ being the Boltzmann constant and $T$ the absolute
temperature,
and
\begin{align}
    g_k(\mu)
    =
    \sum_{i=1}^{N}
    J_{k, i} s_{i; \mu}
    +
    h_k
\end{align}
is the local field or input signal to spin $k$ at state $\mu$.
If the interaction matrix is asymmetric, the parameter $\beta$ as
inverse temperature is ill-defined, but we interpret it as the
parameter controlling the stochastic fluctuation of the interactions.
We can write the term inside the exponential function of
equation~\eqref{eq:transition-rate-arrhenius} and
equation~\eqref{eq:transition-rate-glauber} using the vector notation:
\begin{equation}
    s_{k; \mu}
    g_k(\mu)
    =
    -
    \frac{1}{2}
    (
        \bm{s}_{\mu^{(k)}}
        -
        \bm{s}_\mu
    )^\top
    (
        \bm{J}
        \bm{s}_\mu
        +
        \bm{h}
    )
    .
\end{equation}
These transition rates imply that spin $k$ is likely to flip if its
state is not aligned with its local field;
that is, spin $k$ tends to align with the local field.
Below, we use the Arrhenius transition rate as an example; the extension to the
Glauber transition rate is given in
Supplementary Note~7.

Note that both transition rates
(equations~\eqref{eq:transition-rate-arrhenius}
and~\eqref{eq:transition-rate-glauber}) satisfy the detailed balance condition
\begin{equation}
    w(\mu \mid \nu)
    \pi(\nu)
    =
    w(\nu \mid \mu)
    \pi(\mu)
    ,
    \quad
    \forall (\mu, \nu)
    \label{eq:detailed-balance-condition}
    ,
\end{equation}
if the interaction matrix is symmetric, $\bm{J} = \bm{J}^\top$.
Here,
$
\pi(\mu)
\coloneqq
(1/\mathcal{Z})
\exp
\bm{(}
    -
    \beta
    \mathcal{H}(\bm{s}_\mu)
\bm{)}
$
is the canonical probability of state $\mu$,
where
$
\mathcal{Z}
\coloneqq
\sum_{\mu=1}^{2^N}
\exp
\bm{(}
    -
    \beta
    \mathcal{H}(\bm{s}_\mu)
\bm{)}
$
is the normalization constant, or partition function, of
statistical mechanics.
If the interaction matrix is asymmetric, $\bm{J} \neq \bm{J}^\top$,
the detailed balance is broken as
\begin{equation}
    \frac{
        w(\mu^{(k)} \mid \mu)
    }{
        w(\mu \mid \mu^{(k)})
    }
    =
    \frac{
        \pi(\mu^{(k)})
    }{
        \pi(\mu)
    }{
    }
    \exp
    \Bigg(
        -
        \beta
        s_{k; \mu}
        \sum_{i=1}^{N}
        (
            J_{k, i}
            -
            J_{i, k}
        )
        s_{i; \mu}
    \Bigg)
    \label{eq:broken-detailed-balance}
    .
\end{equation}
Here, the factor
$
\exp
\bm{(}
    -
    \beta
    s_{k;\mu}
    \sum_{i=1}^{N}
    (
        J_{k, i}
        -
        J_{i, k}
    )
    s_{i; \mu}
\bm{)}
$
breaks the detailed balance.
See Supplementary Note~7 for the derivation of
equation~\eqref{eq:broken-detailed-balance}.
Indeed, the detailed balance is broken if the interaction matrix is
asymmetric, $\bm{J} \neq \bm{J}^\top$,
and the magnitude of its antisymmetry $J_{i, j} - J_{j, i}$ corresponds
to the degree of detailed balance breaking.

We then fit the model parameters of interaction $\bm{J}$,
external input $\bm{h}$, and transition rate constant $A$ to the
empirical observations.
Using the estimated transition rates of
equation~\eqref{eq:transition-rate-estimated},
we define the loss function as the squared difference between the
logarithmic transition rates of the model and those of the data,
$
\sum
[
    \ln(w_{\mathrm{model}})
    -
    \ln(w_{\mathrm{data}})
]^2
$,
as a function of these parameters,
\begin{align}
    &
    L
    (
        \beta \bm{J}, \beta \bm{h}, A
    )
    \coloneqq
    \sum_{(\mu^{(k)}, \mu)}
    \bigg[
        2 \ln (A)
        +
        \bigg.
        \nonumber
        \\&
        \quad\quad\quad
        \bigg.
        (
            \bm{s}_{\mu^{(k)}}
            -
            \bm{s}_\mu
        )^\top
        (
            \beta
            \bm{J}
            \bm{s}_\mu
            +
            \beta
            \bm{h}
        )
        ~
        -
        2
        \ln
        \bigg(
            \frac{1}{\tau}
            \frac{n(\mu^{(k)}, \mu)}{p(\mu)}
        \bigg)
    \bigg]^2
    \label{eq:loss-function}
    ,
\end{align}
where the inverse temperature $\beta$ is a nuisance parameter and
the sum $\sum_{(\mu^{(k)}, \mu)}$ is over all pairs of
states with a single spin-flip difference; there are $N2^N$ such pairs.
The loss function is minimized when the logarithms of the transition
rates of the model
(equations~\eqref{eq:transition-rate-arrhenius}
or~\eqref{eq:transition-rate-glauber}) match those of the empirical
observations (equation~\eqref{eq:transition-rate-estimated}).
The parameters minimizing the loss function of
equation~\eqref{eq:loss-function}
give the fitted parameters:
$\beta \tilde{\bm{J}}$, $\beta \tilde{\bm{h}}$, and $\tilde{A}$.
\begin{equation}
    \beta \tilde{\bm{J}}, \beta \tilde{\bm{h}},  \tilde{A}
    =
    \operatorname*{arg\,min}_{\{\beta \bm{J}, \beta \bm{h}, A\}}
    \bm{(}
        L(\beta \bm{J}, \beta \bm{h}, A)
    \bm{)}
    \label{eq:parameter-inference}
    .
\end{equation}
The estimated interaction matrix is in general asymmetric,
$\tilde{\bm{J}} \neq \tilde{\bm{J}}^\top$.
The minimization problem of equation~\eqref{eq:parameter-inference}
can be mapped to a linear matrix equation (Supplementary
Note~6).
Note that the number of inferred parameters,
$(N^2 - N) + N + 1 = N^2 + 1$,
is smaller than the number of constraints, $N2^N$,
so the problem is sufficiently constrained.
The rate constant $\tilde{A}$ can be spin-dependent,
$\tilde{A}_k$, which is the rate constant for spin $k$,
but the spin-dependent rate constant does not significantly affect
the inference (Supplementary Note~8).

\subsection*{Comparing interaction networks among tasks}
\label{sec:matrix-comparison}
To compare the interaction networks among tasks,
we employ several methods to quantify the similarity
between the interaction networks~\cite{Cole2014}.
First, we calculate the pairwise cosine similarity between the
interaction matrices of different tasks.
For the interaction matrix $\bm{J}$,
we simply vectorize the off-diagonal elements and calculate the
cosine similarity (Fig.~\ref{fig:fig-05}a).
For the symmetric part $\bm{J}^{\mathrm{(s)}}$ and the antisymmetric
part $\bm{J}^{\mathrm{(a)}}$,
we use the upper triangular part of the matrices and calculate the
cosine similarity because the matrices are symmetric or antisymmetric
(Fig.~\ref{fig:fig-05}b--c).
For each pairwise cosine similarity matrix
(Fig.~\ref{fig:fig-05}a--c),
we perform hierarchical clustering
(`Spatial coarse-graining through hierarchical clustering')
and sort the rows and columns by the hierarchical clustering result.

\subsection*{Reconstructing flux diagrams from the inferred model}
\label{sec:reconstruction}
The steady-state hypercubic probability flux diagrams are
reconstructed from the inferred model as follows.
Using the estimated parameters of equation~\eqref{eq:parameter-inference},
we first calculate the transition rates
(equations~\eqref{eq:transition-rate-arrhenius}
or~\eqref{eq:transition-rate-glauber}).
We then rewrite the master equation (equation~\eqref{eq:master-equation})
in matrix--vector form:
\begin{equation}
    \frac{\mathrm{d}}{\mathrm{d}t}
    \bm{p}(t)
    =
    \bm{W}
    \bm{p}(t)
    .
\end{equation}
Here,
$
\bm{p}(t)
\coloneqq
[
    p(1; t) \; \cdots \; p(\mu; t) \; \cdots \; p(2^N; t)
]^\top
\in
[0, 1]^{2^N}
$
is the probability vector of the system at time $t$,
which is the statistical state of the system.
The element of the transition rate matrix
$
\bm{W}
\in
\mathbb{R}^{2^N \times 2^N}
$
is defined by
\begin{equation}
    W_{\mu, \nu}
    \coloneqq
    \begin{cases}
        w(\mu \mid \nu)
        & \mathrm{if } \nu \in \{\mu^{(k)}\}_{k=1}^{N} \\
        -\sum_{\nu=1}^{2^N} w(\nu \mid \mu)
        & \mathrm{if } \nu = \mu\\
        0
        & \mathrm{otherwise }
    \end{cases}
    .
\end{equation}
Because the transition rate matrix is a stochastic rate matrix,
by the Perron--Frobenius theorem,
the eigenvalue with the largest real part is zero
for the transition rate matrix and the
corresponding right eigenvector $\bm{p}_0$ is
the stationary distribution vector, that is,
$\bm{W} \bm{p}_0 = 0 \bm{p}_0$.
Using the stationary distribution $\bm{p}_0$, we calculate the probability flux
$
\mathcal{J}(\mu, \nu)
\coloneqq
w(\mu \mid \nu) p(\nu) - w(\nu \mid \mu) p(\mu)
$
to validate the inferred model with the empirical data in Fig.~\ref{fig:fig-06}.
Note that if the inferred effective interaction matrix is symmetric,
$\tilde{\bm{J}} = \tilde{\bm{J}}^\top$,
the stationary distribution is given by the Boltzmann distribution
$\bm{p}_0 = \bm{\pi}$
and the probability flux vanishes,
$\mathcal{J}(\mu, \nu) = 0$ for all pairs of states $(\mu, \nu)$.
We compare the model stationary distribution $\bm{p}_0$ with the empirical
brain state distribution in Supplementary Note~2.

\subsection*{Entropy production rate}
\label{sec:entropy-production-rate}
The entropy production rate quantifies the irreversibility of the
dynamics of the system and can be calculated from the joint transition rates as
\begin{equation}
    \dot{\mathcal{S}}_{\mathrm{tot}} (t)
    \nonumber
    =
    \frac{k_\mathrm{B}}{2}
    \sum_{\mu=1}^{2^N} \sum_{\nu=1}^{2^N}
    \mathcal{J}(\mu, \nu; t)
    \ln
    \bigg(
        \frac{
            w(\mu \mid \nu) p(\nu; t)
        }{
            w(\nu \mid \mu) p(\mu; t)
        }
    \bigg)
    .
\end{equation}
In nonequilibrium steady state,
the entropy production rate arises from the cycles of the probability fluxes.
In Supplementary Note~7,
we show that the entropy production per cycle arises from the
antisymmetric interactions between the spins flipped
irreversibly (not reversibly) in the cycle.

Although the entropy production rate is calculated from the probability
fluxes and joint transition rates,
we find that the entropy production rate is less task-specific than the
sequential patterns of
the probability fluxes
(Supplementary Note~4 and Supplementary
Fig.~7).

\subsection*{Software}
\label{sec:software}
Calculations and visualizations in this work were performed using
open-source Python~\cite{Rossum2010} libraries:
BrainSpace~\cite{VosDeWael2020},
JAX~\cite{Frostig2019},
Matplotlib~\cite{Hunter2007},
NetworkX~\cite{Hagberg2008},
neuromaps~\cite{Markello2022},
nilearn~\cite{Abraham2014},
Numba~\cite{Lam2015},
NumPy~\cite{Harris2020},
pandas~\cite{McKinney2010},
scikit-learn~\cite{Pedregosa2011},
SciPy~\cite{Virtanen2020},
and
seaborn~\cite{Waskom2021}.
The colour maps of some figures were generated with
ColorCET~\cite{Kovesi2015}.

\subsection*{Use of large language models}
Claude (Anthropic) was used to assist with condensing and reorganizing
the manuscript text,
converting it to the journal template,
and refactoring the analysis code.
The authors reviewed and verified all changes
and take full responsibility for the content of the manuscript.

\section*{Data availability}
This study uses previously collected~\cite{VanEssen2013,Smith2013,Barch2013}
data from the Human Connectome Project,
which are available at \url{https://www.humanconnectome.org}
under the WU-Minn HCP Open Access Data Use Terms.
The data were preprocessed and parcellated as described
previously~\cite{Lynn2021}.
The derived data needed to reproduce the figures are deposited in
Zenodo~\cite{Horiike2026}
together with the code
(\url{https://doi.org/10.5281/zenodo.23530108})
and are redistributed under the same data use terms.

\section*{Code availability}
The code used in this study is available at Zenodo~\cite{Horiike2026}
(\url{https://doi.org/10.5281/zenodo.23530108}).

\section*{Supplementary information}
Supplementary Information is available for this paper.

\section*{Acknowledgements}
Y.H.~thanks Mizuho Aoki and Henri Schmidt for their helpful comments
on an early version of the manuscript,
and Ramón Nartallo-Kaluarachchi for insightful discussions.
Data were provided by the Human Connectome Project, WU-Minn Consortium
(Principal Investigators: David Van Essen and Kamil Ugurbil; 1U54MH091657)
funded by the 16 NIH Institutes and Centers that support the NIH Blueprint
for Neuroscience Research; and by the McDonnell Center for Systems
Neuroscience at Washington University.

\section*{Funding}
Y.H.~discloses support for the research of this work from
the Japan Science and Technology Agency
[SPRING Grant No.~JPMJSP2125, ``THERS Make New Standards Program for the
Next Generation Researchers''],
ANRI Inc.~[8th ANRI Fellowship]
and the Japan Society for the Promotion of Science
[Research Fellowships for Young Scientists; KAKENHI Grant No.~JP26KJ1295].
S.F.~discloses support for the research of this work from
the Japan Society for the Promotion of Science
[KAKENHI Grants No.~JP22H00406 and No.~JP24H00061].

\section*{Author contributions}
Y.H.~performed
conceptualization, data curation, formal analysis, investigation,
software development, validation and visualization;
Y.H.~partially contributed to the funding acquisition;
Y.H.~and S.F.~developed the methodology;
S.F.~provided computational resources;
Y.H.~wrote the original draft;
and Y.H.~and S.F.~reviewed and edited the manuscript.

\section*{Competing interests}
The authors declare no competing interests.

\section*{Correspondence}
Correspondence and requests for materials should be addressed to
Yoshiaki Horiike.

\clearpage
\captionsetup[figure]{labelformat=extendeddata}
\setcounter{figure}{0}

\begin{figure*}[tb]
    \centering
    \includegraphics{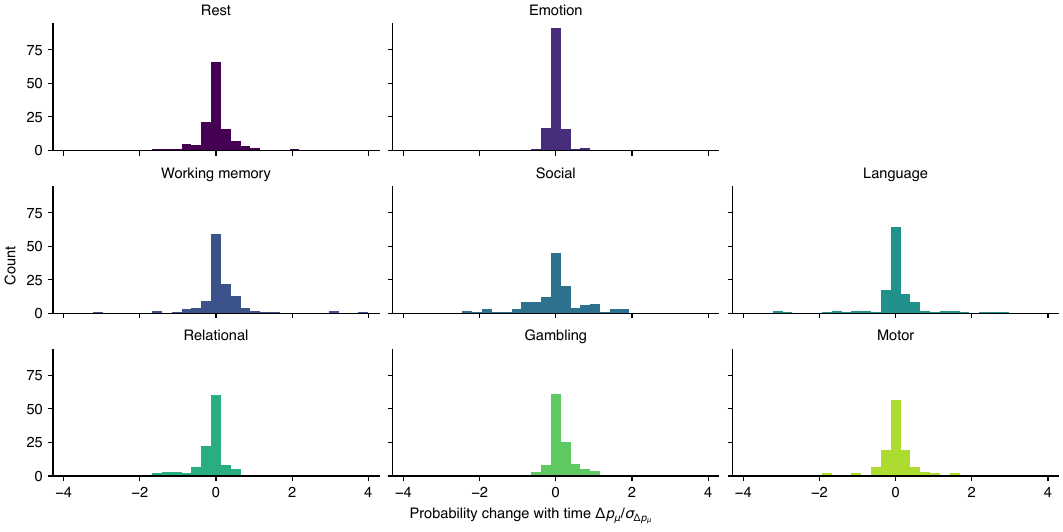}
    \caption{\textbf{The stationarity of the probability distribution.}
        In each panel, the probability changes of
        equation~\eqref{eq:probability-change} for all states are
        shown as a histogram.
        The values are normalized by the standard deviation,
    calculated from all data.}
    \label{fig:fig-s01}
\end{figure*}

\begin{figure*}[tb]
    \centering
    \includegraphics{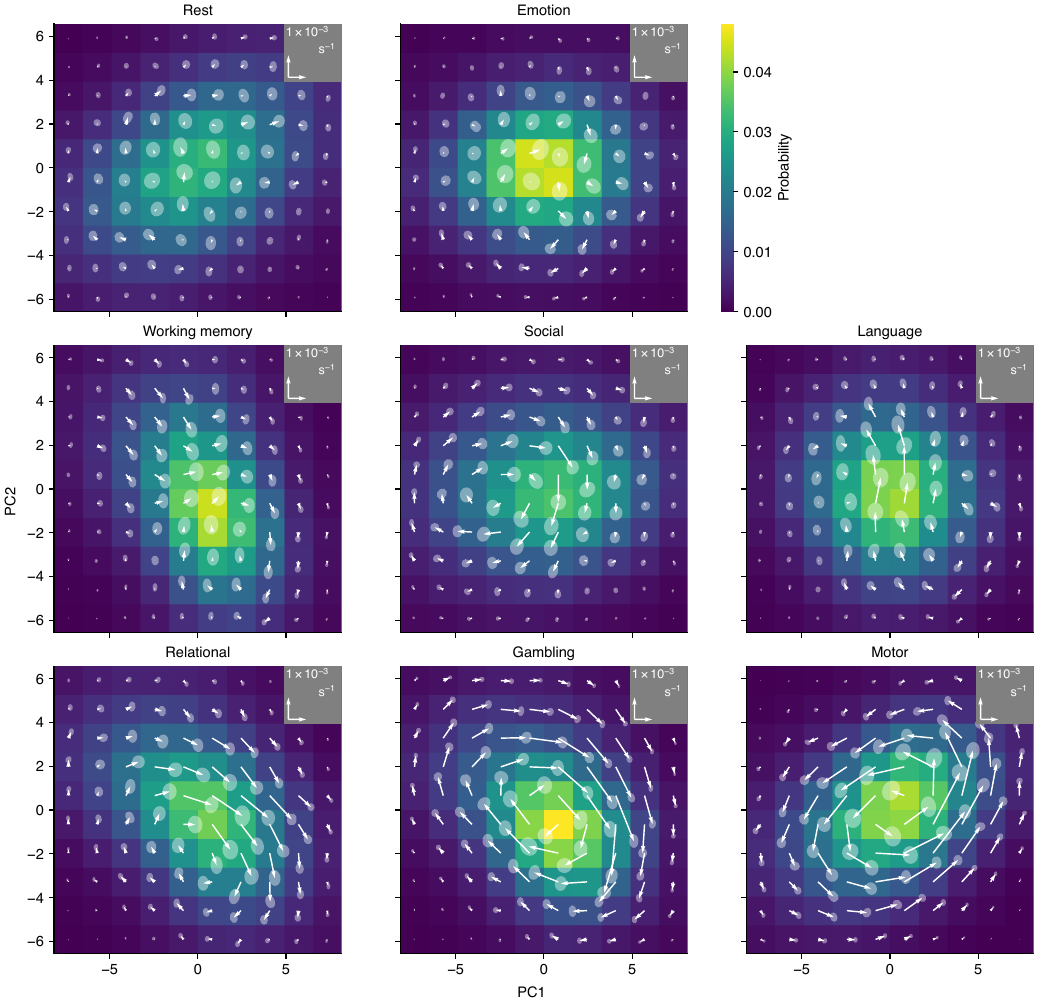}
    \caption{\textbf{Probability flux analysis on the state space
        defined by the first two PC scores.}
        In each panel, we show the probability distribution (colour)
        and probability flux (arrows) on the state space defined by the
        first two PC scores of the neural activity of brain
        parcels~\cite{Lynn2021}.
        In the top right corner of each panel, we show the scale of the arrows.
    The ellipses indicate the error ranges~\cite{Battle2016,Lynn2021}.}
    \label{fig:fig-s02}
\end{figure*}

\begin{figure*}[tb]
    \centering
    \includegraphics{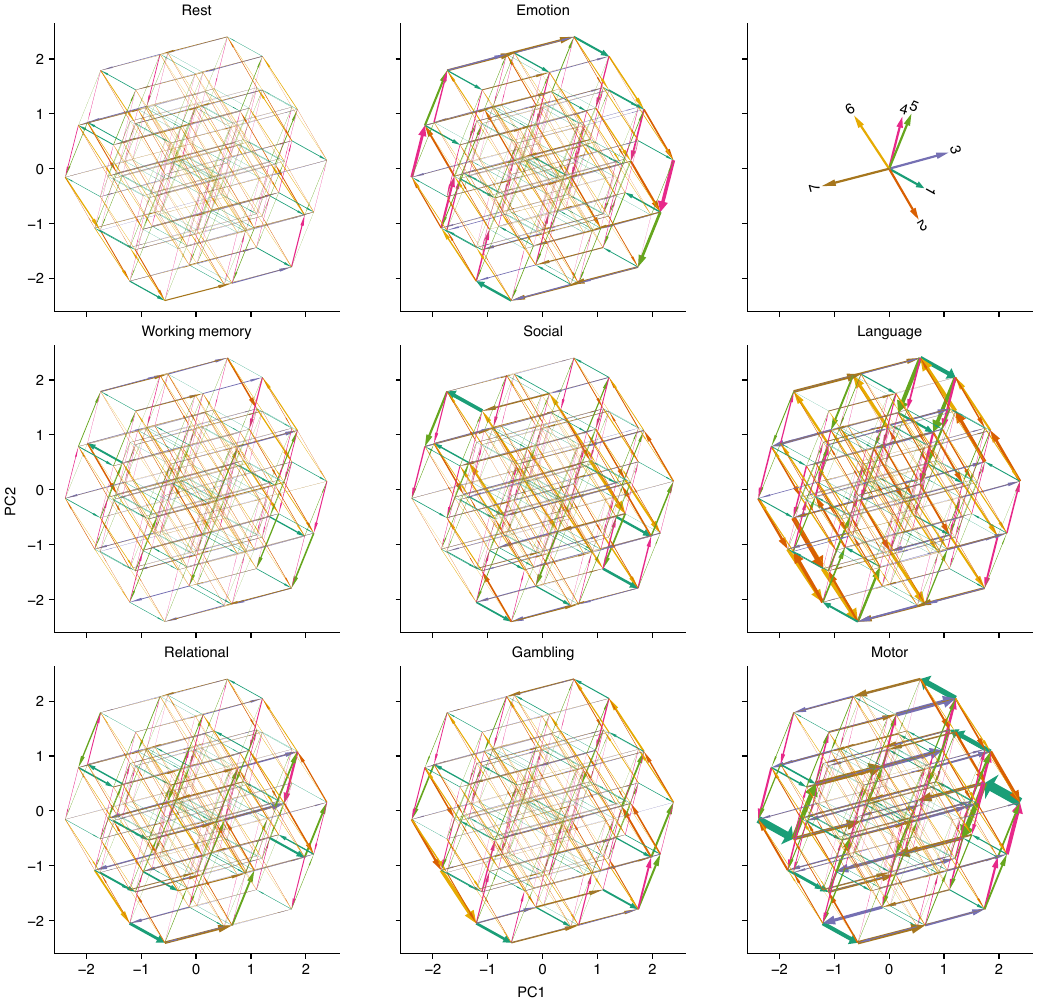}
    \caption{\textbf{Probability fluxes estimated from the known
        functional clustering.}
        Same as Fig.~\ref{fig:fig-02}, but the clusters of brain regions are
        defined by the known functional clusters~\cite{ThomasYeo2011}
    (Fig.~\ref{fig:fig-01}d).}
    \label{fig:fig-s03}
\end{figure*}

\begin{figure*}[tb]
    \centering
    \includegraphics{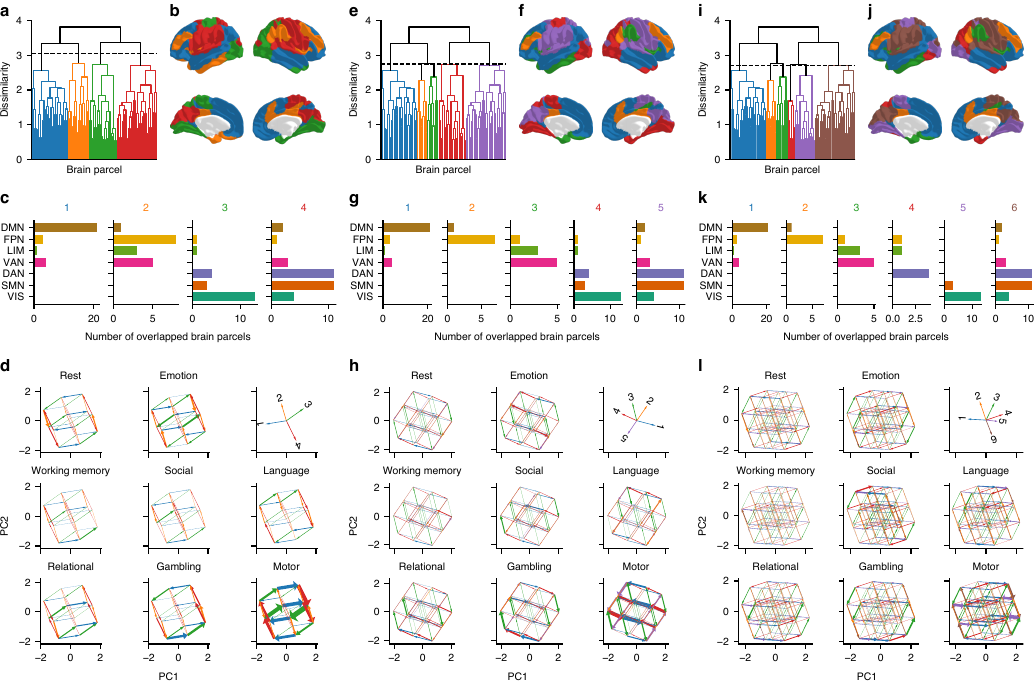}
    \caption{\textbf{Changing the number of clusters for hypercubic
        probability flux analysis.}
        Same as Fig.~\ref{fig:fig-01}b--d,
        and Fig.~\ref{fig:fig-02}, but the number of clusters is
        $N=4$ (\textbf{a--d}), $N=5$ (\textbf{e--h}),
    and $N=6$ (\textbf{i--l}).}
    \label{fig:fig-s04}
\end{figure*}

\begin{figure*}[tb]
    \centering
    \includegraphics{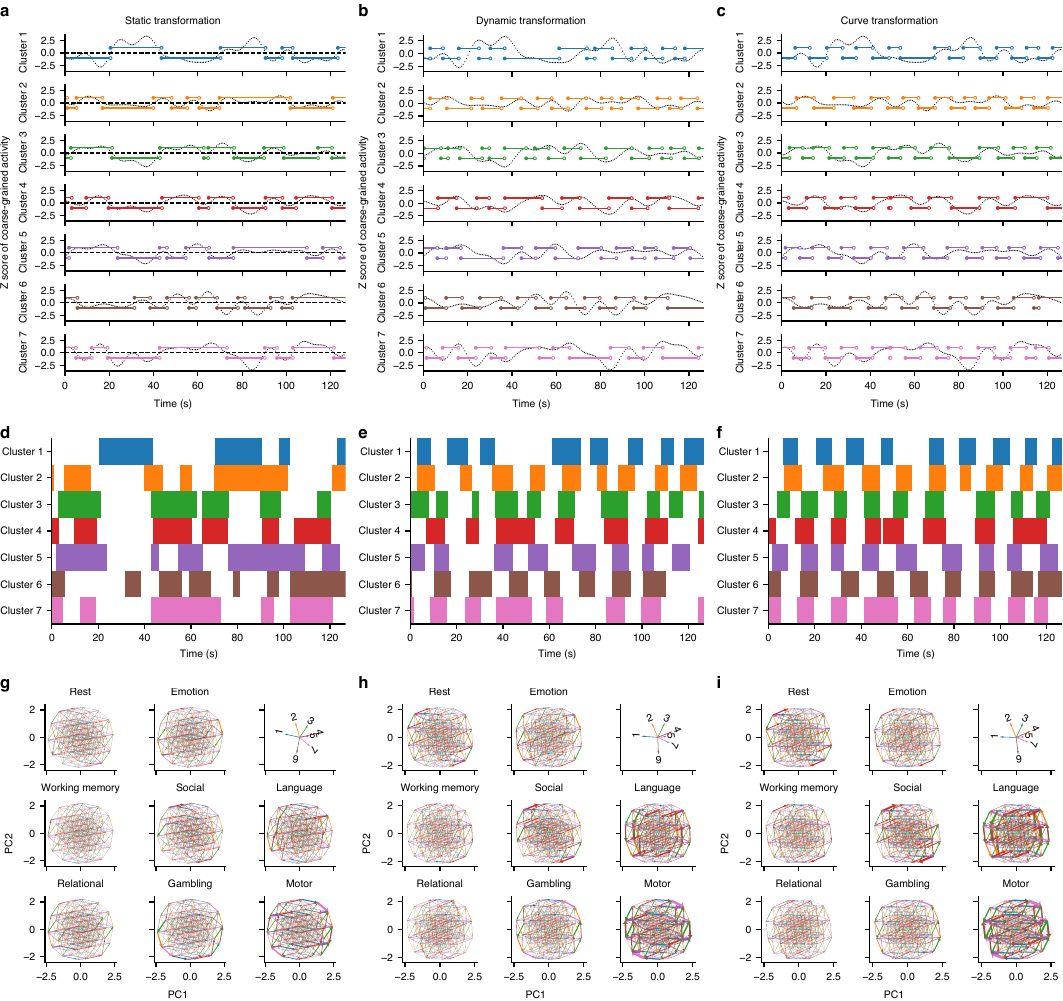}
    \caption{\textbf{Various methods of transformation applied to the
        fMRI signals and the estimated probability flux.}
        \textbf{a--c},
        Examples of the static, dynamic,
        and curve transformations.
        \textbf{d--f},
        The temporally coarse-grained (binarized) time series of panels a--c.
        Panels a and d are the same as
        Fig.~\ref{fig:fig-01}f,g, respectively.
        \textbf{g--i},
    The estimated probability flux of each task.}
    \label{fig:fig-s05}
\end{figure*}

\begin{figure*}[tb]
    \centering
    \includegraphics{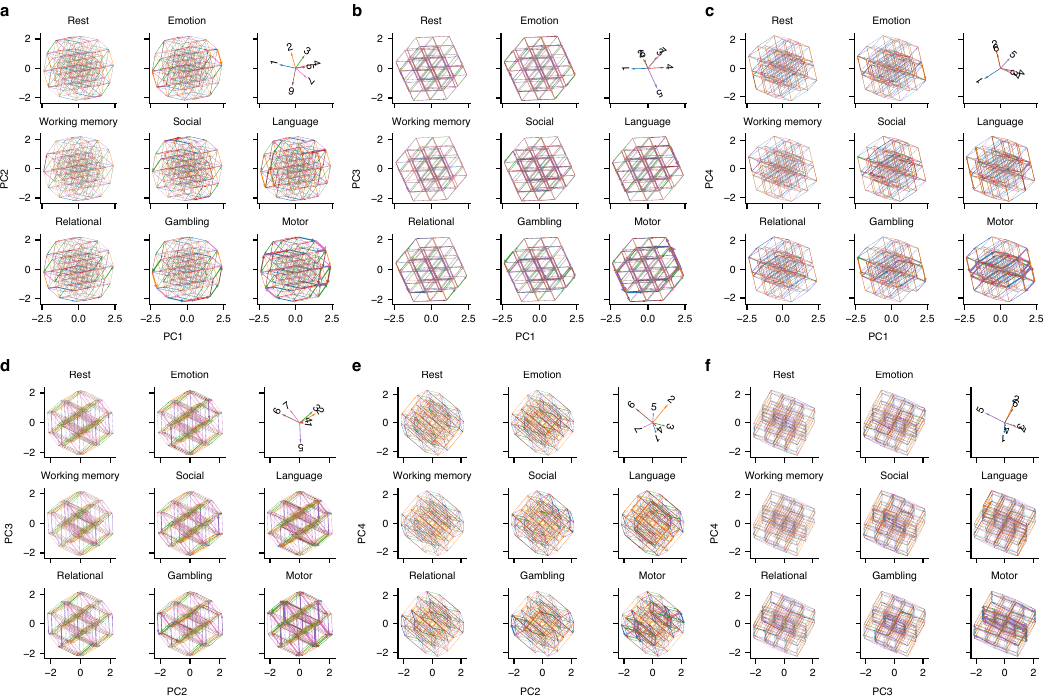}
    \caption{\textbf{Observing probability fluxes through other PCs.}
        Each panel is the same as Fig.~\ref{fig:fig-02}
    but the probability flux diagrams are visualized using other PCs.}
    \label{fig:fig-s06}
\end{figure*}

\begin{figure*}[tb]
    \centering
    \includegraphics{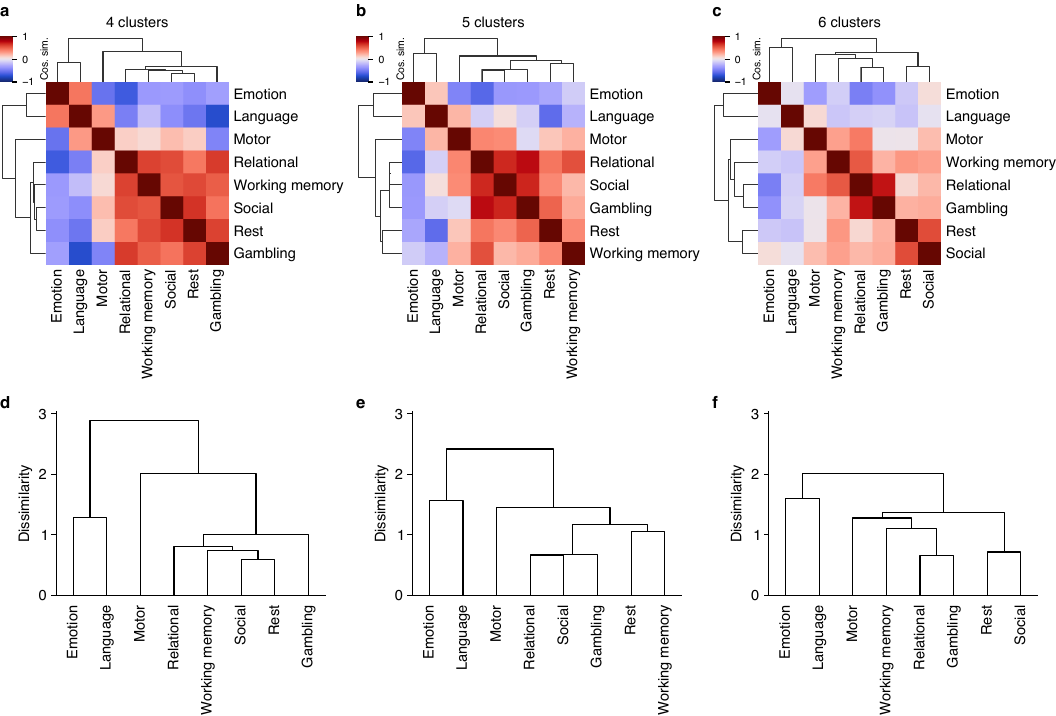}
    \caption{\textbf{Cosine similarity of probability flux diagrams
        among tasks for different numbers of clusters.}
        \textbf{a--c},
        Pairwise cosine similarity matrix of probability flux diagrams among
        tasks, computed from the
        vectorized probability flux, for $\{4, 5, 6\}$
        coarse-grained clusters.
        The matrices are sorted by hierarchical clustering.
        \textbf{d--f},
        The corresponding dendrograms of the hierarchical clustering
    of panels a--c.}
    \label{fig:fig-s07}
\end{figure*}

\begin{figure*}[tb]
    \centering
    \includegraphics{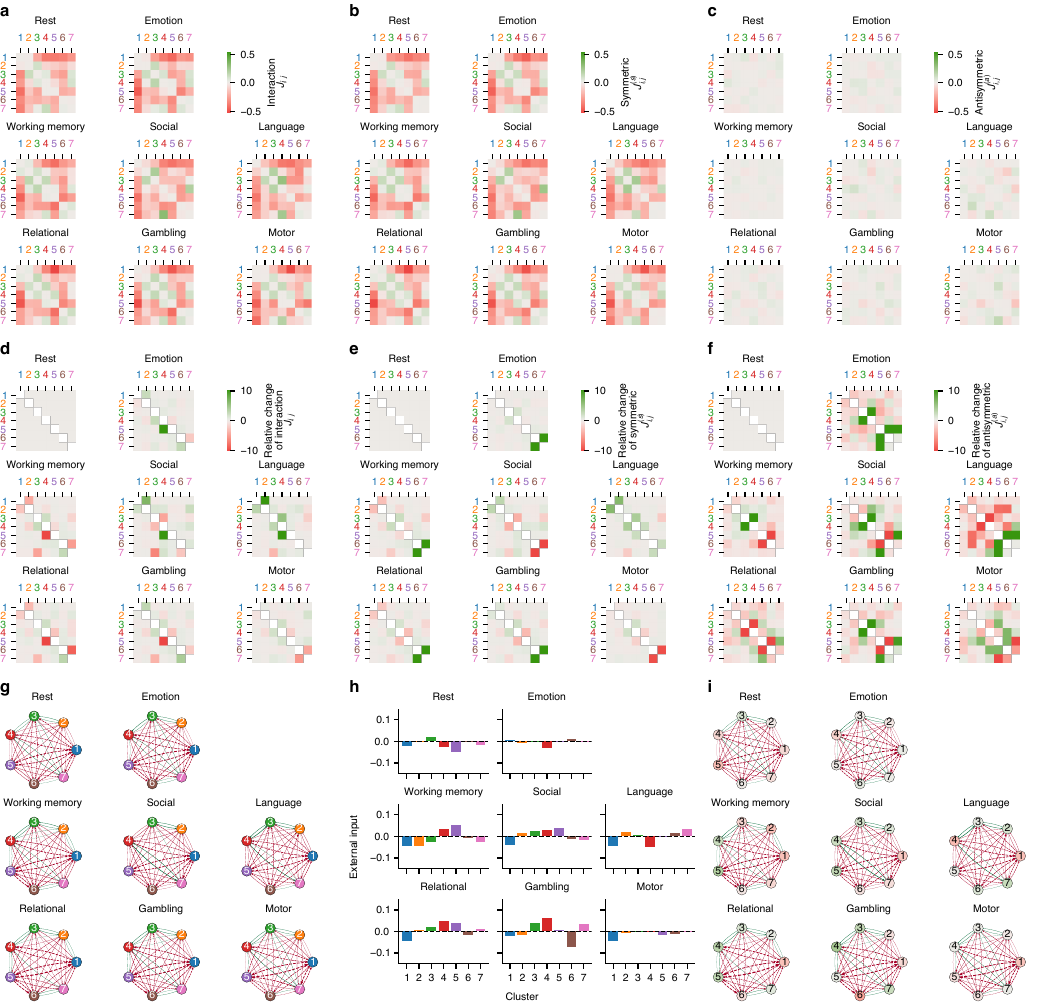}
    \caption{\textbf{Inferring the Ising spin system using the
        Glauber transition rate.}
        Same as Fig.~\ref{fig:fig-04}, but the
        Glauber transition rate is used instead of the Arrhenius
        transition rate.
        The inferred rate constants $A$ are
        1.776 (rest), 2.919 (emotion), 1.522 (working memory), 1.978
        (social), 2.156 (language), 1.781 (relational), 1.815
    (gambling), 1.804 (motor) $\mathrm{s}^{-1}$.}
    \label{fig:fig-s08}
\end{figure*}

\begin{figure*}[tb]
    \centering
    \includegraphics{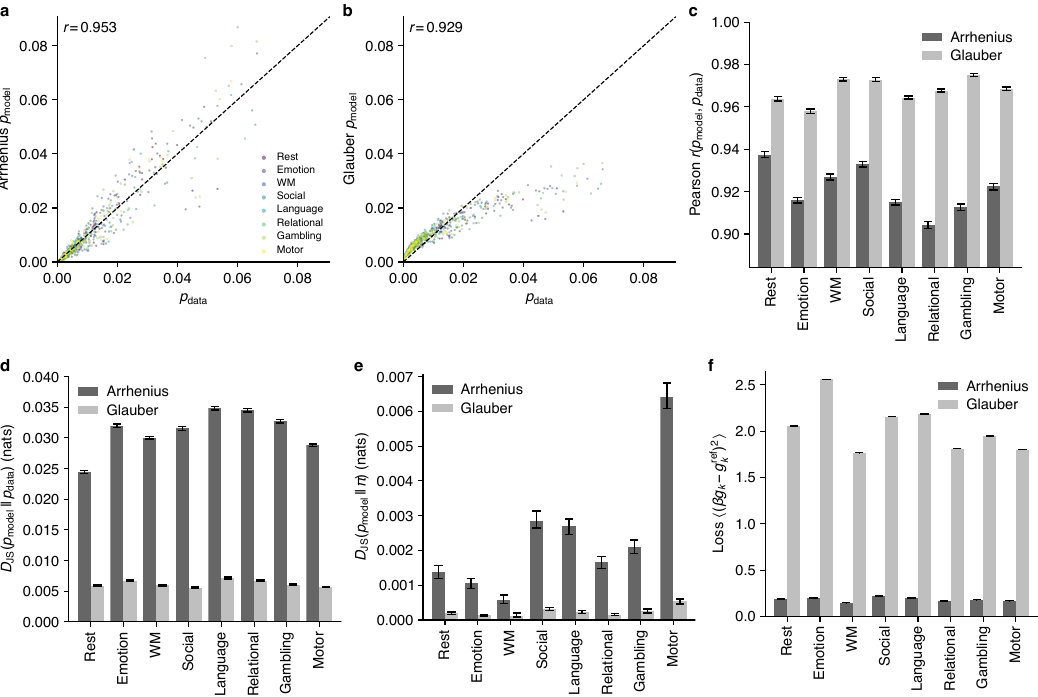}
    \caption{\textbf{Comparing the stationary distribution and the
        loss of the Arrhenius and Glauber rate models.}
        \textbf{a,b},
        Scatter plots of the model stationary distribution $p_{\mathrm{model}}$
        (right null vector of $\bm{W}$) versus the empirical brain-state
        distribution $p_{\mathrm{data}}$ across all 128 states and eight
        cognitive tasks (colours) for the Arrhenius (panel a) and Glauber
        (panel b) models.
        Both scatter panels share the same axis scale.
        The dashed line is the identity.
        \textbf{c},
        Pearson correlation
        $r(p_{\mathrm{model}}, p_{\mathrm{data}})$
        for both models across tasks.
        Error bars show the interquartile range from 200 trajectory
        bootstrap resamples (brain-state sequences resampled with replacement).
        \textbf{d},
        Jensen--Shannon divergence between the model and data
        stationary distributions
        $D_\mathrm{JS}(p_{\mathrm{model}} \| p_{\mathrm{data}})$
        for both models.
        Error bars as in panel c.
        \textbf{e},
        Jensen--Shannon divergence between the model and equilibrium
        stationary distributions,
        $D_\mathrm{JS}(p_{\mathrm{model}} \| \pi_{\mathrm{eq}})$,
        where $\pi_{\mathrm{eq}}$ is the Boltzmann
        equilibrium distribution.
        Error bars show the interquartile range from 100 parameter bootstrap
        resamples
        (the network inference was repeated for 100 trajectory resamples;
            $\beta\bm{J}$, $\beta\bm{h}$, and $A$
        are recomputed for each resample).
        \textbf{f},
        Inference fitting loss: mean squared error between the inferred local
        field $\beta g_k(\bm{s}_\mu)$ and the reference field derived from
        the empirical transition rates
        ($-s_{k;\mu}\ln(w/A)$ for Arrhenius;
        $(s_{k;\mu}/2)\ln(A/w - 1)$ for Glauber),
        averaged over all feasible one-spin-flip transitions.
    Error bars as in panel e.}
    \label{fig:fig-s10}
\end{figure*}

\begin{figure*}[tb]
    \centering
    \includegraphics{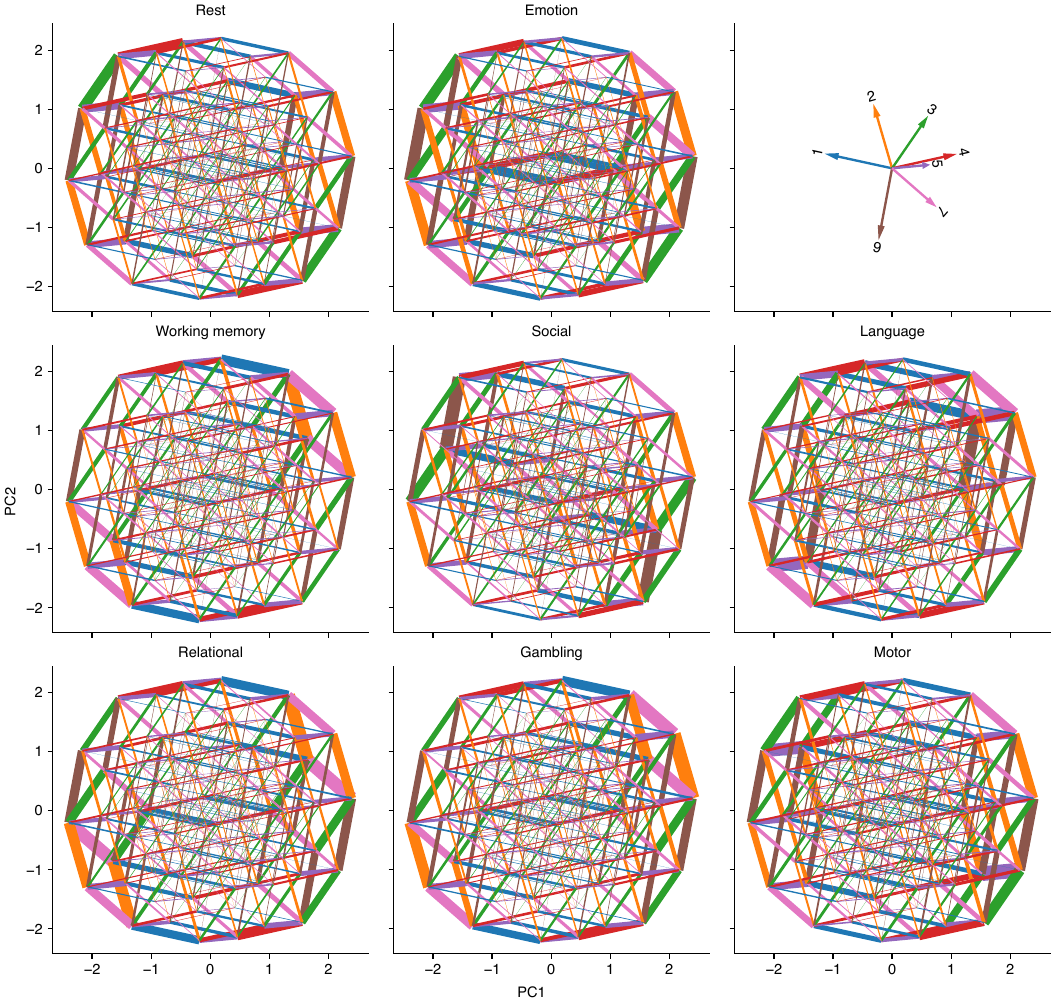}
    \caption{\textbf{Estimated probability traffic of the human brain.}
        Each panel, except for the top right,
        shows the hypercubic probability traffic diagram of
        the task being performed.
        The top right panel shows the biplot vectors, which are the projected
        unit-vector representations of the original dimensions.
        The colour of the biplot vectors corresponds to the colour of
        the coarse-grained brain regions of Fig.~\ref{fig:fig-01}b,c
        of the main text.
        The hypercubic probability traffic diagrams visualize the
        probability traffic on the hypercubic edges,
        where the edge width is proportional to the magnitude of the traffic
        $|\mathcal{T}(\mu, \nu)|$.
        The edge width on the top right panel indicates the
        magnitude $0.002$~s$^{-1}$.
        The colour of the edge corresponds to the colour of the
        coarse-grained brain regions of Fig.~\ref{fig:fig-01}b,c of
        the main text and biplot on the top right.
        The annotation indicates the Ising state vector,
        where the filled (empty) square means active (inactive) state.
        The colour of the filling indicates the brain cluster as
        indicated in the top right panel.
        The hypercubic vertices (binary states) are projected through
    PCA of the task-averaged probability distribution.}
    \label{fig:fig-s17}
\end{figure*}

\end{document}


\title{
    Supplementary Information for\\
    ``Distinct weak antisymmetric interactions shape human brain
    functions as probability fluxes''
}

\author{Yoshiaki Horiike\orcidlink{0009-0000-2010-2598}}
\email{yoshi.h@nagoya-u.jp}
\affiliation{
    Department of Applied Physics, Nagoya University, Nagoya, Japan
}
\affiliation{
    Department of Neuroscience, University of Copenhagen, Copenhagen, Denmark
}

\author{Shin Fujishiro\orcidlink{0000-0002-0127-0761}}
\affiliation{
    Fukui Institute for Fundamental Chemistry, Kyoto University, Kyoto, Japan
}

\renewcommand{\thesection}{\arabic{section}}
\renewcommand{\theequation}{S\arabic{equation}}
\captionsetup[figure]{labelformat=suppfig}
\makeatletter
\let\np@seccntformat\@seccntformat
\def\@seccntformat#1{%
    \ifnum\pdfstrcmp{#1}{section}=0 Supplementary Note~\fi
\np@seccntformat{#1}}
\makeatother

\date{21 October 2026}

\begingroup
\let\clearpage\relax
\let\vfil\relax
\maketitle
\endgroup
\tableofcontents

\newpage

\section{Dependence of the results on the number of clusters}
\label{sec:vary-num-clusters}
In the main text, we fix the number of coarse-grained brain clusters at
seven on the basis of the dendrogram threshold
(Fig.~1b of the main text).
With the seven clusters,
we show that the human brain function corresponds to probability fluxes
(Fig.~3 of the main text)
and such probability fluxes arise from the weak antisymmetric part of
the interaction network (Fig.~4 of the main text).
Here, we verify that the qualitative findings are robust to the
number of clusters.

\subsection{Human brain function as probability fluxes}
First, we verify that the qualitative finding in
Fig.~3a of the main text is robust to the number of clusters.
We repeat the hypercubic probability flux analysis (HyPFA) for
numbers of clusters in $\{4, 5, 6\}$ by rethresholding the same hierarchical
linkage matrix and rerunning the full HyPFA pipeline.
Extended Data Fig.~7 shows the results of the
clustering of the tasks
by probability flux.
We find one cluster
(consisting of Gambling, Motor, Relational, Rest, Social, and Working memory)
and two tasks (Emotion and Language) that have negative cosine
similarity with it
(Extended Data Fig.~7a--c).
This large-scale pattern of clustering is robust to the choice of the
number of clusters
(Extended Data Fig.~7a--c and Fig.~3a).
While the large-scale pattern of clustering is robust,
the hierarchical structure of the largest cluster depends on the
number of clusters (Extended Data Fig.~7d--f).

\subsection{Weak antisymmetric interaction}
We then verify that the results on the weak antisymmetric interaction are
robust to the number of clusters.
Same as Fig.~4a of the main text,
Supplementary Fig.~\ref{fig:fig-s13}a shows the asymmetry ratio
$\|\bm{J}^{(\mathrm{a})}\|_\mathrm{F} / \|\bm{J}^{(\mathrm{s})}\|_\mathrm{F}$
of the inferred interaction matrix per cognitive task at the same
number of clusters,
confirming that the interaction network remains dominated by its symmetric part.
Although the asymmetry ratios show that the interaction network is
dominated by its symmetric part,
the ratio increases as the number of clusters increases
(Supplementary Fig.~\ref{fig:fig-s13}b),
as pointed out in previous
work~\cite{Egolf2000,Esposito2012,Lynn2021,Nartallo-Kaluarachchi2026}.
Nevertheless, the task-dependent brain dynamics at this scale are still
distinct, as shown in Fig.~2 and Fig.~6
of the main text.

\section{Comparing the stationary distribution and the loss of the Arrhenius and
Glauber rate models}
\label{sec:statdist}
Here, we compare the stationary distribution of the inferred model
with the empirical distribution.
The stationary distribution of a transition rate matrix $\bm{W}$ is its
right null vector (the right eigenvector corresponding to eigenvalue zero),
that is, the vector $\bm{p}_0$ satisfying $\bm{W}\bm{p}_0 = 0\bm{p}_0$.
When the inferred interaction matrix $\bm{J}$ is symmetric,
detailed balance
(equation~(23))
ensures that this right null vector equals the canonical distribution
\begin{align}
    \pi(\bm{s})
    \propto
    \exp
    \bm{(}
        -\beta \mathcal{H}(\bm{s})
    \bm{)}
    =
    \exp
    \bigg(
        \frac{1}{2}
        \bm{s}^\top \beta\bm{J} \bm{s}
        +
        \bm{s}^\top \beta\bm{h}
    \bigg)
    .
    \label{eq:boltzmann}
\end{align}
However, the inferred $\bm{J}$ is in general asymmetric,
so detailed balance is broken
(Supplementary Note~\ref{sec:broken-detailed-balance})
and the model stationary distribution
$p_{\mathrm{model}}$ is obtained from the right null vector of
$\bm{W}$ constructed from the inferred parameters
$\{\beta\tilde{\bm{J}}, \beta\tilde{\bm{h}}, \tilde{A}\}$.

We compare the model stationary distribution $p_{\mathrm{model}}$
with the empirically observed brain-state distribution
$p_{\mathrm{data}}$ using the Pearson correlation coefficient $r$ and
the Jensen--Shannon divergence
\begin{align}
    D_\mathrm{JS}(p_{\mathrm{model}} \| p_{\mathrm{data}})
    =
    \frac{1}{2}
    D_{\mathrm{KL}}(p_{\mathrm{model}} \| Q)
    +
    \frac{1}{2}
    D_{\mathrm{KL}}(p_{\mathrm{data}} \| Q)
    ,
    \quad
    Q \coloneqq \frac{p_{\mathrm{model}} + p_{\mathrm{data}}}{2},
    \label{eq:jsd}
\end{align}
which is symmetric
$
D_\mathrm{JS}(p_{\mathrm{model}} \| p_{\mathrm{data}})
=
D_\mathrm{JS}(p_{\mathrm{data}} \| p_{\mathrm{model}})
$.
Note that the Kullback--Leibler divergence is defined as
\begin{align}
    D_{\mathrm{KL}}(p \| q)
    =
    \sum_{i} p_i \log \frac{p_i}{q_i}
    ,
\end{align}
where $p_i$ and $q_i$ are the probabilities of the $i$th state.
We apply trajectory bootstrapping (100 samples):
transition events are resampled
(`Probability flux analysis' in Methods),
the full model
$\{\beta\tilde{\bm{J}}, \beta\tilde{\bm{h}}, \tilde{A}\}$
is re-inferred from each resample, and the interquartile range (IQR) of
the resulting statistics is reported as error bars.

Extended Data Fig.~9a,b compare
$p_{\mathrm{model}}$ against $p_{\mathrm{data}}$ across all 128
states and eight tasks.
Both models reproduce the empirical distribution well
(Arrhenius: $r > 0.95$; Glauber: $r > 0.92$ for all tasks;
Extended Data Fig.~9a,b),
but the task-wise correlation coefficients of the Arrhenius rate
are consistently lower than those of the Glauber rate.

In terms of the Jensen--Shannon divergence,
the Glauber rate achieves a consistently lower Jensen--Shannon divergence
($D_\mathrm{JS} \approx 0.005$~nats; Extended Data Fig.~9c)
than the Arrhenius rate
($D_\mathrm{JS} \approx 0.025$--$0.035$~nats; Extended Data
Fig.~9d),
indicating that the Glauber rate stationary distribution is a
closer approximation to the empirical distribution.
We find that the Glauber rate, as shown in Extended Data
Fig.~9e,
is closer to the canonical distribution (equation~\eqref{eq:boltzmann})
than the Arrhenius rate.

We then examine the converged loss function
(equation~(25))
of the Arrhenius and Glauber rate models.
The final loss of the Arrhenius rate model is typically 10\% of
that of the Glauber rate model, indicating that the Arrhenius rate
model fits better than the Glauber rate model in terms of the transition rates.

\section{Limitations of the data processing and model}
\label{sec:limitations}
The model we introduced has several assumptions and limitations.
The first assumption is that the coarse-graining of the brain regions
does not change the quantitative properties of the data.
We define the seven clusters of parcels based on the hierarchical
clustering to reduce the dimensionality of the data.
In practice, this reduces the computational cost and the effect of
limited data availability; biologically, each of the seven clusters is
related to the specific functions of the brain~\cite{ThomasYeo2011}.
As we show in Fig.~1e of the main text,
our clusters are roughly aligned with the known functional
clusters~\cite{ThomasYeo2011}.

In the spirit of the renormalization group of statistical
physics~\cite{Wilson1979},
we seek an alternative description that reduces the dimensionality of
the data and the state space at once
and that is simple but captures the essence of the system.
The standard block spin
transformation~\cite{Wilson1979}---coarse-graining the system
based on the periodic regular structure of the interaction network---is not
valid for neural networks:
we need to consider the highly irregular and heterogeneous structure
of the interaction networks.
The core idea is to start from empirical correlations rather than
the actual interaction network, which is unknown beforehand~\cite{Meshulam2019}.
If the interactions of the system are local (as expected in the cortex),
the strongest correlation likely arises from the interaction.

The second assumption is that the binarization captures the essential features
of the data.
Considering the binary nature of the underlying neurons of the human brain,
grasping the trend of the activity by binarization seems reasonable,
but further examination is needed to validate what features are not captured.

The third assumption is that the probability distribution is
independent of time, that is, stationary.
The timescale of the neuronal spiking and BOLD fMRI signal is
several orders of magnitude smaller than the duration of scanning,
and we reasonably assume the underlying dynamics to be approximately stationary
over the scanning period (`Examining the stationarity of probability
    distribution' in Methods and
Extended Data Fig.~1).
Assessing the non-stationarity of data is one of the future directions.

Our fourth assumption is the Markov property,
which seems to be a reasonable first step.
Indeed, we successfully reconstruct the pattern of probability fluxes
from the inferred Ising spin system, but the reconstruction is not perfect,
and in the working-memory task, it fails to capture the features of the dynamics
(Fig.~6 of the main text).
Our method may be improved by introducing the higher-order Markov
process, which reduces the error of the probability flux
analysis~\cite{Battle2016,Lynn2021}.

Fifthly, the inferred Ising spin system is based on the assumption of
pairwise interactions between brain regions,
which may be justified by previous work~\cite{Geli2025}.
As we show in Fig.~6 of the main text,
the model struggles to reconstruct the probability flux of the
working-memory task.
This may be due to the simplification of the underlying network structure.
It may be improved by considering higher-order interactions
beyond pairwise interactions~\cite{Schneidman2006,Ganmor2011,Battiston2021},
or by anatomically constrained interaction network
inference~\cite{Pulvermuller2021,Greaves2025,Luppi2026}.

Finally, the biological origin of the external input term of the
inferred model is not clear.
We expect some sensory or perceptual signals to contribute to the
external input, but this needs further investigation.
Nevertheless, as shown in Fig.~4h,i of the main text,
the external input suppresses the brain regions that are active during
the default mode,
which is consistent with the known features of the default mode.

Overcoming these limitations might improve the present method and may
even reveal new aspects of human brain dynamics and underlying
network structure.

\section{Entropy production rate in the human brain}
\label{sec:epr-brain}
We estimated the entropy production rate in the human brain
using the coarse-grained data.
Supplementary Fig.~\ref{fig:fig-s16}a shows the fraction of all
possible single-spin-flip
pairs for which state transitions are observed.
We find that this fraction decreases as the number of clusters increases.
With seven clusters, roughly 95\% of the possible state transitions
are observed.
Supplementary Fig.~\ref{fig:fig-s16}b shows the estimated entropy
production rate
as a function of the number of clusters.
Depending on the number of clusters,
the estimated entropy production rate varies across tasks,
indicating that the entropy production rate depends on the method of
coarse-graining.
We then show the entropy production rate of the seven clusters
in Supplementary Fig.~\ref{fig:fig-s16}c and perform the
Kolmogorov--Smirnov test
between the bootstrapped distributions of the entropy production rate in
Supplementary Fig.~\ref{fig:fig-s16}d.
We find that all pairwise comparisons between tasks are statistically
significant.
Finally, we show the relation between the response
rate~\cite{Lynn2021} and the entropy production rate in
Supplementary Fig.~\ref{fig:fig-s16}e.
Here, the response rate is the frequency (in s$^{-1}$) at which
participants are required to make a physical motor response during a given
task~\cite{Lynn2021}.
The correlation is positive but falls just outside the conventional threshold
($r = 0.689$, $p = 0.059$).
These results indicate that, while tasks differ in their entropy production
rate, the entropy production rate alone does not capture the task-specific
sequential structure of the probability flux reported in the main text.
Note that the entropy production rate reported here is a lower bound
on the \emph{true} entropy production
rate~\cite{Esposito2012,Lynn2022,Lynn2022a},
so it is less meaningful to compare the task-dependent dynamics
through the entropy production
rate~\cite{Nartallo-Kaluarachchi2026}.
The entropy production rate is, however, still useful to determine whether
the system breaks detailed balance~\cite{Nartallo-Kaluarachchi2026a},
and we confirm that the human brain breaks detailed balance at
rest and in all tasks.

\section{Probability traffic diagrams}
\label{sec:traffic}

Here, we introduce the probability traffic diagram,
which shows the usage of the hypercubic edges by the joint transitions.
The probability traffic~\cite{Maes2020} is defined as the sum of the
joint transition
rates between two spin states, $\mu$ and $\nu$,
\begin{align}
    \mathcal{T}(\mu, \nu)
    \coloneqq
    w(\mu \mid \nu)
    p(\nu)
    +
    w(\nu \mid \mu)
    p(\mu)
    ,
\end{align}
which is reversible and symmetric,
$\mathcal{T}(\mu, \nu) = \mathcal{T}(\nu, \mu)$,
and indicates the total joint transition between two adjacent states
rather than the irreversible flux.

We visually examine the probability traffic diagram in Extended Data
Fig.~10.
Unlike the probability flux diagram in Fig.~2 of
the main text,
the probability traffic diagram does not exhibit a clear task-dependent pattern.
Although some hypercubic edges are more likely to be used than others,
qualitative features of the probability traffic are largely not task-dependent.

We quantitatively compare the probability traffic diagrams by
applying hierarchical clustering to them
(Supplementary Fig.~\ref{fig:fig-s18}).
Because probability traffic is symmetric,
we retain only one representative direction per hypercubic edge.
The tasks are clustered with uniformly high cosine similarity
(Supplementary Fig.~\ref{fig:fig-s18}a),
and the dissimilarity between the tasks is smaller ($\sim 0.2$) than
that of probability flux ($\sim 2.0$) in Supplementary Fig.~\ref{fig:fig-s18}b.
The hypercubic edges also show uniform cosine similarity across tasks
(Supplementary Fig.~\ref{fig:fig-s18}c).
Similar to the clustering results of the probability flux diagram
(Fig.~3d of the main text),
the probability traffic diagram also shows that some hypercubic edges
carry consistently more traffic than others, independent of task.
This quantitatively confirms that
probability traffic carries less task-specific information than
probability flux.

We confirm that this lack of task-specificity is robust to the same
methodological choices probed for probability flux.
Using the known functional clustering~\cite{ThomasYeo2011}
(Fig.~1d of the main text) in place of the
hierarchical clustering does not change the result (Supplementary
Fig.~\ref{fig:fig-s19}),
nor does varying the number of clusters (Supplementary Fig.~\ref{fig:fig-s20}),
the trajectory discretization method (Supplementary Fig.~\ref{fig:fig-s21}),
or the choice of PCs (Supplementary Fig.~\ref{fig:fig-s22}).
In every case the probability traffic diagram remains qualitatively
similar across tasks,
in contrast to the corresponding probability flux diagrams
(Extended Data Figs.~3--6).

As a further validation of the inferred model, we reconstruct the
probability traffic diagram from the fitted Ising spin system and
compare it with the empirical probability traffic diagram of
Extended Data Fig.~10,
analogous to the probability flux
reconstruction of Fig.~6 of the main text.
Supplementary Figs.~\ref{fig:fig-s23} and~\ref{fig:fig-s24} show the
reconstructed
probability traffic diagrams for the Arrhenius and Glauber rate models,
respectively.
For both rate models, the reconstructed traffic diagrams are highly
correlated with the empirical ones across all tasks
(Arrhenius: $r = 0.94$--$0.96$; Glauber: $r = 0.93$--$0.95$).
As with the empirical traffic diagrams, the reconstructed traffic
patterns do not exhibit a strong task dependence,
confirming that the probability traffic is less task-dependent than
the probability flux.

\section{Method to minimize loss function}
\label{sec:loss-minimization}
Inference methods for asymmetric interaction matrices have been
proposed in the
literature~\cite{Mezard2011,Roudi2011,Zeng2013,Hoang2019}.
Our method is conceptually different because it uses the inter-state
transition rates rather than the inter-spin correlations.
A weakness of our method is its worse scaling with the system size
due to the exponential growth of the state space,
whereas the previous methods~\cite{Mezard2011,Roudi2011,Zeng2013,Hoang2019}
scale linearly with the system size.
Our method is, however, not restricted to the Hamiltonian of the
(asymmetric) Ising spin systems,
and it can be applied to other interacting systems.
Therefore, our method is more general and applicable to a wider range
of nonequilibrium many-body systems.

The loss
$L = \sum[\ln(w_{\mathrm{model}}) - \ln(w_{\mathrm{data}})]^2$
is minimized to infer $\beta \bm{J}$, $\beta \bm{h}$,
and the rate constant $A$ from the empirical transition rates.
For the Arrhenius rate (equation~(19))
this loss reduces to a single linear least-squares problem for all of
$\{\beta \bm{J}, \beta \bm{h}, A\}$ at once.
For the Glauber rate (equation~(20)), it does
not: the rate constant $A$ must be fixed first, and only then does the
remaining fit for $\{\beta \bm{J}, \beta \bm{h}\}$ become linear.
We first address, in Section~\ref{sec:rate-identifiability} below, the
question of why the rate constant $A$ can (Arrhenius) or cannot
(Glauber) be estimated jointly with the fields, and then give the
explicit linear-algebra formulation for each case.

\subsection{Identifiability of the transition-rate constant}
\label{sec:rate-identifiability}
The Arrhenius and Glauber transition rates
(equations~(19)
and~(20))
are each fitted by minimizing a mean-squared-error loss in the inverted
(reference-field) representation introduced in
equation~(25).
For the Arrhenius rate, the rate constant $A$ is a free parameter,
jointly optimized by gradient descent alongside $\beta \bm{J}$ and
$\beta \bm{h}$.
For the Glauber rate, by contrast, $A$ must be fixed before optimization
begins, or the fit fails to converge.
We explain this asymmetry here, before generalizing both rates to a
spin-dependent constant $A_k$ in Section~\ref{sec:spin-dependent-rate-constant}.

\subsubsection{Arrhenius transition rate: an unbounded rate function}
Inverting the Arrhenius transition rate gives the reference field
\begin{align}
    \beta g_k (\bm{s}_\mu)
    =
    -
    s_{k;\mu}
    \ln
    \bigg(
        \frac{w(\mu^{(k)} \mid \mu)}{A}
    \bigg)
    ,
    \label{eq:arrhenius-reference-field-spinindep}
\end{align}
using $s_{k;\mu}^2 = 1$.
Because $w = A\exp(-\beta s_{k;\mu} g_k)$ is a strictly monotonic bijection
from $\beta g_k \in (-\infty, \infty)$ onto $w \in (0, \infty)$ for every
$A > 0$, rescaling $A \rightarrow \kappa A$ is compensated exactly, for
any $\kappa > 0$ and every transition, by the finite shift
$\beta g_k \rightarrow \beta g_k + s_{k;\mu} \ln \kappa$.
This is a genuine one-parameter symmetry of the raw rate function at every
scale, not only asymptotically.
Nevertheless, the required shift is proportional to $s_{k;\mu}$, the sign of
the very spin whose row is being fitted, and $s_{k;\mu}$ is excluded from
that row's own predictors by construction ($J_{k,k}=0$).
A change in $A$ therefore cannot be absorbed into $\beta \bm{J}_{k,\cdot}$
or $\beta h_k$ without changing every fitted field by an amount that
generically fails to match $s_{k;\mu}\ln(\kappa)$ pointwise across the
many transitions used to fit row $k$.
Away from the correct $A$, the least-squares residual strictly increases,
so the joint gradient descent over $\{\beta \bm{J}, \beta \bm{h}, A\}$
converges to a unique, well-identified rate constant, consistent with
its treatment as a free parameter.

\subsubsection{Glauber: a saturating rate function}
Inverting the Glauber transition rate gives the analogous reference field
(equation~\eqref{eq:glauber-reference-field} below), defined only for
$w(\mu^{(k)} \mid \mu) < A$, using $s_{k;\mu}^2=1$.
Unlike the Arrhenius rate, the Glauber rate
$w = A / \bm{(} 1+\exp(2\beta s_{k;\mu}g_k) \bm{)}$
is bounded: it saturates at $0$ and at $A$ as
$\beta s_{k;\mu} g_k \rightarrow \pm \infty$.
In the regime $A \gg w$,
equation~\eqref{eq:glauber-reference-field} reduces to
\begin{align}
    \beta g_k (\bm{s}_\mu)
    \approx
    \frac{s_{k;\mu}}{2}
    \ln (A)
    -
    \frac{s_{k;\mu}}{2}
    \ln \bm{(}w(\mu^{(k)} \mid \mu)\bm{)}
    ,
    \label{eq:glauber-reference-field-asymptotic}
\end{align}
the same $s_{k;\mu}$-proportional dependence on $\ln (A)$ found for the
Arrhenius rate above, but here only as an asymptotic approximation, valid
once $A$ is much larger than the observed rate.
Because the underlying rate function saturates, nothing prevents the
fitted fields from being pushed arbitrarily far into this regime: as
$A \rightarrow \infty$, the reference field
(equation~\eqref{eq:glauber-reference-field-asymptotic}) diverges for
every transition, and the achievable loss approaches a flat asymptote
rather than increasing without bound, so gradient descent has no
incentive to stop increasing $A$.
This asymptotically flat direction, absent for the unbounded Arrhenius
rate, is why $A$ cannot be left free for the Glauber rate.

\paragraph{Fixing the rate constant.}
To break this degeneracy, $A$ is not optimized but is fixed at
$A = 2\hat{w}_{99}$, where $\hat{w}_{99}$ is the 99th percentile of the
observed transition rates.
The 99th-percentile ceiling excludes rare-state outliers: $w$ diverges
when the source state $\mu$ has near-zero probability, and these
large-rate transitions would otherwise inflate $A$ and distort the
inferred fields.
Denoting the $n$th observed transition as a flip of spin $k(n)$ from
source state $\bm{s}_n$ with empirical rate
$w_n = n(\mu^{(k)}, \mu) / [\tau\, p(\mu)]$, the loss is
\begin{align}
    L(\beta\bm{J}, \beta\bm{h})
    =
    \frac{1}{N_\mathrm{feas}}
    \sum_{n:\, w_n < A}
    \bigg[
        \beta\bm{J}_{k(n),\cdot} \bm{s}_n + \beta h_{k(n)}
        -
        \frac{s_{k(n);\bm{s}_n}}{2}
        \ln
        \bigg(
            \frac{A}{w_n} - 1
        \bigg)
    \bigg]^2
    ,
    \label{eq:loss-glauber-spinindep}
\end{align}
where $N_\mathrm{feas}$ is the number of feasible transitions (those
satisfying $w_n < A$); transitions above $A$ contribute negligibly to
the probability flux and are excluded from the loss via this
feasibility mask.

\paragraph{Optimization.}
With $A$ fixed, $\beta\bm{J}$ and $\beta\bm{h}$ are updated by gradient
descent
\begin{align}
    \beta\bm{J}^{(t+1)}
    &=
    \beta\bm{J}^{(t)}
    -
    \alpha
    \,
    \nabla_{\beta\bm{J}}\,
    L
    (
        \beta\bm{J}^{(t)},
        \beta\bm{h}^{(t)}
    )
    ,
    \label{eq:glauber-gd-J}
    \\
    \beta\bm{h}^{(t+1)}
    &=
    \beta\bm{h}^{(t)}
    -
    \alpha
    \,
    \nabla_{\beta\bm{h}}\,
    L
    (
        \beta\bm{J}^{(t)},
        \beta\bm{h}^{(t)}
    )
    ,
    \label{eq:glauber-gd-h}
\end{align}
with learning rate $\alpha = 10^{-3}$ throughout.
The number of steps $N_\mathrm{steps}$ depends on the fitting context:
the point estimate reported for each task uses
$N_\mathrm{steps} = 4\times 10^{4}$ steps;
each of the $100$ trajectory-bootstrap replicates
(Extended Data Fig.~9e,f) uses
$N_\mathrm{steps} = 2\times 10^{4}$ steps,
fewer per replicate for computational tractability;
and the synthetic-data validation against a known ground truth
uses $N_\mathrm{steps} = 2\times 10^{5}$ steps.
Gradients are computed by automatic differentiation (JAX\@).
Parameters are initialized at zero.

\subsection{Arrhenius: a single linear system}
Using $(\bm{s}_{\mu^{(k)}} - \bm{s}_\mu)^\top = -2 s_{k;\mu} \bm{e}_k^\top$
for a spin-$k$ flip ($\bm{e}_k$ the $k$th standard basis vector),
the loss function of equation~(25) becomes
\begin{align}
    L
    (
        \beta \bm{J}, \beta \bm{h}, A
    )
    \coloneqq
    \sum_{(\mu^{(k)}, \mu)}
    \bigg[
        2 \ln (A)
        +
        (
            -2
            s_{k;\mu}
        )
        \bm{e}_k^\top
        (
            \beta
            \bm{J}
            \bm{s}_\mu
            +
            \beta
            \bm{h}
        )
        -
        2
        \ln
        \bigg(
            \frac{1}{\tau}
            \frac{n(\mu^{(k)}, \mu)}{p(\mu)}
        \bigg)
    \bigg]^2
    ,
\end{align}
which is an affine function of $\beta \bm{J}$, $\beta \bm{h}$,
and $\eta \coloneqq 2\ln (A)$ for every transition.
Denoting the $m$th observed transition as a flip of spin $k(m)$ from
source state $\bm{s}_m$ with empirical rate $w_m$, each transition
contributes one row of a linear matrix equation
\begin{align}
    \bm{C}\bm{x} = \bm{b}
    ,
\end{align}
where
\begin{align}
    \bm{x}
    \coloneqq
    [
        \mathrm{vec}(\beta \bm{J}) \;
        \eta
    ]^\top
    ,
    \qquad
    b_m
    \coloneqq
    2 \ln(w_m)
    ,
    \label{eq:arrhenius-lstsq-system}
\end{align}
where row $m$ of the matrix $\bm{C}$ has entries
$-2 s_{k(m);\bm{s}_m} s_{j;\bm{s}_m}$ at the position $(J_{k(m),j})$
($j \neq k(m)$),
$-2 s_{k(m);\bm{s}_m}$ at the position $(h_{k(m)})$
(the diagonal position $(J_{k(m),k(m)})$, otherwise unused since
$J_{k,k}=0$, is reused to store this coefficient), and $1$ at the position
for $\eta$.

\subsection{Glauber: fixing $A$ linearizes the remaining fit}
Because the Glauber rate is
$\ln(w) = \ln (A) - \ln\bm{(}1 + \exp(2\beta s_{k;\mu} g_k)\bm{)}$,
the second term is a nonlinear function of $g_k$ that cannot be
rearranged to isolate $\ln (A)$ as an additive constant the way the
Arrhenius exponential can; there is no analogue of
equation~\eqref{eq:arrhenius-lstsq-system} with $A$ included as a free
unknown.
$A$ is therefore fixed before the fields are fit, at
$A = 2\hat{w}_{99}$, as derived in
Section~\ref{sec:rate-identifiability} above.
With $A$ fixed, inverting the rate equation gives a fully-determined
reference field for every feasible ($w < A$) transition,
\begin{align}
    \beta g_k(\bm{s}_\mu)
    =
    \frac{s_{k;\mu}}{2}
    \ln
    \bigg(
        \frac{A}{w(\mu^{(k)} \mid \mu)} - 1
    \bigg)
    ,
    \label{eq:glauber-reference-field}
\end{align}
using $s_{k;\mu}^2=1$.
Because $\beta g_k(\bm{s}_\mu) = \beta \bm{J}_{k,\cdot} \bm{s}_\mu + \beta h_k$
involves only row $k$'s own parameters, and $A$ is no longer an unknown
linking the spins together, the fit decouples into $N$ independent
per-spin linear systems $\bm{C}_k \bm{x}_k = \bm{b}_k$: for each feasible
transition flipping spin $k$ from state $\bm{s}_n$ with rate $w_n$, one
row has entries $s_{j;\bm{s}_n}$ at column $J_{k,j}$ ($j \neq k$), $1$ at
the column for $h_k$, and target
\begin{align}
    b_n
    \coloneqq
    \frac{s_{k;\bm{s}_n}}{2}
    \ln
    \bigg(
        \frac{A}{w_n} - 1
    \bigg)
    .
\end{align}
Each of the $N$ systems is solved independently,
in contrast to the single combined system in
equation~\eqref{eq:arrhenius-lstsq-system} for the Arrhenius rate.
On synthetic data generated from known $\{\beta \bm{J}, \beta \bm{h}, A\}$,
this closed-form solve recovers the ground truth to machine precision when
$A$ is known exactly, and agrees with the existing gradient-descent
inference
to within the optimizer's convergence tolerance
when both use the same fixed $A$.

\section{Broken detailed balance and entropy production arising from
antisymmetric interactions}
\label{sec:broken-detailed-balance}

\subsection{Broken detailed balance and antisymmetric interactions}
Here, we derive the broken detailed balance of
equation~(24).
We recall that the (pseudo-)Hamiltonian of the Ising spin system is
given by equation~(15):
\begin{equation}
    \mathcal{H} (\bm{s})
    \coloneqq
    -
    \frac{1}{2}
    \sum_{i=1}^{N} \sum_{j=1}^{N}
    s_i J_{i,j} s_j
    -
    \sum_{i=1}^{N}
    s_i
    h_i
    =
    -
    \frac{1}{2}
    \bm{s}^\top \bm{J} \bm{s}
    -
    \bm{s}^\top \bm{h}
    \nonumber
    ,
\end{equation}
and the transition rate of flipping the $k$th spin is given by either
the Arrhenius transition rate
[equation~(19)]
\begin{align}
    w(\mu^{(k)} \mid \mu)
    =
    A
    \exp
    \bm{(}
        -
        \beta
        s_{k; \mu}
        g_{k}(\bm{s}_\mu)
    \bm{)}
    ,
\end{align}
or the Glauber transition rate
[equation~(20)]
\begin{align}
    w(\mu^{(k)} \mid \mu)
    =
    A
    \frac{1}{
        1
        +
        \exp
        \bm{(}
            2
            \beta
            s_{k; \mu}
            g_{k}(\bm{s}_\mu)
        \bm{)}
    }
    .
\end{align}
To show that the detailed balance is broken if the interaction matrix is
asymmetric, we decompose the (pseudo-)Hamiltonian
(equation~(15)) into the spin $k$ independent and
dependent parts as follows:
\begin{align}
    \mathcal{H} (\bm{s}_\mu)
    &=
    -
    \frac{1}{2}
    \sum_{i=1, i \neq k}^{N}
    s_{i; \mu}
    \sum_{j=1, j \neq k}^{N}
    J_{i, j} s_{j; \mu}
    \underbrace{
        -
        \frac{1}{2}
        s_{k; \mu}
        \sum_{j=1}^{N}
        J_{k, j}
        s_{j; \mu}
    }_{
        =
        \,
        -
        s_{k; \mu}
        \sum_{j=1}^{N}
        J_{k, j}
        s_{j; \mu}
        +
        \frac{1}{2}
        s_{k; \mu}
        \sum_{j=1}^{N}
        J_{k, j}
        s_{j; \mu}
    }
    -
    \frac{1}{2}
    \sum_{i=1}^{N}
    s_{i; \mu}
    J_{i, k}
    s_{k; \mu}
    -
    \sum_{i=1, i \neq k}^{N}
    s_{i; \mu} h_i
    -
    s_{k; \mu} h_k
    \nonumber
    \\&=
    -
    \frac{1}{2}
    \sum_{i=1, i \neq k}^{N}
    \sum_{j=1, j \neq k}^{N}
    s_{i; \mu}
    J_{i, j} s_{j; \mu}
    -
    \sum_{i=1, i \neq k}^{N}
    s_{i; \mu} h_i
    -
    \frac{1}{2}
    s_{k; \mu}
    \sum_{i=1}^{N}
    J_{i, k}
    s_{i; \mu}
    +
    \frac{1}{2}
    s_{k; \mu}
    \sum_{i=1}^{N}
    J_{k, i}
    s_{i; \mu}
    +
    \underbrace{
        \Bigg(
            -
            s_{k; \mu}
            \sum_{i=1}^{N}
            J_{k, i}
            s_{i; \mu}
            -
            s_{k; \mu} h_k
        \Bigg)
    }_{
        =
        \,
        -
        s_{k; \mu}
        g_k (\bm{s}_\mu)
    }
    \nonumber
    \\&=
    \underbrace{
        -
        \frac{1}{2}
        \sum_{i=1, i \neq k}^{N}
        \sum_{j=1, j \neq k}^{N}
        s_{i; \mu}
        J_{i, j}
        s_{j; \mu}
        -
        \sum_{i=1, i \neq k}^{N}
        s_{i; \mu} h_i
    }_{\textrm{independent of } s_{k; \mu}}
    \underbrace{
        -
        s_{k; \mu}
        g_k (\bm{s}_\mu)
        +
        \frac{1}{2}
        s_{k; \mu}
        \sum_{i=1}^{N}
        (
            J_{k, i}
            -
            J_{i, k}
        )
        s_{i; \mu}
    }_{\textrm{dependent on } s_{k; \mu}}
    .
\end{align}
Thus, we have the input signal as
\begin{equation}
    s_{k; \mu}
    g_k (\bm{s}_\mu)
    =
    -
    \mathcal{H} (\bm{s}_\mu)
    -
    \frac{1}{2}
    \sum_{i=1, i \neq k}^{N}
    \sum_{j=1, j \neq k}^{N}
    s_{i; \mu}
    J_{i, j}
    s_{j; \mu}
    -
    \sum_{i=1, i \neq k}^{N}
    s_{i; \mu} h_i
    +
    \frac{1}{2}
    s_{k; \mu}
    \sum_{i=1}^{N}
    (
        J_{k, i}
        -
        J_{i, k}
    )
    s_{i; \mu}
    .
\end{equation}
Similarly, we have
\begin{equation}
    \mathcal{H} (\bm{F}_k \bm{s}_\mu)
    =
    \underbrace{
        -
        \frac{1}{2}
        \sum_{i=1, i \neq k}^{N}
        \sum_{j=1, j \neq k}^{N}
        s_{i; \mu}
        J_{i, j}
        s_{j; \mu}
        -
        \sum_{i=1, i \neq k}^{N}
        s_{i; \mu} h_i
    }_{\textrm{independent of } s_{k; \mu}}
    \underbrace{
        +
        s_{k; \mu}
        g_k (\bm{s}_\mu)
        -
        \frac{1}{2}
        s_{k; \mu}
        \sum_{i=1}^{N}
        (
            J_{k, i}
            -
            J_{i, k}
        )
        s_{i; \mu}
    }_{\textrm{dependent on } s_{k; \mu}}
    ,
\end{equation}
and therefore,
\begin{equation}
    -
    s_{k; \mu}
    g_k (\bm{s}_\mu)
    =
    -
    \mathcal{H} (\bm{F}_k \bm{s}_\mu)
    -
    \frac{1}{2}
    \sum_{i=1, i \neq k}^{N}
    \sum_{j=1, j \neq k}^{N}
    s_{i; \mu}
    J_{i, j} s_{j; \mu}
    -
    \sum_{i=1, i \neq k}^{N}
    s_{i; \mu} h_i
    -
    \frac{1}{2}
    s_{k; \mu}
    \sum_{i=1}^{N}
    (
        J_{k, i}
        -
        J_{i, k}
    )
    s_{i; \mu}
    .
\end{equation}
Then the ratio of the transition rates is calculated as
\begin{align}
    \frac{
        w(\mu^{(k)} \mid \mu)
    }{
        w(\mu \mid \mu^{(k)})
    }
    =
    \frac{
        \exp
        \bm{(}
            -
            \beta
            s_{k; \mu}
            g_{k} (\bm{s}_\mu)
        \bm{)}
    }{
        \exp
        \bm{(}
            +
            \beta
            s_{k; \mu}
            g_{k} (\bm{s}_\mu)
        \bm{)}
    }
    =
    \frac{
        \exp
        \bm{(}
            -
            \beta
            \mathcal{H}
            (\bm{F}_k \bm{s}_\mu)
        \bm{)}
    }{
        \exp
        \bm{(}
            -
            \beta
            \mathcal{H}
            (\bm{s}_\mu)
        \bm{)}
    }
    \exp
    \Bigg(
        -
        \beta
        s_{k; \mu}
        \sum_{i=1}^{N}
        (
            J_{k, i}
            -
            J_{i, k}
        )
        s_{i; \mu}
    \Bigg)
    \label{eq:transition-rate-ratio-ising}
    ,
\end{align}
which is equivalent to equation~(24).
Therefore, the antisymmetric interaction breaks the detailed balance, that is,
the broken detailed balance arises from the antisymmetric part of the
interaction.
See also
Refs.~\cite{Aguilera2021,Fruchart2021,Nartallo-Kaluarachchi2024,Nartallo-Kaluarachchi2026}
for related discussions on the antisymmetric interaction and broken
detailed balance.

\subsection{Entropy production}
Based on the stochastic
thermodynamics~\cite{VanDenBroeck2013,VanDenBroeck2015},
we formulate the entropy production arising from the antisymmetric interactions.
\subsubsection{Trajectory}
We define a path, $ \underset{\rightarrow}{\varPi} $,
as a sequence of states of a system in a time interval
$[t_\mathrm{i}, t_\mathrm{f}]$.
\begin{equation}
    \underset{\rightarrow}{\varPi}
    \coloneqq
    \{
        \mu(t_0 = t_\mathrm{i})
        \rightarrow
        \mu(t_1)
        \rightarrow
        \mu(t_2)
        \rightarrow
        \mu(t_3)
        \rightarrow
        \cdots
        \rightarrow
        \mu(t_M)
    \}
\end{equation}
where $\mu(t)$ is a state of the system at time $t$,
and
$t_\mathrm{i} = t_0 < t_1 < t_2 < t_3 < \cdots < t_M < t_{M+1}= t_\mathrm{f}$.
Note that
$\mu(t_a) \neq \mu(t_{a+1})$
and
$\mu(t_{\mathrm{f}}) = \mu(t_{m+1}) = \mu(t_M)$.
The trajectory is the time evolution of state following the path, and
it has $M$ state transitions during the time interval
$[t_\mathrm{i}, t_\mathrm{f}]$ at time points $\{t_a\}_{a=1}^M$.
Formally, the trajectory is defined as
\begin{align}
    \mu[
        \underset{\rightarrow}{\varPi}
    ]
    \coloneqq
    \begin{cases}
        \mu(t_0)
        &
        \text{if} \quad
        t = t_0 \\
        \mu(t_a)
        &
        \text{if} \quad
        t_a \leq t < t_{a+1}, \, a \in \{1, 2, \ldots, M\} \\
        \mu(t_{M})
        &
        \text{if} \quad
        t = t_{M+1}
    \end{cases}
    .
\end{align}
The corresponding energy trajectory is given as
\begin{align}
    \underset{\rightsquigarrow}{E}
    [
        \underset{\rightarrow}{\varPi}
    ]
    \coloneqq
    \{
        &
        E\bm{(}\mu(t_0); t_0\bm{)} \rightsquigarrow
        E\bm{(}\mu(t_0); t_1\bm{)}
        \rightarrow
        \nonumber
        \\&
        E\bm{(}\mu(t_1); t_1\bm{)} \rightsquigarrow
        E\bm{(}\mu(t_1); t_2\bm{)}
        \rightarrow
        \nonumber
        \\&
        E\bm{(}\mu(t_2); t_2\bm{)} \rightsquigarrow
        E\bm{(}\mu(t_2); t_3\bm{)}
        \rightarrow
        \nonumber
        \\&
        \cdots
        \rightarrow
        \nonumber
        \\&
        E\bm{(}\mu(t_{M-1}); t_{M-1}\bm{)} \rightsquigarrow
        E\bm{(}\mu(t_{M-1}); t_{M}\bm{)}
        \rightarrow
        \nonumber
        \\&
        E\bm{(}\mu(t_M); t_{M}\bm{)} \rightsquigarrow
        E\bm{(}\mu(t_M); t_{M+1}\bm{)}
    \}
    .
\end{align}
Here, the arrow ``$\rightarrow$'' indicates the state transition,
and the squiggly arrow ``$\rightsquigarrow$'' indicates the time-evolution
of the energy level.
Along a trajectory, the stochastic energy of the system is defined as
\begin{equation}
    \varepsilon(t)
    \coloneqq
    \begin{cases}
        E\bm{(}\mu(t); t\bm{)}
        &
        \text{if} \quad
        t \neq t_a, \, a \in \{1, 2, \ldots, M\} \\
        \lim_{t \to t_a + 0} E\bm{(}\mu(t); t\bm{)}
        &
        \text{otherwise}
    \end{cases}
    .
\end{equation}

\subsubsection{First law of trajectory stochastic thermodynamics}
The change of the stochastic energy before and after the trajectory
is defined as
\begin{align}
    \underset{\rightarrow}{\Delta}
    \varepsilon
    &
    \coloneqq
    \varepsilon (t_\mathrm{f})
    -
    \varepsilon (t_\mathrm{i})
    \\&=
    E\bm{(} \mu(t_{\mathrm{f}}); t_{\mathrm{f}} \bm{)}
    -
    E\bm{(} \mu(t_{\mathrm{i}}); t_{\mathrm{i}} \bm{)}
    \\&=
    E\bm{(} \mu(t_{M+1}); t_{M+1} \bm{)}
    -
    E\bm{(} \mu(t_{0}); t_{0} \bm{)}
    \\&=
    E\bm{(} \mu(t_{M}); t_{M+1} \bm{)}
    -
    E\bm{(} \mu(t_{0}); t_{0} \bm{)}
    .
\end{align}
The first law of stochastic thermodynamics states that
the energy change is decomposed into stochastic heat and stochastic work:
\begin{equation}
    \underset{\rightarrow}{\Delta}
    \varepsilon
    =
    \underset{\rightarrow}{q}
    +
    \underset{\rightsquigarrow}{w}
    .
\end{equation}
Heat in trajectory, or stochastic heat, is defined as the sum of the heat
absorbed by the system at each state transition:
\begin{align}
    \underset{\rightarrow}{q}
    &\coloneqq
    \sum_{a=1}^{M}
    [
        \varepsilon (t_{a} + 0)
        -
        \varepsilon (t_{a} - 0)
    ]
    \\&=
    \sum_{a=1}^{M}
    [
        E\bm{(} \mu(t_a); t_a \bm{)}
        -
        E\bm{(} \mu(t_{a-1}); t_a \bm{)}
    ]
    \\&=
    [
        E\bm{(} \mu(t_1); t_1 \bm{)}
        -
        E\bm{(} \mu(t_0); t_1 \bm{)}
    ]
    \nonumber
    \\&\quad\,+
    [
        E\bm{(} \mu(t_2); t_2 \bm{)}
        -
        E\bm{(} \mu(t_1); t_2 \bm{)}
    ]
    \nonumber
    \\&\quad\,+
    [
        E\bm{(} \mu(t_3); t_3 \bm{)}
        -
        E\bm{(} \mu(t_2); t_3 \bm{)}
    ]
    \nonumber
    \\&\quad\,+
    \cdots
    \nonumber
    \\&\quad\,+
    [
        E\bm{(} \mu(t_M); t_M \bm{)}
        -
        E\bm{(} \mu(t_{M-1}); t_M \bm{)}
    ]
    \nonumber
    .
\end{align}
This trajectory heat is the sum of the energy change of the system at
each state transition.
The trajectory heat can be written using the transition rates of the
master equation.
\begin{equation}
    E(\mu; t)
    -
    E(\nu; t)
    =
    -
    k_\mathrm{B} T
    \ln
    \bigg(
        \frac{
            w(\mu \mid \nu; t)
        }{
            w(\nu \mid \mu; t)
        }
    \bigg)
    \label{eq:energy-transition-rate-relation}
    .
\end{equation}
Using equation~\eqref{eq:energy-transition-rate-relation},
stochastic heat is rewritten as
\begin{equation}
    \underset{\rightarrow}{q}
    =
    -
    k_\mathrm{B}
    T
    \sum_{a=1}^{M}
    \ln
    \bigg(
        \frac{
            w\bm{(}\mu(t_a) \mid \mu(t_{a-1}); t_a\bm{)}
        }{
            w\bm{(}\mu(t_{a-1}) \mid \mu(t_a); t_a\bm{)}
        }
    \bigg)
    .
\end{equation}
Stochastic work is defined as the sum of the work done on the system
during the time-evolution of the energy level:
\begin{align}
    \underset{\rightsquigarrow}{w}
    &\coloneqq
    \sum_{a=0}^{M}
    [
        \varepsilon (t_{a+1} - 0)
        -
        \varepsilon (t_{a} + 0)
    ]
    \\&=
    \sum_{a=0}^{M}
    [
        E\bm{(}\mu(t_{a}); t_{a+1}\bm{)}
        -
        E\bm{(}\mu(t_{a}); t_{a}\bm{)}
    ]
    \\&=
    [
        E\bm{(}\mu(t_{0}); t_{1}\bm{)}
        -
        E\bm{(}\mu(t_{0}); t_{0}\bm{)}
    ]
    \nonumber
    \\&\quad\,+
    [
        E\bm{(}\mu(t_{1}); t_{2}\bm{)}
        -
        E\bm{(}\mu(t_{1}); t_{1}\bm{)}
    ]
    \nonumber
    \\&\quad\,+
    [
        E\bm{(}\mu(t_{2}); t_{3}\bm{)}
        -
        E\bm{(}\mu(t_{2}); t_{2}\bm{)}
    ]
    \nonumber
    \\&\quad\,+
    \cdots
    \nonumber
    \\&\quad\,+
    [
        E\bm{(}\mu(t_{M-1}); t_{M}\bm{)}
        -
        E\bm{(}\mu(t_{M-1}); t_{M-1}\bm{)}
    ]
    \nonumber
    \\&\quad\,+
    [
        E\bm{(}\mu(t_{M}); t_{M+1}\bm{)}
        -
        E\bm{(}\mu(t_{M}); t_{M}\bm{)}
    ]
    \nonumber
    .
\end{align}

\subsubsection{Second law of trajectory stochastic thermodynamics}
Along a trajectory, the entropy of the system is defined as the
stochastic entropy.
\begin{equation}
    \varsigma(t)
    \coloneqq
    S\bm{(}\mu(t); t\bm{)}
    =
    -k_\mathrm{B}
    \ln
    [ p\bm(\mu(t); t\bm{)} ]
    .
\end{equation}
Entropy change of system before and after the trajectory,
or trajectory entropy production of the system,
is defined as
\begin{align}
    \underset{\rightarrow}{\Delta}
    \varsigma
    &
    \coloneqq
    \varsigma (t_\mathrm{f})
    -
    \varsigma (t_\mathrm{i})
    \\&=
    -k_\mathrm{B}
    \ln
    [
        p\bm{(}\mu(t_\mathrm{f}); t_\mathrm{f}\bm{)}
    ]
    +
    k_\mathrm{B}
    \ln
    [
        p\bm{(}\mu(t_\mathrm{i}); t_\mathrm{i}\bm{)}
    ]
    .
\end{align}
The trajectory entropy production of the reservoir is defined as
\begin{align}
    \underset{\rightarrow}{\Delta}
    \varsigma_\mathrm{res}
    &\coloneqq
    -
    \frac{1}{T}
    \underset{\rightarrow}{q}
    \\&=
    k_\mathrm{B}
    \sum_{a=1}^{M}
    \ln
    \bigg(
        \frac{
            w\bm{(}\mu(t_a) \mid \mu(t_{a-1}); t_a\bm{)}
        }{
            w\bm{(}\mu(t_{a-1}) \mid \mu(t_a); t_a\bm{)}
        }
    \bigg)
    .
\end{align}
Thus, the total trajectory entropy production is
\begin{align}
    \underset{\rightarrow}{\Delta}
    \varsigma_\mathrm{tot}
    &=
    \underset{\rightarrow}{\Delta}
    \varsigma
    +
    \underset{\rightarrow}{\Delta}
    \varsigma_\mathrm{res}
    \\&=
    -k_\mathrm{B}
    \ln
    [
        p\bm{(}\mu(t_\mathrm{f}); t_\mathrm{f}\bm{)}
    ]
    +
    k_\mathrm{B}
    \ln
    [
        p\bm{(}\mu(t_\mathrm{i}); t_\mathrm{i}\bm{)}
    ]
    +
    k_\mathrm{B}
    \sum_{a=1}^{M}
    \ln
    \bigg(
        \frac{
            w\bm{(}\mu(t_a) \mid \mu(t_{a-1}); t_a\bm{)}
        }{
            w\bm{(}\mu(t_{a-1}) \mid \mu(t_a); t_a\bm{)}
        }
    \bigg)
    \label{eq:entropy-production-trajectory}
    .
\end{align}
As we discuss later, the total trajectory entropy
production can be negative.

\subsubsection{Cycle entropy production}
In the nonequilibrium steady state (NESS),
the probability fluxes form cycles in the state space.
Here we consider the entropy production along a cycle.
As a special case of the path, we consider a cycle
\begin{equation}
    \underset{\circlearrowleft}{\varPi}
    \coloneqq
    \{
        \mu(t_0 = t_\mathrm{i}) \to
        \mu(t_1) \to
        \mu(t_2) \to
        \mu(t_3) \to
        \cdots \to
        \mu(t_{M} = t_\mathrm{f}) = \mu(t_0 = t_\mathrm{i})
    \}
    \label{eq:cycle}
    .
\end{equation}
The total cycle entropy production
[equation~\eqref{eq:entropy-production-trajectory}] is then
\begin{align}
    \underset{\circlearrowleft}{\Delta}
    \varsigma_\mathrm{tot}
    &=
    -
    k_\mathrm{B}
    \ln
    [
        p\bm{(}\mu(t_{M}); t_\mathrm{f}\bm{)}
    ]
    +
    k_\mathrm{B}
    \ln
    [
        p\bm{(}\mu(t_{0}); t_\mathrm{i}\bm{)}
    ]
    +
    k_\mathrm{B}
    \sum_{a=1}^{M}
    \ln
    \bigg(
        \frac{
            w\bm{(}\mu(t_a) \mid \mu(t_{a-1}); t\bm{)}
        }{
            w\bm{(}\mu(t_{a-1}) \mid \mu(t_a); t\bm{)}
        }
    \bigg)
    .
\end{align}
In the NESS, the probability and the transition rate
are independent of time; we thus have
\begin{align}
    \underset{\circlearrowleft}{\Delta}
    \varsigma_\mathrm{tot}
    &\stackrel{\mathrm{NESS}}{=}
    -
    k_\mathrm{B}
    \ln
    [
        p\bm{(}\mu(t_{0})\bm{)}
    ]
    +
    k_\mathrm{B}
    \ln
    [
        p\bm{(}\mu(t_{0})\bm{)}
    ]
    +
    k_\mathrm{B}
    \sum_{a=1}^{M}
    \ln
    \bigg(
        \frac{
            w\bm{(}\mu(t_a) \mid \mu(t_{a-1})\bm{)}
        }{
            w\bm{(}\mu(t_{a-1}) \mid \mu(t_a)\bm{)}
        }
    \bigg)
    \nonumber
    \\&\stackrel{\phantom{\mathrm{NESS}}}{=}
    k_\mathrm{B}
    \sum_{a=1}^{M}
    \ln
    \bigg(
        \frac{
            w\bm{(}\mu(t_a) \mid \mu(t_{a-1})\bm{)}
        }{
            w\bm{(}\mu(t_{a-1}) \mid \mu(t_a)\bm{)}
        }
    \bigg)
    \label{eq:cycle-entropy-production-sum}
    \\&\stackrel{\phantom{\mathrm{NESS}}}{=}
    k_\mathrm{B}
    \ln
    \Bigg(
        \prod_{a=1}^{M}
        \frac{
            w\bm{(}\mu(t_a) \mid \mu(t_{a-1})\bm{)}
        }{
            w\bm{(}\mu(t_{a-1}) \mid \mu(t_a)\bm{)}
        }
    \Bigg)
    \label{eq:cycle-entropy-production-product}
    ,
\end{align}
where the system entropy production is zero because the initial and final
states are the same.
Equations~\eqref{eq:cycle-entropy-production-sum}
and~\eqref{eq:cycle-entropy-production-product} give the entropy
production per state transition along the cycle.
Note that the backward cycle produces the negative of entropy
production produced by forward cycle:
\begin{equation}
    \underset{\circlearrowright}{\Delta}
    \varsigma_\mathrm{tot}
    =
    -
    \underset{\circlearrowleft}{\Delta}
    \varsigma_\mathrm{tot}
    .
\end{equation}

\subsubsection{Cycle entropy production arising from antisymmetric interactions}
We then calculate the cycle entropy production
[equations~\eqref{eq:cycle-entropy-production-sum}
and~\eqref{eq:cycle-entropy-production-product}]
in the Ising spin system with asymmetric interaction.
Rather than specifying the cycle as a sequence of states
[equation~\eqref{eq:cycle}],
we specify the cycle by a reference state $\bm{s}_\mu$ in the cycle and sequence
of spins flipped from the state:
\begin{equation}
    \underset{\circlearrowleft}{\varPi}
    =
    \Bigg\{
        \bm{s}_\mu
        ;
        k \rightarrow l \rightarrow m \rightarrow \cdots
        \mid
        \# k = 2, \forall k; \sum_k \# k = M
    \Bigg\}
\end{equation}
where $\# k$ is the number of times the spin $k$ is flipped in the cycle.
By applying equation~\eqref{eq:transition-rate-ratio-ising},
into equation~\eqref{eq:cycle-entropy-production-sum}, we obtain
\begin{align}
    \underset{\circlearrowleft}{\Delta}
    \varsigma_\mathrm{tot}
    &=
    k_\mathrm{B}
    \sum_{a=1}^{M}
    \ln
    \bigg(
        \frac{
            w\bm{(} \mu(a+1) \mid \mu(a) \bm{)}
        }{
            w\bm{(} \mu(a) \mid \mu(a+1) \bm{)}
        }
    \bigg)
    \nonumber
    \\&=
    k_\mathrm{B}
    \sum_{a=1}^{M}
    \Bigg(
        -
        \beta
        \mathcal{H}
        (\bm{F}_{k(a)} \bm{s}_{\mu(a)})
        +
        \beta
        \mathcal{H}
        (\bm{s}_{\mu(a)})
        -
        \beta
        s_{k(a); \mu(a)}
        \sum_{i=1}^{N}
        (
            J_{k(a), i}
            -
            J_{i, k(a)}
        )
        s_{i; \mu(a)}
    \Bigg)
    \nonumber
    \\&=
    -
    \frac{
        1
    }{
        T
    }
    \sum_{a=1}^{M}
    s_{k(a); \mu(a)}
    \sum_{i=1}^{N}
    (
        J_{k(a), i}
        -
        J_{i, k(a)}
    )
    s_{i; \mu(a)}
    \label{eq:cycle-entropy-production-ising-general}
    .
\end{align}
The entropy production per cycle arises from the antisymmetric interaction
between the spins involved in the cycle.

For example,
we consider a cycle consisting of two-spin flips:
\begin{equation}
    \underset{\circlearrowleft}{\varPi}
    =
    \{
        \bm{s}_\mu
        ;
        k \rightarrow l \rightarrow k \rightarrow l
    \}
    .
\end{equation}
The cycle entropy production of
equation~\eqref{eq:cycle-entropy-production-product} is then
\begin{align}
    \underset{\circlearrowleft}{\Delta}
    \varsigma_\mathrm{tot}
    &=
    k_\mathrm{B}
    \ln
    \bigg(
        \frac{
            w (\mu^{(k)} \mid \mu)
        }{
            w (\mu \mid \mu^{(k)})
        }
        \frac{
            w (\mu^{(k,l)} \mid \mu^{(k)})
        }{
            w (\mu^{(k)} \mid \mu^{(k,l)})
        }
        \frac{
            w (\mu^{(k,l,k)} \mid \mu^{(k,l)})
        }{
            w (\mu^{(k,l)} \mid \mu^{(k,l,k)})
        }
        \frac{
            w (\mu^{(k,l,k,l)} \mid \mu^{(k,l,k)})
        }{
            w (\mu^{(k,l,k)} \mid \mu^{(k,l,k,l)})
        }
    \bigg)
    \nonumber
    \\&=
    k_\mathrm{B}
    \ln
    \bigg(
        \frac{
            w (\mu^{(k)} \mid \mu)
        }{
            w (\mu \mid \mu^{(k)})
        }
        \frac{
            w (\mu^{(k,l)} \mid \mu^{(k)})
        }{
            w (\mu^{(k)} \mid \mu^{(k,l)})
        }
        \frac{
            w (\mu^{(l)} \mid \mu^{(k,l)})
        }{
            w (\mu^{(k,l)} \mid \mu^{(l)})
        }
        \frac{
            w (\mu \mid \mu^{(l)})
        }{
            w (\mu^{(l)} \mid \mu)
        }
    \bigg)
    \nonumber
    \\&=
    k_\mathrm{B}
    \Bigg[
        \Bigg(
            \cancel{
                -
                \beta
                s_{k; \mu}
                \sum_{i=1, i\neq l}^{N}
                (
                    J_{k, i}
                    -
                    J_{i, k}
                )
                s_{i; \mu}
            }
            -
            \beta
            s_{k; \mu}
            (
                J_{k, l}
                -
                J_{l, k}
            )
            s_{l; \mu}
        \Bigg)
        \nonumber
        \\&\quad\quad\quad+
        \Bigg(
            \bcancel{
                -
                \beta
                s_{l; \mu}
                \sum_{i=1, i\neq k}^{N}
                (
                    J_{l, i}
                    -
                    J_{i, l}
                )
                s_{i; \mu}
            }
            -
            \beta
            s_{l; \mu}
            (
                J_{l, k}
                -
                J_{k, l}
            )
            (-s_{k; \mu})
        \Bigg)
        \nonumber
        \\&\quad\quad\quad+
        \Bigg(
            \cancel{
                -
                \beta
                (-s_{k; \mu})
                \sum_{i=1, i\neq l}^{N}
                (
                    J_{k, i}
                    -
                    J_{i, k}
                )
                s_{i; \mu}
            }
            -
            \beta
            (-s_{k; \mu})
            (
                J_{k, l}
                -
                J_{l, k}
            )
            (-s_{l; \mu})
        \Bigg)
        \nonumber
        \\&\quad\quad\quad+
        \Bigg(
            \bcancel{
                -
                \beta
                (-s_{l; \mu})
                \sum_{i=1, i\neq k}^{N}
                (
                    J_{l, i}
                    -
                    J_{i, l}
                )
                s_{i; \mu}
            }
            -
            \beta
            (-s_{l; \mu})
            (
                J_{l, k}
                -
                J_{k, l}
            )
            s_{k; \mu}
        \Bigg)
    \Bigg]
    \nonumber
    \\&=
    -
    \frac{1}{T}
    4
    s_{k; \mu}
    s_{l; \mu}
    (
        J_{k, l}
        -
        J_{l, k}
    )
    .
\end{align}
Thus, the cycle entropy production arises from the antisymmetric interaction
between the two spins involved in the cycle.
The reservoir receives the heat of
\begin{equation}
    \underset{\circlearrowleft}{q}
    =
    4
    s_{k; \mu}
    s_{l; \mu}
    (
        J_{k, l}
        -
        J_{l, k}
    )
    .
\end{equation}

We then consider another example: a cycle consisting of three-spin flips:
\begin{equation}
    \underset{\circlearrowleft}{\varPi}
    =
    \{
        \bm{s}_\mu
        ;
        k \rightarrow l \rightarrow m \rightarrow k \rightarrow l
        \rightarrow m
    \}
    \label{eq:3spins-cycle-example1}
    .
\end{equation}
The cycle entropy production is calculated as
\begin{align}
    \underset{\circlearrowleft}{\Delta}
    \varsigma_\mathrm{tot}
    &=
    k_\mathrm{B}
    \ln
    \bigg(
        \frac{
            w (\mu^{(k)} \mid \mu)
        }{
            w (\mu \mid \mu^{(k)})
        }
        \frac{
            w (\mu^{(k,l)} \mid \mu^{(k)})
        }{
            w (\mu^{(k)} \mid \mu^{(k,l)})
        }
        \frac{
            w (\mu^{(k,l,m)} \mid \mu^{(k,l)})
        }{
            w (\mu^{(k,l)} \mid \mu^{(k,l,m)})
        }
        \nonumber
        \\&\quad\quad\quad\quad\times
        \frac{
            w (\mu^{(k,l,m,k)} \mid \mu^{(k,l,m)})
        }{
            w (\mu^{(k,l,m)} \mid \mu^{(k,l,m,k)})
        }
        \frac{
            w (\mu^{(k,l,m,k,l)} \mid \mu^{(k,l,m,k)})
        }{
            w (\mu^{(k,l,m,k)} \mid \mu^{(k,l,m,k,l)})
        }
        \frac{
            w (\mu^{(k,l,m,k,l,m)} \mid \mu^{(k,l,m,k,l)})
        }{
            w (\mu^{(k,l,m,k,l)} \mid \mu^{(k,l,m,k,l,m)})
        }
    \bigg)
    \nonumber
    \\&=
    k_\mathrm{B}
    \ln
    \bigg(
        \frac{
            w (\mu^{(k)} \mid \mu)
        }{
            w (\mu \mid \mu^{(k)})
        }
        \frac{
            w (\mu^{(k,l)} \mid \mu^{(k)})
        }{
            w (\mu^{(k)} \mid \mu^{(k,l)})
        }
        \frac{
            w (\mu^{(k,l,m)} \mid \mu^{(k,l)})
        }{
            w (\mu^{(k,l)} \mid \mu^{(k,l,m)})
        }
        \frac{
            w (\mu^{(l,m)} \mid \mu^{(k,l,m)})
        }{
            w (\mu^{(k,l,m)} \mid \mu^{(l,m)})
        }
        \frac{
            w (\mu^{(m)} \mid \mu^{(l,m)})
        }{
            w (\mu^{(l,m)} \mid \mu^{(m)})
        }
        \frac{
            w (\mu \mid \mu^{(m)})
        }{
            w (\mu^{(m)} \mid \mu)
        }
    \bigg)
    \nonumber
    \\&=
    k_\mathrm{B}
    [
        -
        \beta
        s_{k; \mu}
        (
            J_{k, l}
            -
            J_{l, k}
        )
        s_{l; \mu}
        -
        \beta
        s_{k; \mu}
        (
            J_{k, m}
            -
            J_{m, k}
        )
        s_{m; \mu}
        \nonumber
        \\&\quad\quad\;\;
        -
        \beta
        s_{l; \mu}
        (
            J_{l, k}
            -
            J_{k, l}
        )
        (-s_{k; \mu})
        -
        \beta
        s_{l; \mu}
        (
            J_{l, m}
            -
            J_{m, l}
        )
        s_{m; \mu}
        \nonumber
        \\&\quad\quad\;\;
        -
        \beta
        s_{m; \mu}
        (
            J_{m, k}
            -
            J_{k, m}
        )
        (-s_{k; \mu})
        -
        \beta
        s_{m; \mu}
        (
            J_{m, l}
            -
            J_{l, m}
        )
        (-s_{l; \mu})
        \nonumber
        \\&\quad\quad\;\;
        -
        \beta
        (-s_{k; \mu})
        (
            J_{k, l}
            -
            J_{l, k}
        )
        (-s_{l; \mu})
        -
        \beta
        (-s_{k; \mu})
        (
            J_{k, m}
            -
            J_{m, k}
        )
        (-s_{m; \mu})
        \nonumber
        \\&\quad\quad\;\;
        -
        \beta
        (-s_{l; \mu})
        (
            J_{l, k}
            -
            J_{k, l}
        )
        s_{k; \mu}
        -
        \beta
        (-s_{l; \mu})
        (
            J_{l, m}
            -
            J_{m, l}
        )
        (-s_{m; \mu})
        \nonumber
        \\&\quad\quad\;\;
        -
        \beta
        (-s_{m; \mu})
        (
            J_{m, k}
            -
            J_{k, m}
        )
        s_{k; \mu}
        -
        \beta
        (-s_{m; \mu})
        (
            J_{m, l}
            -
            J_{l, m}
        )
        s_{l; \mu}
        \nonumber
    ]
    \\&=
    -
    \frac{1}{T}
    [
        4
        s_{k; \mu} s_{l; \mu}
        (
            J_{k, l}
            -
            J_{l, k}
        )
        +
        4
        s_{k; \mu} s_{m; \mu}
        (
            J_{k, m}
            -
            J_{m, k}
        )
        +
        4
        s_{l; \mu} s_{m; \mu}
        (
            J_{l, m}
            -
            J_{m, l}
        )
    ]
    .
\end{align}
Note that flipping the same spin twice returns the spin to its original state,
for example, $\mu^{(k, k)} = \mu$.
Thus, the cycle entropy production arises from the antisymmetric interactions
between the three spins involved in the cycle.
We then consider the cycle,
\begin{equation}
    \underset{\circlearrowleft}{\varPi}
    =
    \{
        \bm{s}_\mu
        ;
        k \rightarrow l \rightarrow m \rightarrow
        k \rightarrow m \rightarrow l
    \}
    \label{eq:3spins-cycle-example2}
    ,
\end{equation}
to highlight the role of the order of spin flips in the cycle.
The cycle entropy production is calculated as
\begin{align}
    \underset{\circlearrowleft}{\Delta}
    \varsigma_\mathrm{tot}
    &=
    k_\mathrm{B}
    \ln
    \bigg(
        \frac{
            w (\mu^{(k)} \mid \mu)
        }{
            w (\mu \mid \mu^{(k)})
        }
        \frac{
            w (\mu^{(k,l)} \mid \mu^{(k)})
        }{
            w (\mu^{(k)} \mid \mu^{(k,l)})
        }
        \frac{
            w (\mu^{(k,l,m)} \mid \mu^{(k,l)})
        }{
            w (\mu^{(k,l)} \mid \mu^{(k,l,m)})
        }
        \nonumber
        \\&\quad\quad\quad\quad\times
        \frac{
            w (\mu^{(k,l,m,k)} \mid \mu^{(k,l,m)})
        }{
            w (\mu^{(k,l,m)} \mid \mu^{(k,l,m,k)})
        }
        \frac{
            w (\mu^{(k,l,m,k,m)} \mid \mu^{(k,l,m,k)})
        }{
            w (\mu^{(k,l,m,k)} \mid \mu^{(k,l,m,k,m)})
        }
        \frac{
            w (\mu^{(k,l,m,k,m,l)} \mid \mu^{(k,l,m,k,m)})
        }{
            w (\mu^{(k,l,m,k,m)} \mid \mu^{(k,l,m,k,m,l)})
        }
    \bigg)
    \nonumber
    \\&=
    k_\mathrm{B}
    \ln
    \bigg(
        \frac{
            w (\mu^{(k)} \mid \mu)
        }{
            w (\mu \mid \mu^{(k)})
        }
        \frac{
            w (\mu^{(k,l)} \mid \mu^{(k)})
        }{
            w (\mu^{(k)} \mid \mu^{(k,l)})
        }
        \frac{
            w (\mu^{(k,l,m)} \mid \mu^{(k,l)})
        }{
            w (\mu^{(k,l)} \mid \mu^{(k,l,m)})
        }
        \frac{
            w (\mu^{(l,m)} \mid \mu^{(k,l,m)})
        }{
            w (\mu^{(k,l,m)} \mid \mu^{(l,m)})
        }
        \frac{
            w (\mu^{(l)} \mid \mu^{(l,m)})
        }{
            w (\mu^{(l,m)} \mid \mu^{(l)})
        }
        \frac{
            w (\mu \mid \mu^{(l)})
        }{
            w (\mu^{(l)} \mid \mu)
        }
    \bigg)
    \nonumber
    \\&=
    k_\mathrm{B}
    [
        -
        \beta
        s_{k; \mu}
        (
            J_{k, l}
            -
            J_{l, k}
        )
        s_{l; \mu}
        -
        \beta
        s_{k; \mu}
        (
            J_{k, m}
            -
            J_{m, k}
        )
        s_{m; \mu}
        \nonumber
        \\&\quad\quad\;\;
        -
        \beta
        s_{l; \mu}
        (
            J_{l, k}
            -
            J_{k, l}
        )
        (-s_{k; \mu})
        \cancel{
            -
            \beta
            s_{l; \mu}
            (
                J_{l, m}
                -
                J_{m, l}
            )
            s_{m; \mu}
        }
        \nonumber
        \\&\quad\quad\;\;
        -
        \beta
        s_{m; \mu}
        (
            J_{m, k}
            -
            J_{k, m}
        )
        (-s_{k; \mu})
        \bcancel{
            -
            \beta
            s_{m; \mu}
            (
                J_{m, l}
                -
                J_{l, m}
            )
            (-s_{l; \mu})
        }
        \nonumber
        \\&\quad\quad\;\;
        -
        \beta
        (-s_{k; \mu})
        (
            J_{k, l}
            -
            J_{l, k}
        )
        (-s_{l; \mu})
        -
        \beta
        (-s_{k; \mu})
        (
            J_{k, m}
            -
            J_{m, k}
        )
        (-s_{m; \mu})
        \nonumber
        \\&\quad\quad\;\;
        -
        \beta
        (-s_{m; \mu})
        (
            J_{m, k}
            -
            J_{k, m}
        )
        s_{k; \mu}
        \cancel{
            -
            \beta
            (-s_{m; \mu})
            (
                J_{m, l}
                -
                J_{l, m}
            )
            (-s_{l; \mu})
        }
        \nonumber
        \\&\quad\quad\;\;
        -
        \beta
        (-s_{l; \mu})
        (
            J_{l, k}
            -
            J_{k, l}
        )
        s_{k; \mu}
        \bcancel{
            -
            \beta
            (-s_{l; \mu})
            (
                J_{l, m}
                -
                J_{m, l}
            )
            s_{l; \mu}
        }
        \nonumber
    ]
    \\&=
    -
    \frac{1}{T}
    [
        4
        s_{k; \mu} s_{l; \mu}
        (
            J_{k, l}
            -
            J_{l, k}
        )
        +
        4
        s_{k; \mu} s_{m; \mu}
        (
            J_{k, m}
            -
            J_{m, k}
        )
    ]
    ,
\end{align}
where the term involving the interaction between spins $l$ and $m$ is absent.
The difference of the cycles in the sequence of spin flips in
equations~\eqref{eq:3spins-cycle-example1} and~\eqref{eq:3spins-cycle-example2}
is the reversibility of the spin flips between $l$ and $m$.
In equation~\eqref{eq:3spins-cycle-example1},
the spin flips between $l$ and $m$ are irreversible, that is,
in the order of $l \rightarrow m \rightarrow l \rightarrow m$.
On the other hand, in equation~\eqref{eq:3spins-cycle-example2},
the spin flips between $l$ and $m$ are reversible, that is,
in the order of $l \rightarrow m \rightarrow m \rightarrow l$.
From these examples, we simplify the cycle entropy production of
equation~\eqref{eq:cycle-entropy-production-ising-general} as
\begin{align}
    \underset{\circlearrowleft}{\Delta}
    \varsigma_\mathrm{tot}
    &=
    -
    \frac{1}{T}
    \sum_{(k, l) \in \underset{\circlearrowleft}{\varPi}}
    4
    s_{k; \mu}
    s_{l; \mu}
    (
        J_{k, l}
        -
        J_{l, k}
    )
    \label{eq:cycle-entropy-production-ising-simplified}
    \\&=
    -
    \frac{1}{T}
    \sum_{(k, l) \in \mathcal{I}[\underset{\circlearrowleft}{\varPi}]}
    4
    s_{k; \mu}
    s_{l; \mu}
    (
        J_{k, l}
        -
        J_{l, k}
    )
    \label{eq:cycle-entropy-production-ising-simplified-irreversible}
    ,
\end{align}
where the sum is taken over the set of spin pairs in the cycle
in equation~\eqref{eq:cycle-entropy-production-ising-simplified},
and the set of irreversibly flipped spin pairs in the cycle
in equation~\eqref{eq:cycle-entropy-production-ising-simplified-irreversible}.
The set of irreversibly flipped spin pairs in the cycle is defined as
\begin{equation}
    \mathcal{I} [\underset{\circlearrowleft}{\varPi}]
    \coloneqq
    \{
        (k, l)
        \mid
        \textrm{the sequence of spin flips between $k$ and $l$ is
        irreversible in~}
        \underset{\circlearrowleft}{\varPi} \textrm{~as~}
        k \rightarrow l \rightarrow k \rightarrow l
    \}
    .
\end{equation}

\section{Spin-dependent rate constant}
\label{sec:spin-dependent-rate-constant}

\subsection{Arrhenius transition rate}
In equation~(19),
the Arrhenius transition rate shares a single rate constant $A$
across all spins.
Here we generalize the model to allow a spin-dependent rate
constant $A_k > 0$,
\begin{align}
    w(\mu^{(k)} \mid \mu)
    =
    A_k
    \exp
    \bm{(}
        -
        \beta
        s_{k;\mu}
        g_k(\bm{s}_\mu)
    \bm{)}
    \label{eq:transition-rate-arrhenius-spindep}
    ,
\end{align}
and examine whether this extension affects the network inference.
It is straightforward to see that this extension satisfies detailed balance.

\subsubsection{Loss function}
Setting $\eta_k \coloneqq 2\ln (A_k)$,
the log transition rate for a spin-$k$ flip from state $\mu$ is
\begin{align}
    \ln
    \bm{(}
        w(\mu^{(k)} \mid \mu)
    \bm{)}
    =
    \frac{\eta_k}{2}
    -
    \beta
    s_{k;\mu}
    g_k(\bm{s}_\mu)
    .
\end{align}
The parameters associated with spin $k$
(namely $\eta_k$, the $k$th row
$\{\beta J_{k,i}\}_{i=1}^{N}$, and $\beta h_k$)
appear only in transitions that flip spin $k$.
The least-squares problem therefore decouples into $N$ independent
sub-problems, one per spin:
\begin{align}
    &
    L_k
    (
        \beta \bm{J}_{k,\cdot}, \beta h_k, A_k
    )
    \coloneqq
    \sum_{\mu:\, w(\mu^{(k)} \mid \mu)>0}
    \bigg[
        \eta_k
        +
        (
            \bm{s}_{\mu^{(k)}}
            -
            \bm{s}_\mu
        )^\top
        (
            \beta
            \bm{J}
            \bm{s}_\mu
            +
            \beta
            \bm{h}
        )
        -
        2
        \ln
        \bigg(
            \frac{1}{\tau}
            \frac{n(\mu^{(k)}, \mu)}{p(\mu)}
        \bigg)
    \bigg]^2
    \label{eq:loss-function-spindep}
    ,
\end{align}
where $(\bm{s}_{\mu^{(k)}} - \bm{s}_\mu)^\top = -2 s_{k;\mu} \bm{e}_k^\top$
for a spin-$k$ flip (with $\bm{e}_k$ the $k$-th standard basis vector).
The total loss is
$L = \sum_{k=1}^{N} L_k$,
and each $L_k$ is minimized independently by ordinary least squares.
Each sub-problem has $N+1$ unknowns
$(\eta_k,\, \beta J_{k,1}, \ldots, \beta J_{k,N},\, \beta h_k)$
(with $J_{k,k} = 0$ enforced by construction)
fitted to $2^N$ constraints.

\subsubsection{Results}
We applied the spin-dependent inference to all eight cognitive tasks
and compared the results against the spin-independent model.
Supplementary Fig.~\ref{fig:fig-s11}a shows the inferred rate constants
$A_k = \exp(\eta_k/2)$ per spin per task for each of the eight
cognitive tasks.
The spin-dependent constants vary by a factor of up to ${\sim}1.5$
across spins within a task, with cluster~1 (default-mode network)
consistently showing the highest rate.
The relative ordering of $A_k$ across spins is similar across tasks,
though the overall magnitude varies.
Despite this within-task variation, the spin-dependent model yields
essentially the same interaction matrix $\beta\bm{J}$:
the Pearson correlation between the spin-independent and spin-dependent
off-diagonal elements exceeds $r = 0.999$ for all tasks.
The spin-dependent model achieves a modestly lower least-squares
loss (Supplementary Fig.~\ref{fig:fig-s11}b), as expected from the additional
degrees of freedom.
Nevertheless, the Pearson correlation $r$ between the reconstructed
and the empirically observed probability flux
(Supplementary Fig.~\ref{fig:fig-s11}c) is indistinguishable between the two
models ($|\Delta r| < 0.007$ across all tasks).
We therefore conclude that the spin-independent rate constant $A$
is sufficient for the probability flux analysis presented in the
main text.

\subsection{Glauber transition rate}
We now consider the same extension as
section~\ref{sec:spin-dependent-rate-constant} for the
Glauber transition rate
[equation~(20)],
replacing the scalar rate constant $A$ by a spin-dependent constant
$A_k > 0$:
\begin{align}
    w(\mu^{(k)} \mid \mu)
    =
    \frac{
        A_k
    }{
        1
        +
        \exp
        \bm{(}
            2
            \beta
            s_{k;\mu}
            g_k(\bm{s}_\mu)
        \bm{)}
    }
    \label{eq:transition-rate-glauber-spindep}
    .
\end{align}
Again, it is straightforward to see that this extension satisfies
detailed balance.

\subsubsection{Loss function}
With $A_k$ fixed at $2\hat{w}_{99,k}$ (the 99th percentile of the observed
transition rates for spin-$k$ flips), we invert
equation~\eqref{eq:transition-rate-glauber-spindep} for each feasible
observation (those satisfying $w(\mu^{(k)}\mid\mu) < A_k$):
\begin{align}
    \beta g_k(\bm{s}_\mu)
    =
    \frac{s_{k;\mu}}{2}
    \ln
    \bigg(
        \frac{A_k}{w(\mu^{(k)}\mid\mu)} - 1
    \bigg)
    ,
    \label{eq:glauber-sd-reference-field}
\end{align}
using $s_{k;\mu}^2 = 1$.
The parameters associated with spin $k$
(namely the $k$-th row $\beta\bm{J}_{k,\cdot}$ and $\beta h_k$)
appear only in transitions that flip spin $k$,
so the total loss decouples into $N$ independent per-spin sub-problems:
\begin{align}
    L_k
    (
        \beta\bm{J}_{k,\cdot}, \beta h_k
    )
    &\coloneqq
    \sum_{\mu:\, w(\mu^{(k)}\mid\mu) < A_k}
    \bigg[
        \beta\bm{J}_{k,\cdot} \bm{s}_\mu
        +
        \beta h_k
        -
        \frac{s_{k;\mu}}{2}
        \ln
        \bigg(
            \frac{A_k}{w(\mu^{(k)},\mu)} - 1
        \bigg)
    \bigg]^2
    \label{eq:loss-function-glauber-spindep}
    ,
\end{align}
with the total loss $L_{\mathrm{SD,G}} = \sum_{k=1}^{N} L_k$.
Each $L_k$ is linear in $\beta\bm{J}_{k,\cdot}$ and $\beta h_k$
and is minimized independently by ordinary least squares,
exactly as in the Arrhenius spin-dependent case
(equation~\eqref{eq:loss-function-spindep}),
with $N+1$ unknowns ($\beta J_{k,1}, \ldots, \beta J_{k,N}$
and $\beta h_k$, with $J_{k,k}=0$ enforced)
fitted to the feasible observations.

\subsubsection{Results}
We applied the Glauber spin-dependent inference to all eight cognitive
tasks and compared the results against the Glauber spin-independent model.
Supplementary Fig.~\ref{fig:fig-s12}a shows the inferred rate constants
$A_k$ per spin per task for each of the eight cognitive tasks.
Spin~1 (default mode network) consistently shows the highest rate constant,
with a within-task variation of up to ${\sim}2.4$-fold across spins
(Motor task: $A_{\max}/A_{\min} = 0.524/0.222$),
and the relative ordering across spins is consistent across tasks as in
the Arrhenius rate.
The spin-dependent model achieves a lower least-squares loss
(Supplementary Fig.~\ref{fig:fig-s12}b).
The Pearson correlation between the spin-independent and spin-dependent
off-diagonal elements of $\beta\bm{J}$ exceeds $r = 0.995$ for all tasks,
confirming that the inferred interaction network is robust to the
rate-constant parameterization.
The Pearson correlation $r$ between the reconstructed and empirically
observed probability flux (Supplementary Fig.~\ref{fig:fig-s12}c)
differs by at most
$|\Delta r| < 0.07$ across the two models; the spin-dependent model
improves it only for the Relational task and slightly reduces it for the
other tasks, most for Emotion and Working Memory.
We therefore conclude that the spin-independent rate constant $A$ is an
adequate description for all cognitive tasks under the Glauber rate.

\bibliography{references}

\clearpage

\begin{figure*}[bt]
    \centering
    \includegraphics{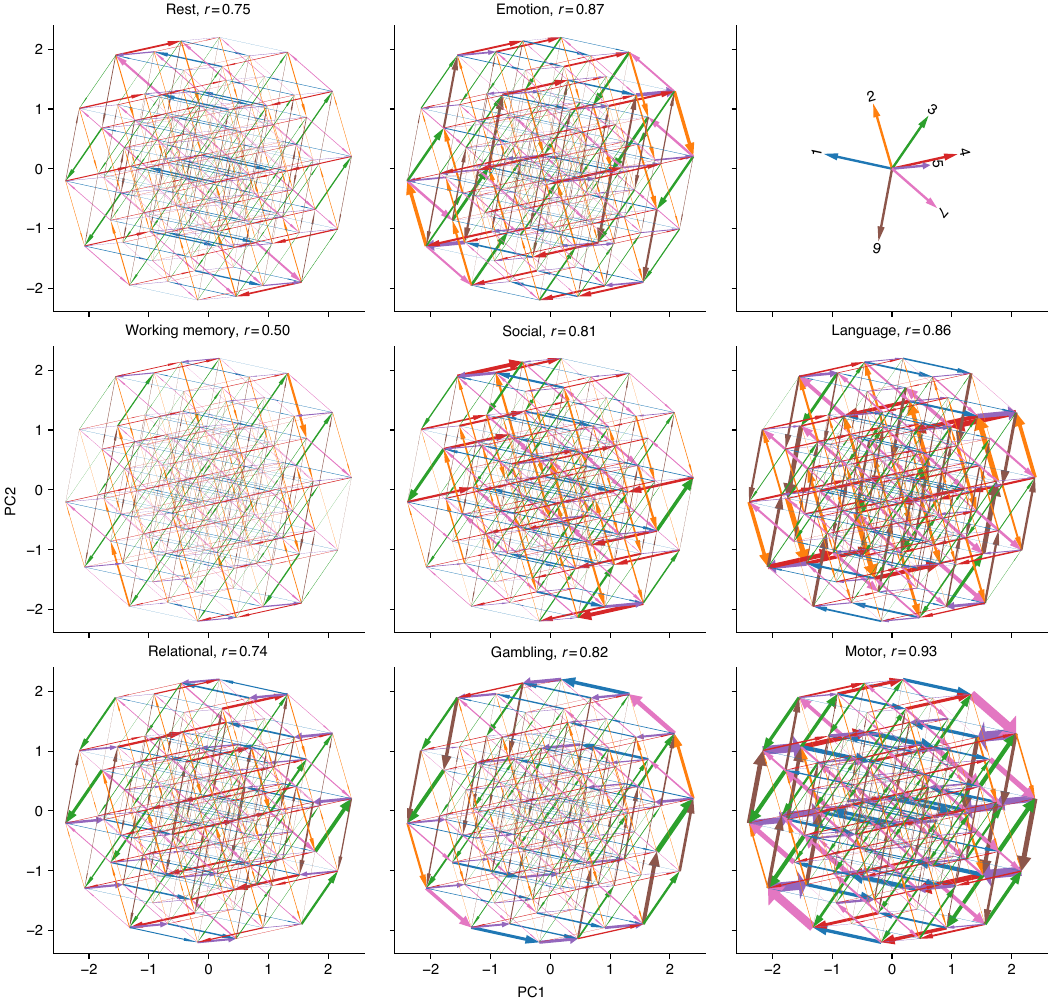}
    \caption{\textbf{Probability flux diagrams reconstructed from the
        inferred Ising spin system using the Glauber transition rate.}
        Same as Fig.~6, but the Glauber rate is used
        instead of the Arrhenius rate.
        Arrow widths are normalized per task to the maximum empirical flux
        to account for the scale offset between the model and empirical
    stationary distributions (Supplementary Note~\ref{sec:statdist}).}
    \label{fig:fig-s09}
\end{figure*}

\begin{figure*}[bt]
    \centering
    \includegraphics{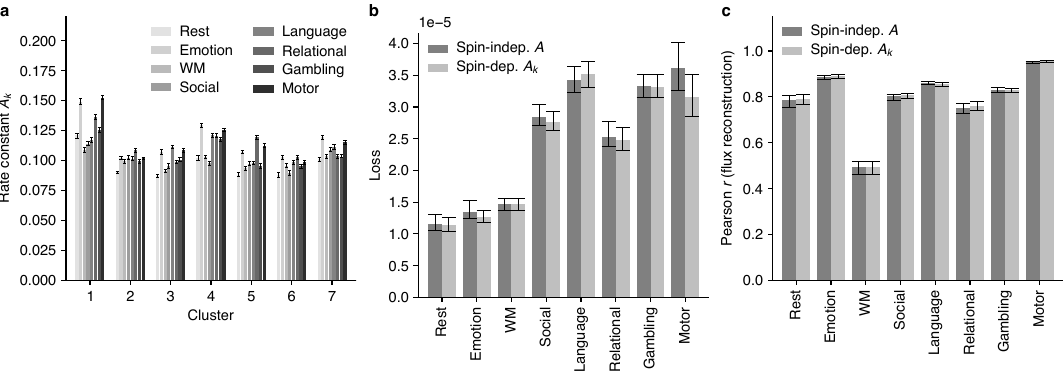}
    \caption{\textbf{Spin-dependent rate constant of the Arrhenius rate.}
        \textbf{a},
        Rate constants $A_k$ per spin (cluster) inferred by the
        spin-dependent Arrhenius rate for each of the eight cognitive
        tasks (colour-coded).
        Cluster indices correspond to the seven brain clusters of
        Fig.~1b.
        Error bars show the interquartile range from 100 trajectory
        bootstrap resamples (transition events resampled with
        replacement), with the full inference
        $\{\beta\bm{J}, \beta\bm{h}, A_k\}$
        repeated for each resample; error bars
        are centred at the point estimate.
        \textbf{b},
        Least-squares loss of the spin-independent ($A$, dark)
        and spin-dependent ($A_k$, light) models.
        The spin-dependent model achieves a lower loss owing to its
        additional degrees of freedom.
        Error bars show the interquartile range from the same 100
        trajectory bootstrap resamples.
        \textbf{c},
        Pearson correlation $r$ between the reconstructed and
        empirically observed probability flux for both models.
        The two models produce statistically indistinguishable flux
        reconstructions ($|\Delta r| < 0.007$).
        Error bars show the interquartile range from the same 100
    trajectory bootstrap resamples.}
    \label{fig:fig-s11}
\end{figure*}

\begin{figure*}[bt]
    \centering
    \includegraphics{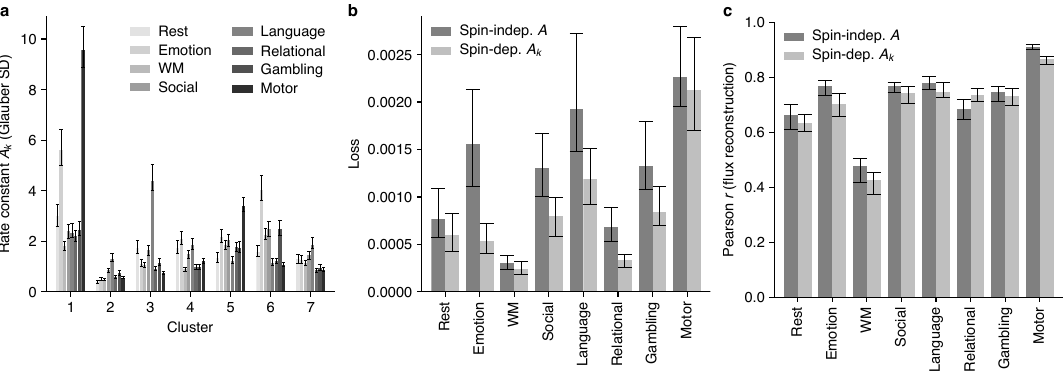}
    \caption{\textbf{Spin-dependent rate constant of the Glauber rate.}
        \textbf{a},
        Rate constants $A_k$ per spin (cluster) inferred by the Glauber
        spin-dependent model for each of the eight cognitive tasks
        (colour-coded).
        Cluster indices correspond to the seven brain clusters of
        Fig.~1b.
        $A_k$ is set to twice the 99th percentile of empirical spin-$k$
        transition rates.
        Error bars show the interquartile range from 100 trajectory
        bootstrap resamples (transition events resampled with replacement),
        with the full inference
        $\{\beta\bm{J}, \beta\bm{h}, A_k\}$
        repeated for each resample; error bars are centred at the point
        estimate.
        \textbf{b},
        Least-squares loss of the Glauber spin-independent ($A$, dark)
        and spin-dependent ($A_k$, light) models.
        The spin-dependent model achieves a lower loss owing to its
        additional degrees of freedom.
        Error bars show the interquartile range from the same 100
        trajectory bootstrap resamples.
        \textbf{c},
        Pearson correlation $r$ between the reconstructed and empirically
        observed probability flux for both models.
        The two models yield comparable flux reconstructions across tasks.
        Error bars show the interquartile range from the same 100
    trajectory bootstrap resamples.}
    \label{fig:fig-s12}
\end{figure*}

\begin{figure*}[bt]
    \centering
    \includegraphics{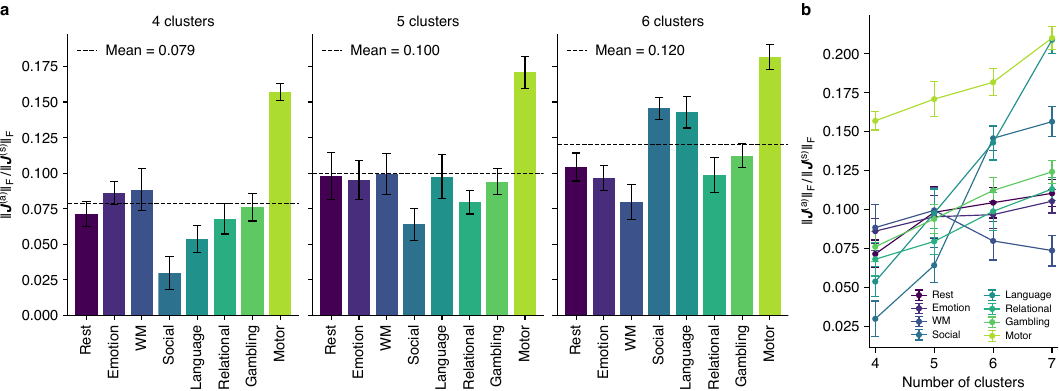}
    \caption{\textbf{Asymmetry ratio of the inferred interaction
        network for different numbers of clusters.}
        \textbf{a},
        Ratio $\|\bm{J}^{(\mathrm{a})}\|_\mathrm{F} /
        \|\bm{J}^{(\mathrm{s})}\|_\mathrm{F}$
        of the antisymmetric to symmetric Frobenius norm of the
        inferred interaction matrix $\bm{J}$ per cognitive task,
        shown for $\{4, 5, 6\}$ coarse-grained clusters
        (panels from left to right).
        The dashed horizontal line indicates the mean across tasks.
        \textbf{b},
        The asymmetry ratio as a function of the number of clusters.
        Error bars indicate the standard deviation across 100 bootstrap
    resamples.}
    \label{fig:fig-s13}
\end{figure*}

\begin{figure*}[bt]
    \centering
    \includegraphics{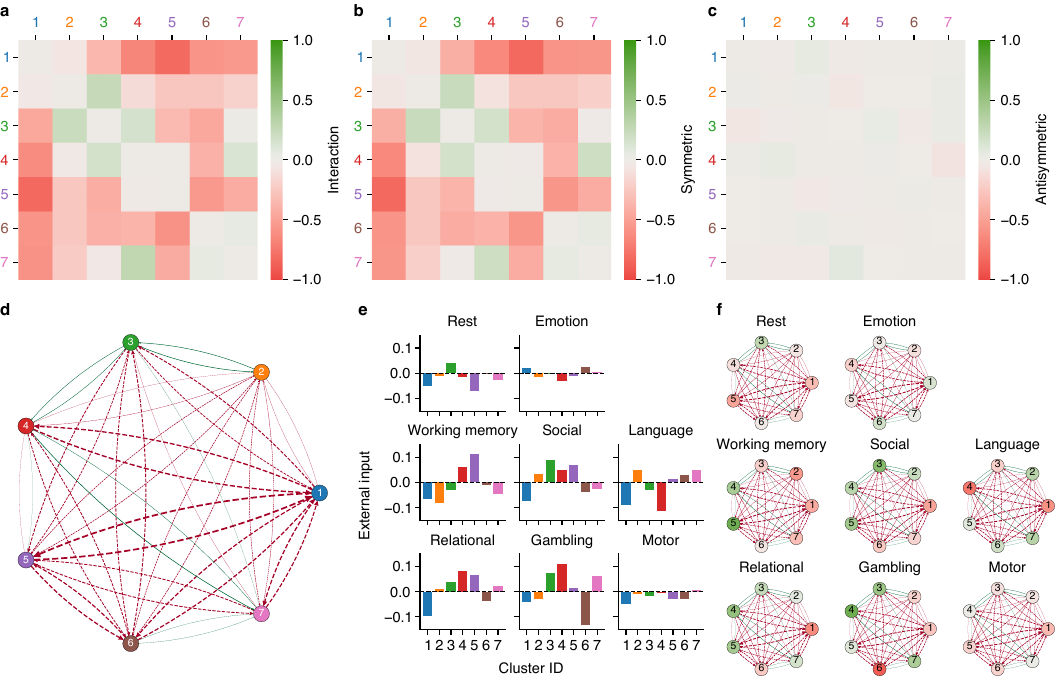}
    \caption{\textbf{Inferring the Ising spin system with a
        task-independent interaction network.}
        Same as Fig.~4, but the interaction network is
    task-independent: we use the same interaction network for all tasks.}
    \label{fig:fig-s14}
\end{figure*}

\begin{figure*}[bt]
    \centering
    \includegraphics{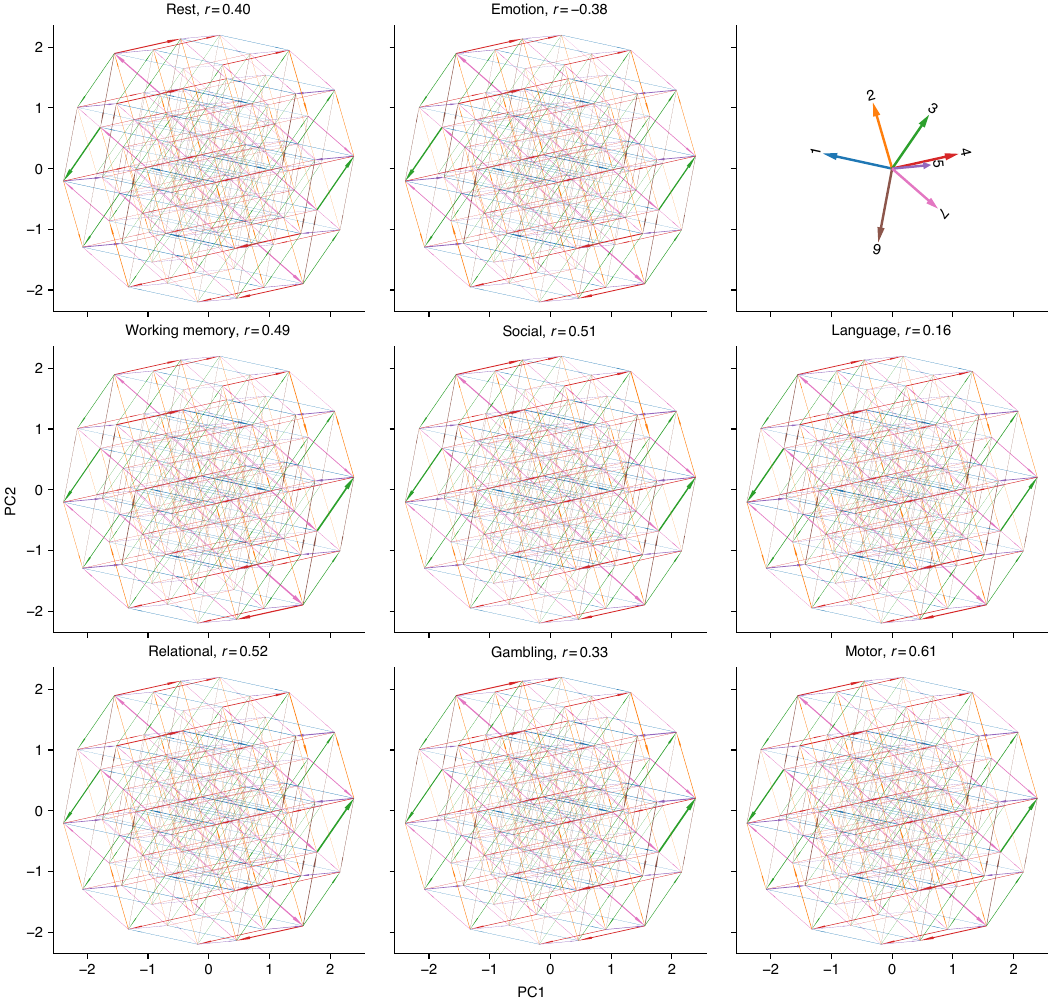}
    \caption{\textbf{Probability flux diagrams reconstructed from the
        inferred Ising spin system with a task-independent interaction network.}
        Same as Fig.~6, but the interaction network is
    task-independent: we use the same interaction network for all tasks.}
    \label{fig:fig-s15}
\end{figure*}

\begin{figure*}[bt]
    \centering
    \includegraphics{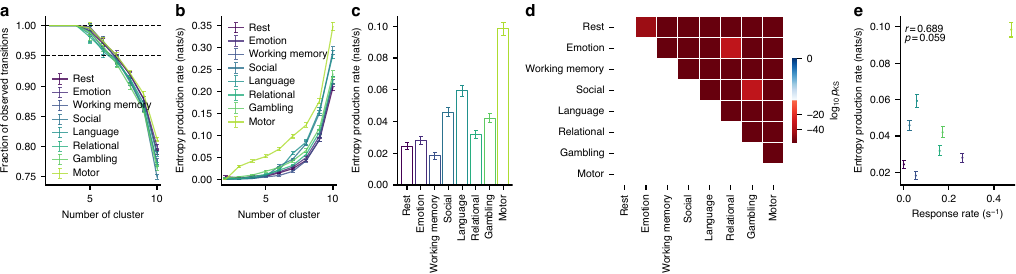}
    \caption{\textbf{Estimated entropy production rate in the human brain.}
        \textbf{a}, The fraction of possible state transitions that
        are observed.
        \textbf{b}, The estimated entropy production rate as a function
        of the number of clusters.
        Error bars represent twice the standard deviation over bootstrap
        resamples of the observed transitions.
        \textbf{c}, The entropy production rate for seven clusters.
        \textbf{d}, The results of the Kolmogorov--Smirnov test
        between the distributions shown in panel c.
        \textbf{e}, Relation between the response rate and the
    entropy production rate~\cite{Lynn2021}.}
    \label{fig:fig-s16}
\end{figure*}

\begin{figure*}[bt]
    \centering
    \includegraphics{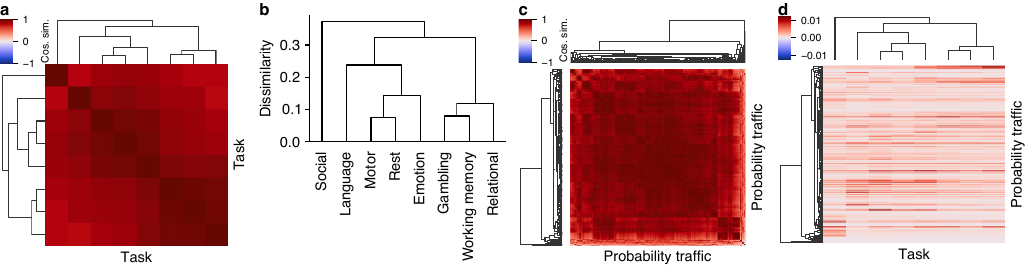}
    \caption{\textbf{Clustering brain function using probability
        traffic patterns.}
        \textbf{a},
        Hierarchically clustered cosine similarity among tasks based on
        probability traffic.
        \textbf{b},
        The dendrogram of panel a.
        \textbf{c},
        Hierarchically clustered cosine similarity among probability
        traffic across tasks.
        \textbf{d},
        The probability traffic as a matrix sorted by the
        hierarchical clustering of panels a and c.
        Unlike Fig.~3c,d, only one
        representative direction per hypercubic edge is shown in
        panels c and d: because probability traffic is
        symmetric, $\mathcal{T}(\mu, \nu) = \mathcal{T}(\nu, \mu)$, the
        two directed copies of each edge are identical, so only the 448
    unique undirected edges are retained.}
    \label{fig:fig-s18}
\end{figure*}

\begin{figure*}[bt]
    \centering
    \includegraphics{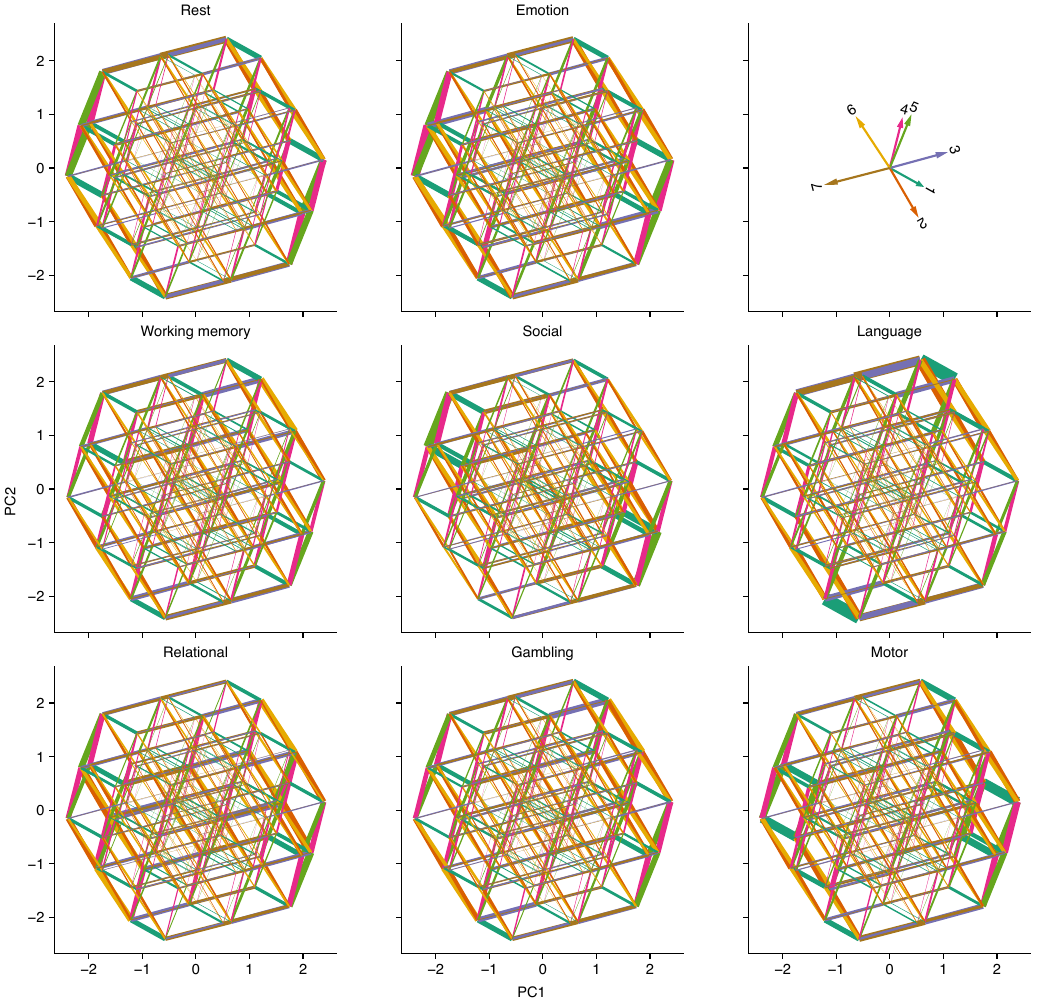}
    \caption{\textbf{Probability traffic estimated from the known
        functional clustering.}
        Same as Extended Data Fig.~10, but the
        clusters of brain regions are
        defined by the known functional clusters~\cite{ThomasYeo2011}
    (Fig.~1d).}
    \label{fig:fig-s19}
\end{figure*}

\begin{figure*}[bt]
    \centering
    \includegraphics{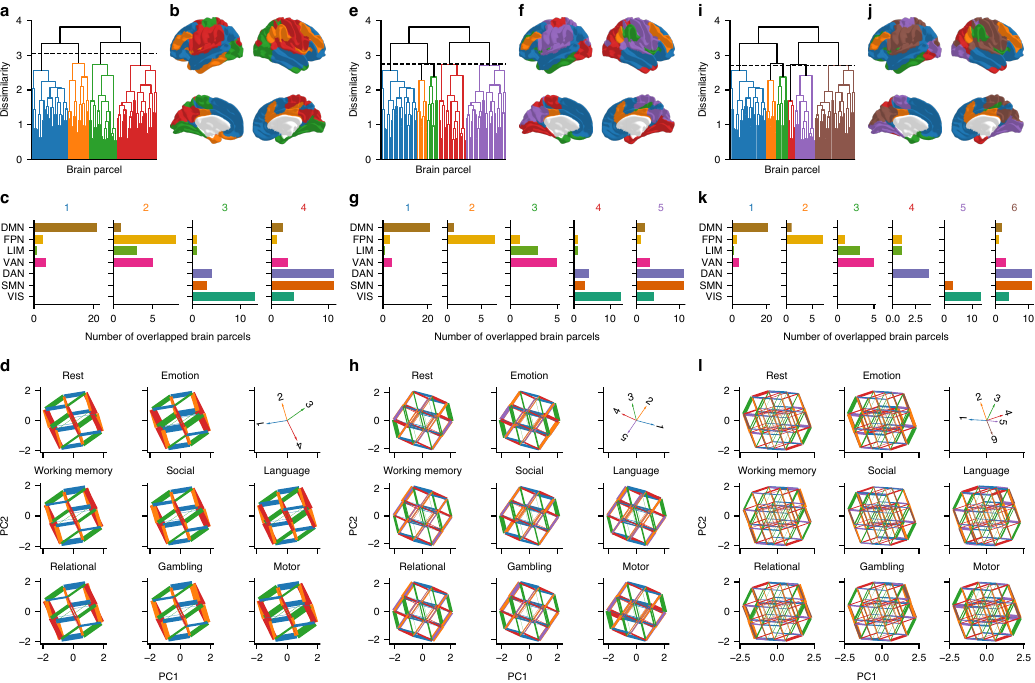}
    \caption{\textbf{Changing the number of clusters for hypercubic
        probability traffic analysis.}
        Same as Fig.~1b--d,
        and Extended Data Fig.~10, but the number of clusters is
        $N=4$ (\textbf{a--d}), $N=5$ (\textbf{e--h}),
    and $N=6$ (\textbf{i--l}).}
    \label{fig:fig-s20}
\end{figure*}

\begin{figure*}[bt]
    \centering
    \includegraphics{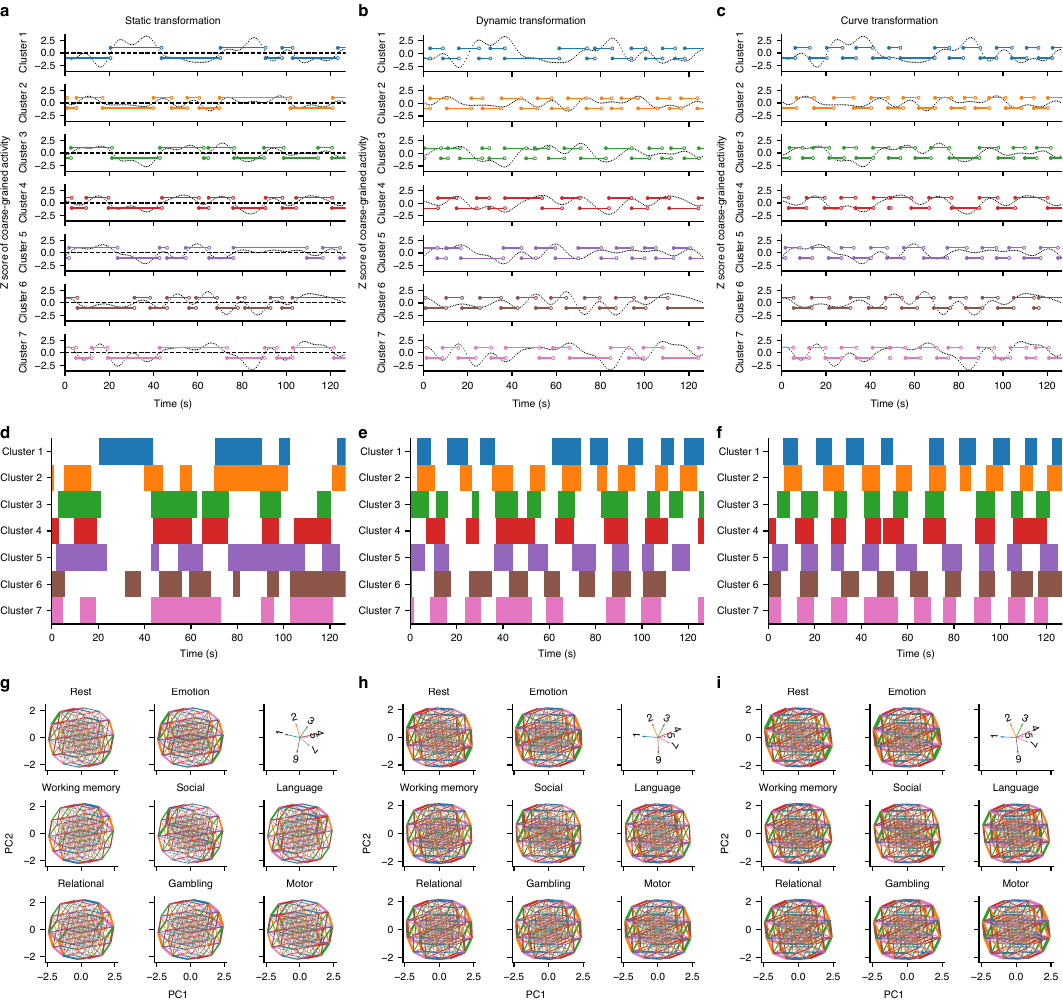}
    \caption{\textbf{Various methods of transformation applied to the
        fMRI signals and the estimated probability traffic.}
        Same as Extended Data Fig.~5, but with the
        estimated probability traffic.
        \textbf{a--c},
        Examples of the static, dynamic,
        and curve transformations.
        \textbf{d--f},
        The temporally coarse-grained (binarized) time series of panels a--c.
        Panels a and d are the same as
        Fig.~1f,g, respectively,
        and panels a--f are the same as in Extended Data Fig.~5.
        \textbf{g--i},
    The estimated probability traffic of each task.}
    \label{fig:fig-s21}
\end{figure*}

\begin{figure*}[bt]
    \centering
    \includegraphics{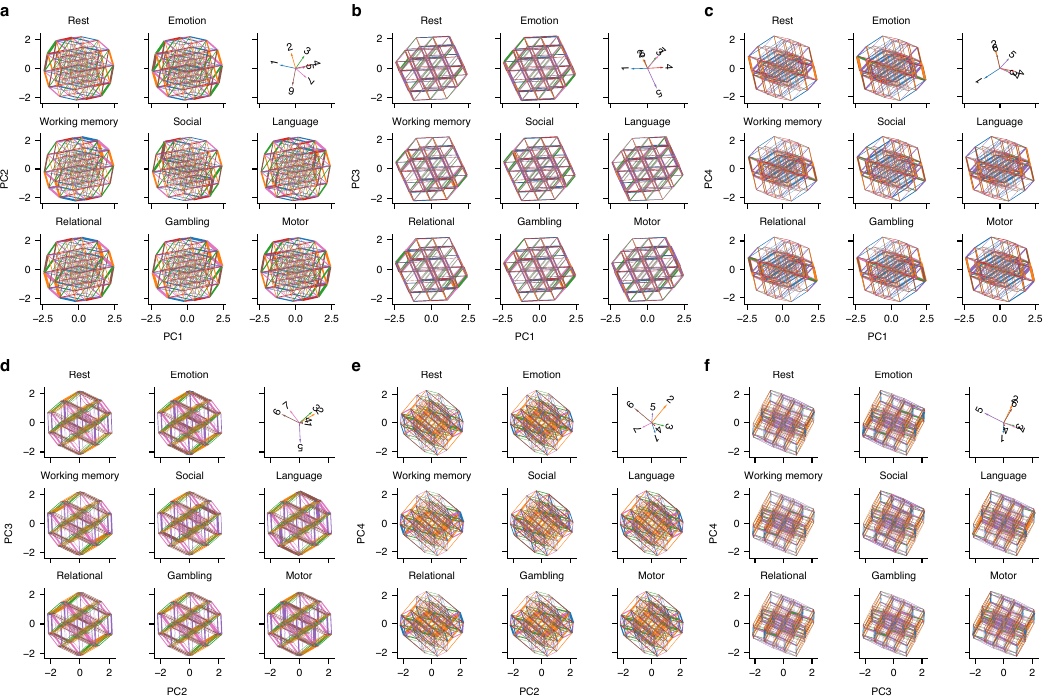}
    \caption{\textbf{Observing probability traffic through other PCs.}
        Same as Extended Data Fig.~10, but the
    probability traffic diagrams are visualized using different PCs.}
    \label{fig:fig-s22}
\end{figure*}

\begin{figure*}[bt]
    \centering
    \includegraphics{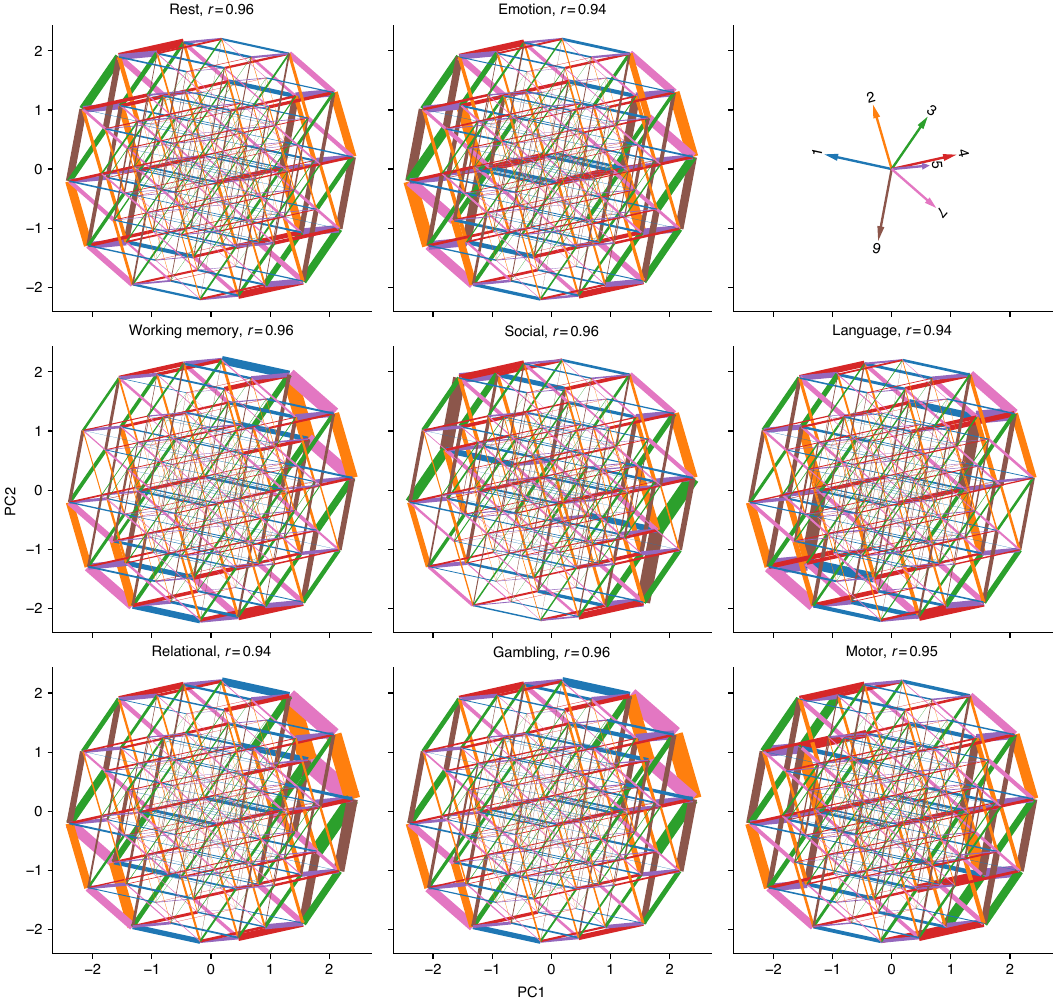}
    \caption{\textbf{Reconstructed probability traffic diagram, Arrhenius rate.}
        Same as Extended Data Fig.~10, but the
        probability traffic is
        calculated from the reconstructed interaction matrix
        $\beta \bm{J}$, reconstructed effective external input
        $\beta \bm{h}$, and reconstructed rate constant $A$ of the
        Arrhenius rate.
        We use the same PC loadings as in Extended Data
        Fig.~10 to
        visualize the probability traffic of the inferred Ising spin
        system.
        The correlation coefficient $r$ between the empirical and
        reconstructed probability traffic is written next to each
    task name.}
    \label{fig:fig-s23}
\end{figure*}

\begin{figure*}[bt]
    \centering
    \includegraphics{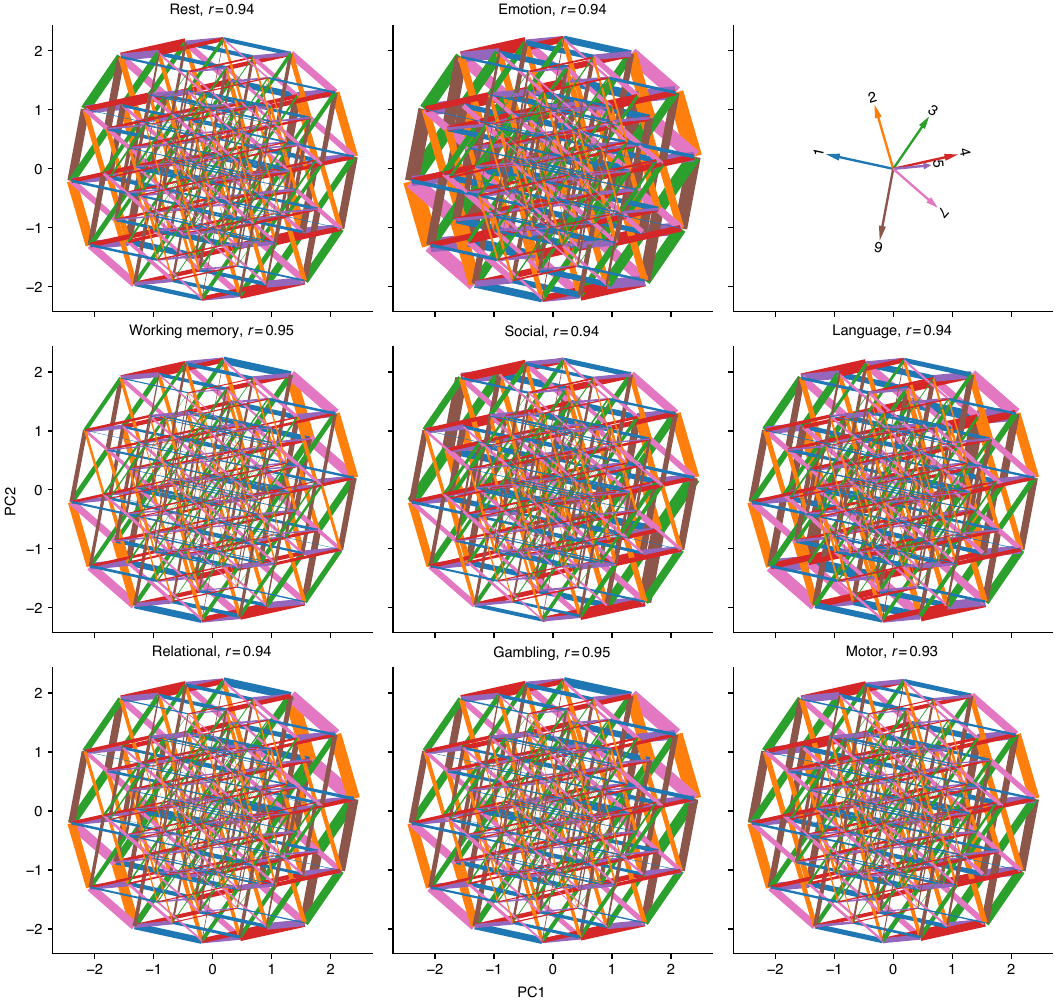}
    \caption{\textbf{Reconstructed probability traffic diagram, Glauber rate.}
        Same as Supplementary Fig.~\ref{fig:fig-s23}, but the Glauber
        rate is used
        (equation~(20)).
        Edge widths are normalized per task to the maximum empirical
        traffic to account for the scale offset between the model and
        empirical stationary distributions
    (Supplementary Note~\ref{sec:statdist}).}
    \label{fig:fig-s24}
\end{figure*}
